\documentclass{aa}

\usepackage{graphicx}
\usepackage{booktabs}
\usepackage{multirow} 
\usepackage{txfonts}
\usepackage{natbib}
\usepackage[breaklinks=true, colorlinks=true, allcolors=blue]{hyperref}
\usepackage{amsmath}	
\usepackage{amssymb}	
\usepackage{multirow}
\usepackage{longtable}
\usepackage{threeparttable}
\usepackage{afterpage}

\usepackage{placeins}
\usepackage{float}

\usepackage{multicol}

\begin{document}

   \title{NO molecule in massive star forming regions}


   \author{Shaomin Su\inst{1},
          Junzhi Wang\inst{1},
          Rui Luo\inst{1},
          Yani Xu\inst{1},
          Fu Mo\inst{1},
          Yijia Liu\inst{1},
          Chang Ruan\inst{1},
          Yingxue Kang\inst{1},
          Yuqiang Li\inst{2},
          \and
          Siqi Zheng\inst{3,4,5}
          }

   \institute{Guangxi Key Laboratory for Relativistic Astrophysics, School of Physical Science and Technology, Guangxi University, Nanning 530004, People’s Republic of China\\
              \email{junzhiwang@gxu.edu.cn}
         \and
             Korea Astronomy and Space Science Institute, No. 776, Daedeok-daero, Yuseong-gu, Daejeon, Republic of Korea
         \and
             State Key Laboratory of Radio Astronomy and Technology, Shanghai Astronomical Observatory, Chinese Academy of Sciences, 80 Nandan Road, Shanghai 200030, China
         \and
             I. Physikalisches Institut, Universit\"{a}t zu K\"{o}ln, Z\"{u}lpicher Str. 77, 50937 K\"{o}ln, Germany
         \and
             School of Astronomy and Space Sciences, University of Chinese Academy of Sciences, No. 19A Yuquan Road, Beijing 100049, People's Republic of China
             }

 \date{Received xx; accepted xxx}

 \authorrunning{Su et al.}


  \abstract
   {Among diatomic molecules composed of the abundant elements C, N and O, NO has been detected far less than the well-studied CN and CO, making it a crucial yet under-observed component in nitrogen-containing chemical networks.
   NO was thought to serve as a potential tracer of shocks, as evidenced with orders  abundance enhancements reported in literature.
   }
   {Large-sample observations for NO molecule in widespread interstellar environments are needed to confirm if  the enhancement of NO is due to shock chemistry or not.
   }
   {Single-point survey for NO $J$ = 3/2 - 1/2 lines around 150 GHz was carried out by Arizona Radio Observatory 12-meter telescope towards a sample of 36 massive star forming regions containing SiO $J$ = 5 - 4 emission, which include three evolutionary stages: 4 IRDCs, 6 protostars and 26 H II regions.}
   {The NO emission was detected in 28 sources with a detection rate of 78\%. Beam-averaged NO column densities and abundances relative to H\textsubscript{2} were derived from integrated intensities of $F$ = 5/2$^+$ - 3/2$^-$ and $F$ = 5/2$^-$ - 3/2$^+$ lines, yielding ranges of $1.5 \times 10^{14}$ cm$^{-2}$ – $3.2 \times 10^{15}$ cm$^{-2}$ and $1.9 \times 10^{-9}$ – $3.2 \times 10^{-8}$, with corresponding medians of $6.0\times 10^{14}$ cm$^{-2}$ and $6.5 \times 10^{-9}$, respectively. Correlations between NO and SiO in integrated intensity and relative abundance (r = 0.63 and 0.61) are similar to the corresponding correlations of $c$-C$_3$H$_2$ with NO (r = 0.70) and SiO (r = 0.66), indicating that NO enrichment may not significantly involve pronounced shock activities, which coincides with the trend in line widths: NO $\approx$ $c$-C$_3$H$_2$ \textless H$_2$CO $\ll$ SiO.}
   {Observational evidence does not strongly support significant NO enhancement by shock chemistry in the observed sources, indicating that the formation of NO does  not necessarily require shocks.
   }


   \maketitle
\nolinenumbers

\section{Introduction} \label{sec:intro}

C, N and O are highly abundant elements after H and He, which can form simple diatomic molecules: carbon monoxide (CO), cyano radical (CN) and nitric oxide (NO).
In contrast to the abundant CO \citep{1970ApJ...161L..43W} and CN \citep{1970ApJ...161L..87J,1993A&A...276..473F}, 
detections of NO remain rarer. 
As the simplest molecule with N-O bond, NO is related to atomic (N) and molecular nitrogen (N$_2$), which lack transitions and cannot be detected directly. NO abundance is mostly determined by neutral-neutral reactions: N + OH $\to$ NO + H produces NO molecules while N + NO $\to$ N$_2$ + O consumes NO molecules \citep{1973ApJ...185..505H, 2010A&A...513A..41H}. Since the reaction N + OH → NO + H can proceed without an activation barrier \citep{1980CPL....69...40H, 2021MNRAS.508.1908B}, the abundance of NO molecule is principally constrained by the amount of nitrogen available for reaction in the gas phase \citep{2018MNRAS.474.5694C}. Another reaction, O + NH $\to$ NO + H, also contributes in part \citep{1973ApJ...185..505H}. Therefore, NO molecule plays a vital role in the chemistry of the interstellar medium (ISM).

After the first detection of hyperfine structure (hfs) lines in Sgr B2 \citep{1978ApJ...224L..73L}, NO radical has  been detected in a variety of  environments, including photon dominated region Orion Bar II \citep{1995A&A...303..541J}, molecular clouds \citep{1986ApJS...60..357B,1990ApJ...359..121M,1991ApJ...373..535Z,1992A&A...266..463G,1993A&A...268..212G,2001ApJS..132..127N,2007A&A...462..221A,2018ApJ...860..145T}, circumstellar envelopes \citep{2013A&A...560L...2Q,2015A&A...575A..84V}, protostellar shock L1157-B1\citep{2018MNRAS.474.5694C}, comet P/Halley \citep{1988ehsc.conf.....G} and extragalactic source NGC 253 \citep{2003A&A...411L.465M}. 
NO molecule is an important participant in the N-bearing chemical network, as demonstrated from starless cores L 1544, L 183, L 1517B, L 310 and Oph D \citep{2010A&A...513A..41H}. Together with $c$-C$_3$H$_2$, NO molecule was simultaneously identified in 16 Galactic edge clouds with distances as far as $R_{\rm GC}$ $\sim$ 23.5 kpc \citep{2025ApJ...986..122K}, which enriched the number of molecular clouds known to contain NO. 

As NO abundance in the protostellar outflow shock L1157-B1 was found to be higher than that in the Class I object SVS13-A envelope \citep{2018MNRAS.474.5694C}, it was thought that NO molecule might be enriched by shocks. 
Since SiO molecule is considered as a good tracer of shock activities in the process of stellar evolution \citep{1992A&A...254..315M}, observations of NO molecule towards the star forming regions (SFRs) with strong SiO emission are required to investigate whether the enhancement of NO formation is related to shocks.

In this paper, observational details are described in Section \ref{sec:obs}. Results of data reduction and derived parameters are given in Section \ref{sec:results}. Section \ref{sec:discussion} presents the discussion and Section \ref{sec:Summary} briefly outlines conclusions.


\section{Observations and data reduction} \label{sec:obs}

\subsection{The sample}
36 massive star formation regions, including 4 infrared dark clouds (IRDCs), 6 protostellar candidates and 26 young H II regions, were selected to perform single-point observations of NO lines based on the criteria of strong SiO $J$ = 5 - 4 emission with $T_{\rm mb} > 0.1$ K using the Submillimeter Telescope (SMT) 10 m and Caltech Submillimeter Observatory (CSO) 10.4 m telescopes \citep{2019ApJ...878...29L}, which is expected to increase the detection rate. 

\subsection{Observations}

The Arizona Radio Observatory (ARO) 12 m telescope, with 2 mm band receiver and AROWS mode 2, was used to conduct observations in 2023 and 2024.
Spectral resolution sections of 0.16 MHz ($\sim$ 0.31 km s$^{-1}$ at 150.5 GHz) and 0.31 MHz ($\sim$ 0.62 km s$^{-1}$ at 150.5 GHz) were adopted, providing bandwidths of 1000 MHz and 2000 MHz to observe 20 and 16 sources, respectively.
The beam size at 150 GHz is $\sim 42^{\prime\prime}$. 
The pointing was checked on a nearby quasar or planet and the focus was checked at the beginning of each observing block. The standard position switching mode was employed with 1min on-source time and 1min off-source time, which typically repeated 40 times for 23 sources, 20 times for 4 sources and 19 times for 2 sources.
The rms level of about 0.01 K 
at 156.25 kHz channel spacing with $T_{\rm sys}$ of 150 K was guaranteed. 
The main-beam brightness temperature was converted from $T_{\rm mb} = T_{\rm A}^{*} / \eta$, where $T_{\rm A}^{*}$ is the antenna temperature and $\eta$ is the average beam efficiency of the 2 mm receiver with a value of 81\% ± 5\% in 2021.

Transitions of NO molecule around 150 GHz are shown in Table \ref{NO lines}, in which the first five spectral lines belong to $\Pi$$^+$ series while the remaining ones belong to $\Pi$$^-$ series. 
Spectroscopic parameters of NO transitions are extracted from Cologne Database for Molecular Spectroscopy (CDMS)\footnote{\url{https://cdms.astro.uni-koeln.de/classic/}}. Upper level energy of NO around 150 GHz is about 7.2 K.
Relative intensities under local thermodynamic equilibrium (LTE) conditions are extracted from \cite{1992A&A...266..463G}. 

The three strongest lines, i.e., $F = 5/2-3/2(+-)$, $F = 5/2-3/2(-+)$ and $F = 3/2-1/2(-+)$ 
, were used for analysis. 

\begin{table}\tiny
\caption{NO $J$ = 3/2 - 1/2, $\Omega$ = 1/2 lines around 150 GHz.}
\centering
\begin{minipage}{\textheight}
\begin{tabular}{cccccc}
\hline\hline
Frequency(GHz)$^1$ & Quantum$^1$   & Log$_{10}$A$_{ij}$$^1$  & $g_{\rm u}$$^1$ & Relative intensity$^2$ \\ 
\hline
150.17648$^*$                        &  $F$ = 5/2$^-$ - 3/2$^+$                        & -6.4802                          & 6 & 0.500 \\
150.19876$^*$                        &  $F$ = 3/2$^-$ - 1/2$^+$                        & -6.7355                          & 4 & 0.186 \\
150.21873                        &  $F$ = 3/2$^-$ - 3/2$^+$                        & -6.8323                          & 4 & 0.148 \\
150.22566                        &  $F$ = 1/2$^-$ - 1/2$^+$                        & -6.5313                           & 2 & 0.148 \\
150.24564                        &  $F$ = 1/2$^-$ - 3/2$^+$                        & -7.4343                         & 2 & 0.018 \\
150.37530                        &  $F$ = 1/2$^+$ - 3/2$^-$                        & -7.4319                         & 2 & 0.018 \\
150.43912                        &  $F$ = 3/2$^+$ - 3/2$^-$                        & -6.8297                           & 4 & 0.148 \\
150.54652$^*$                    &  $F$ = 5/2$^+$ - 3/2$^-$                        &  -6.4771                          & 6 & 0.500 \\
150.58056                        & $F$ = 1/2$^+$ - 1/2$^-$                        &  -6.5282                          & 2 & 0.148 \\
150.64434                        &  $F$ = 3/2$^+$ - 1/2$^-$                        & -6.7321                          & 4 & 0.186 \\
\hline
\end{tabular}

    \end{minipage}
    
\tablefoot{$^*$ Selected for analysis.}

\tablebib{(1)CDMS (2)\cite{1992A&A...266..463G}.}


\label{NO lines}
\end{table}

\subsection{Data reduction}

Data were reduced with the CLASS software in the GILDAS package. 
All scans in each source were averaged using the CLASS programme. A first-order baseline was used to all spectral lines. Based on single-component Gaussian fitting, parameters of each line are obtained, including central velocity, line width, etc. 
Simultaneously obtained lines of H$_2$CO 2$_{11}$-1$_{10}$ at 150.498342 GHz and $c$-C$_3$H$_2$ 4$_{14}$-3$_{03}$ at 150.851899 GHz were also used for scientific analysis, including for confirming the detection of NO lines with comparison of central velocity.


\section{Results} \label{sec:results}
\begin{figure}[!h]
\centering
\includegraphics[width=0.45\textwidth]{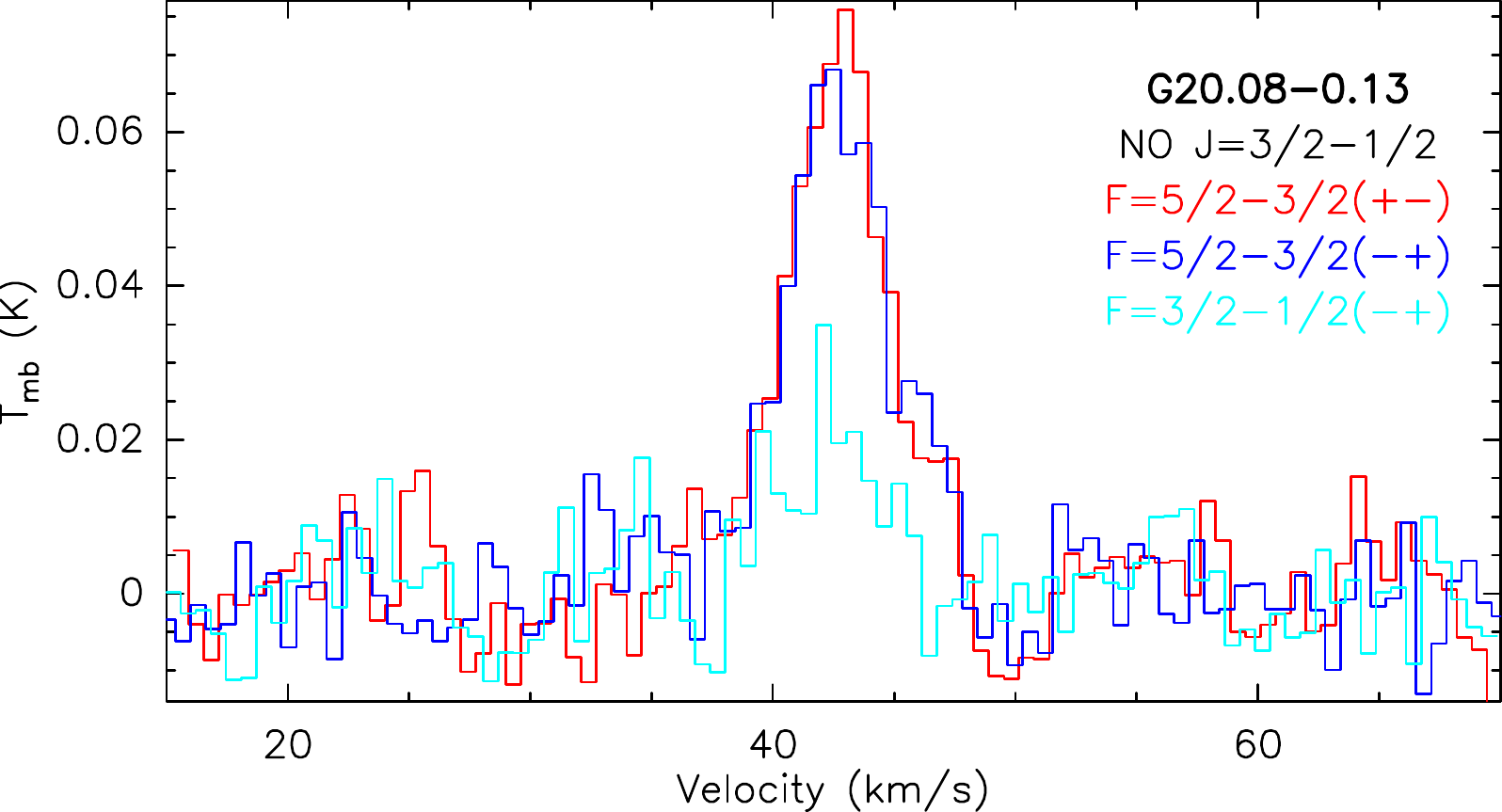}
\includegraphics[width=0.45\textwidth]{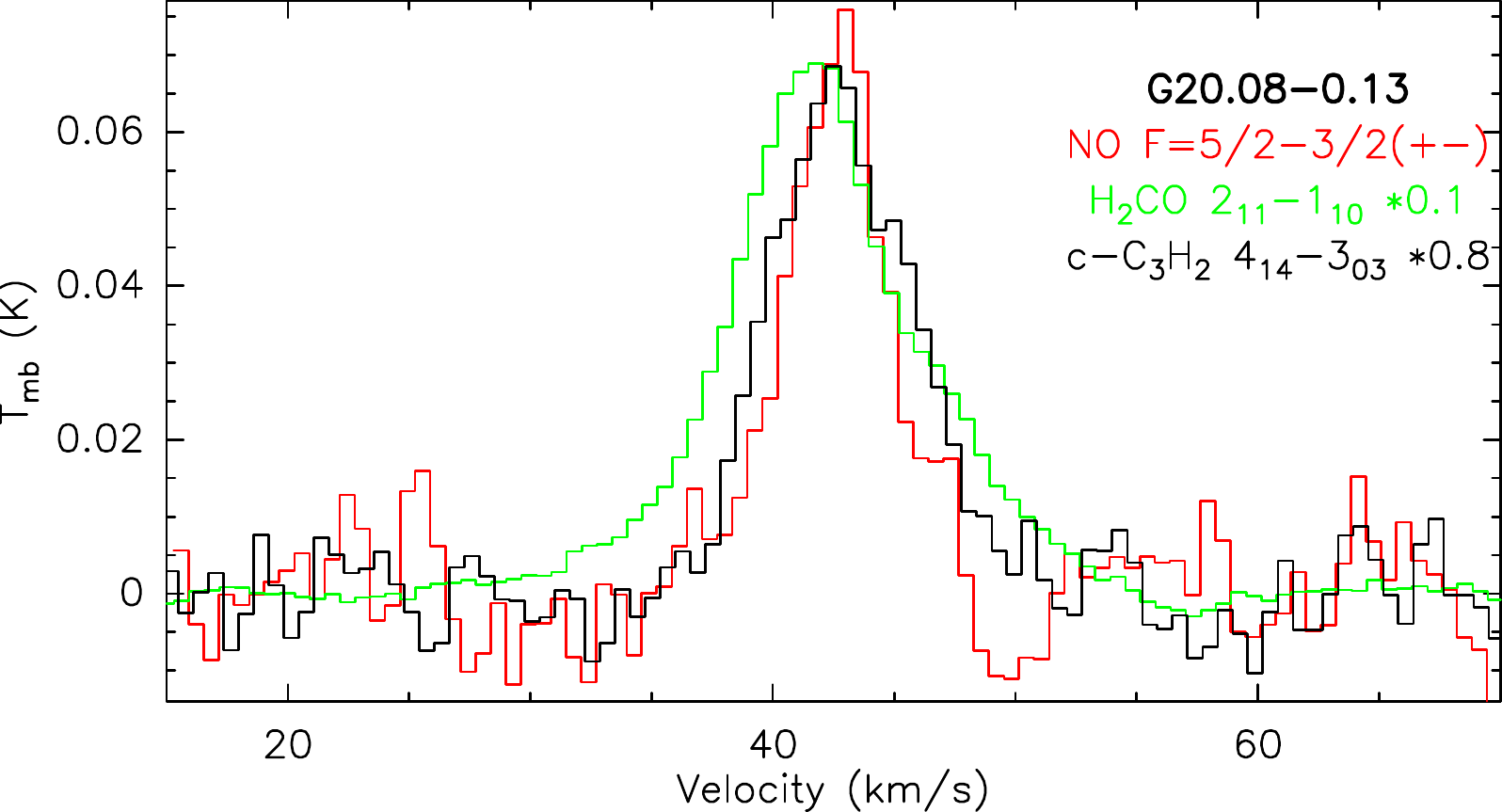}
\caption{Detected NO, $c$-C$_3$H$_2$ and H$_2$CO lines in G20.08-0.13 as an example.}
\label{NO_C3H2_H2CO_G20.08-0.13}
\end{figure}

Detection rates for NO, $c$-C$_3$H$_2$ and H$_2$CO lines are 78\%, 92\% and 100\% in these 36 sources, respectively.
Spectral line parameters from data reduction such as line central velocity ($V_{\rm LSR}$) and velocity integrated intensity ($\int T_{\rm mb}{\rm d}v$) are listed in Table \ref{parameters}, covering three NO $J$ = 3/2-1/2 spectral lines in Figure \ref{spectra of NO}. 
Figure \ref{NO_C3H2_H2CO_G20.08-0.13} demonstrates spectra of three detected species towards source G20.08-0.13 as an example, where H$_2$CO exhibits the strongest spectral line, $c$-C$_3$H$_2$ shows an intermediate intensity and NO presents the weakest signal.

\begin{table*}\tiny

    \centering
    \renewcommand{\arraystretch}{1}  
    \rotatebox{90}{
    \begin{minipage}{\textheight}  
    \centering
    \caption{Sources with NO detections.}  
    \label{parameters}
    \resizebox{\textheight}{!}{  
    
\begin{tabular}{ccccccccccc}
\hline\hline
\multirow{3}{*}{No} & \multirow{3}{*}{Source} & R.A.                        & Decl.                       & \multicolumn{2}{c}{$F$ = 5/2$^+$ - 3/2$^-$}         & \multicolumn{2}{c}{$F$ = 5/2$^-$ - 3/2$^+$}         & \multicolumn{2}{c}{$F$ = 3/2$^-$ - 1/2$^+$}         & Line            \\
\cmidrule(lr){5-6} \cmidrule(lr){7-8} \cmidrule(lr){9-10}
                    &                         & \multirow{2}{*}{(hh:mm:ss)} & \multirow{2}{*}{(dd:mm:ss)} & $\int T_{\rm mb}{\rm d}v$ & $V_{\rm LSR}$   & $\int T_{\rm mb}{\rm d}v$ & $V_{\rm LSR}$   & $\int T_{\rm mb}{\rm d}v$ & $V_{\rm LSR}$   & intensity        \\
                    &                         &                             &                             & (K·km\,s$^{-1}$)          & (km\,s$^{-1}$)  & (K·km\,s$^{-1}$)          & (km\,s$^{-1}$)  & (K·km\,s$^{-1}$)          & (km\,s$^{-1}$)  & ratio           \\
\hline
1  & G121.30+0.66      & 00:36:47.5 & +63:29:02 & 0.29 $\pm$ 0.02 & -17.4 $\pm$ 0.1 & 0.34 $\pm$ 0.02 & -17.3 $\pm$ 0.1 & 0.18 $\pm$ 0.03 & -17.5 $\pm$ 0.3 & 0.61  $\pm$ 0.11 \\
2  & G123.07-6.31      & 00:52:25.2 & +56:33:53 & $\leq$ 0.09 & -30.1 $\pm$ 0.1 & 0.08 $\pm$ 0.01 & -29.7 $\pm$ 0.1 & $\cdots$        & $\cdots$        & $\cdots$         \\
3  & G133.6945+01.2166 & 02:25:30.0 & +62:06:20 & 0.31 $\pm$ 0.03 & -42.7 $\pm$ 0.2 & 0.33 $\pm$ 0.02 & -42.7 $\pm$ 0.1 & $\leq$ 0.07     & -42.8 $\pm$ 0.9 & $\cdots$         \\
4  & G133.7150+01.2155 & 02:25:40.0 & +62:05:52 & $\leq$ 0.09 & -39.3 $\pm$ 0.3 & 0.16 $\pm$ 0.03 & -39.0 $\pm$ 0.3 & $\cdots$        & $\cdots$        & $\cdots$         \\
5  & S231              & 05:39:12.9 & +35:45:54 & 0.30 $\pm$ 0.02 & -16.7 $\pm$ 0.1 & 0.26 $\pm$ 0.02 & -16.5 $\pm$ 0.1 & $\cdots$        & $\cdots$        & $\cdots$         \\
6  & S255              & 06:12:53.7 & +17:59:22 & 0.18 $\pm$ 0.03 & 7.2 $\pm$ 0.2   & 0.23 $\pm$ 0.03 & 7.3 $\pm$ 0.2   & $\cdots$        & $\cdots$        & $\cdots$         \\
7  & G5.89-0.39        & 18:00:30.3 & -24:04:00 & 0.46 $\pm$ 0.02 & 9.1 $\pm$ 0.1   & 0.46 $\pm$ 0.02 & 9.2 $\pm$ 0.1   & 0.12 $\pm$ 0.02 & 8.5 $\pm$ 0.2   & 0.26  $\pm$ 0.05 \\
8  & G10.6-0.4         & 18:10:28.7 & -19:55:48 & 0.71 $\pm$ 0.03 & -2.4 $\pm$ 0.1  & 0.58 $\pm$ 0.03 & -2.4 $\pm$ 0.2  & 0.29 $\pm$ 0.04 & -3.2 $\pm$ 0.5  & 0.41  $\pm$ 0.06 \\
9  & W33IRS3           & 18:14:13.3 & -17:55:40 & 0.83 $\pm$ 0.02 & 35.6 $\pm$ 0.1  & 0.72 $\pm$ 0.02 & 35.6 $\pm$ 0.1  & 0.23 $\pm$ 0.02 & 34.9 $\pm$ 0.3  & 0.28  $\pm$ 0.03 \\
10 & I18182-1433MM1    & 18:21:09.0 & -14:31:57 & 0.26 $\pm$ 0.01 & 59.1 $\pm$ 0.1  & 0.25 $\pm$ 0.02 & 59.5 $\pm$ 0.1  & 0.09 $\pm$ 0.02 & 58.8 $\pm$ 0.3  & 0.34  $\pm$ 0.07 \\
11 & G019.27+00.07MM1  & 18:25:58.0 & -12:03:59 & 0.23 $\pm$ 0.04 & 26.2 $\pm$ 0.3  & 0.17 $\pm$ 0.02 & 25.9 $\pm$ 0.2  & $\cdots$        & $\cdots$        & $\cdots$         \\
12 & G19.61-0.23       & 18:27:38.0 & -11:56:28 & 0.50 $\pm$ 0.03 & 42.7 $\pm$ 0.3  & 0.40 $\pm$ 0.02 & 43.1 $\pm$ 0.1  & 0.13 $\pm$ 0.02 & 43.3 $\pm$ 0.4  & 0.25  $\pm$ 0.04 \\
13 & G20.08-0.13       & 18:28:10.0 & -11:28:48 & 0.37 $\pm$ 0.02 & 42.6 $\pm$ 0.1  & 0.35 $\pm$ 0.02 & 42.7 $\pm$ 0.1  & 0.10 $\pm$ 0.02 & 42.4 $\pm$ 0.3  & 0.29  $\pm$ 0.04 \\
14 & G022.35+00.41MM1  & 18:30:24.0 & -09:10:34 & 0.19 $\pm$ 0.02 & 52.5 $\pm$ 0.1  & 0.15 $\pm$ 0.01 & 53.0 $\pm$ 0.1  & 0.08 $\pm$ 0.01 & 53.6 $\pm$ 0.1  & 0.42  $\pm$ 0.06 \\
15 & G023.60+00.00MM1  & 18:34:12.0 & -08:19:06 & 0.23 $\pm$ 0.01 & 106.1 $\pm$ 0.1 & 0.29 $\pm$ 0.02 & 106.2 $\pm$ 0.1 & 0.12 $\pm$ 0.01 & 106.7 $\pm$ 0.2 & 0.50  $\pm$ 0.06 \\
16 & G24.49-0.04       & 18:36:05.3 & -07:31:23 & 0.26 $\pm$ 0.03 & 110.3 $\pm$ 0.2 & 0.18 $\pm$ 0.02 & 110.7 $\pm$ 0.2 & 0.14 $\pm$ 0.02 & 109.9 $\pm$ 0.4 & 0.54  $\pm$ 0.08 \\
17 & W43S              & 18:46:04.0 & -02:39:26 & 0.28 $\pm$ 0.02 & 97.5 $\pm$ 0.2  & 0.31 $\pm$ 0.02 & 97.8 $\pm$ 0.1  & $\leq$ 0.06 & 97.8 $\pm$ 0.3  & $\cdots$ \\
18 & G31.41+0.31       & 18:47:35.0 & -01:12:46 & 0.83 $\pm$ 0.03 & 97.1 $\pm$ 0.1  & 0.53 $\pm$ 0.02 & 98.0 $\pm$ 0.1  & 0.24 $\pm$ 0.02 & 97.7 $\pm$ 0.2  & 0.28  $\pm$ 0.02 \\
19 & W43Main3          & 18:47:47.0 & -01:54:35 & 0.91 $\pm$ 0.02 & 97.6 $\pm$ 0.1  & 0.81 $\pm$ 0.02 & 98.3 $\pm$ 0.1  & 0.24 $\pm$ 0.02 & 97.6 $\pm$ 0.2  & 0.27  $\pm$ 0.02 \\
20 & G34.26+0.15       & 18:53:18.5 & +01:14:57 & 0.86 $\pm$ 0.01 & 57.9 $\pm$ 0.1  & 0.73 $\pm$ 0.01 & 58.5 $\pm$ 0.1  & 0.25 $\pm$ 0.02 & 58.7 $\pm$ 0.2  & 0.29  $\pm$ 0.02 \\
21 & G35.20-0.74       & 18:58:12.7 & +01:40:36 & 0.37 $\pm$ 0.02 & 33.8 $\pm$ 0.1  & 0.34 $\pm$ 0.02 & 33.8 $\pm$ 0.1  & 0.17 $\pm$ 0.01 & 34.0 $\pm$ 0.2  & 0.46  $\pm$ 0.05 \\
22 & W51D              & 19:23:39.9 & +14:31:06 & 0.79 $\pm$ 0.09 & 61.4 $\pm$ 0.6  & 0.76 $\pm$ 0.07 & 62.3 $\pm$ 0.4  & $\leq$ 0.10 & 60.3 $\pm$ 0.4  & $\cdots$ \\
23 & W51M              & 19:23:43.8 & +14:30:29 & 1.35 $\pm$ 0.04 & 54.7 $\pm$ 0.1  & 1.50 $\pm$ 0.05 & 56.6 $\pm$ 0.2  & 0.55 $\pm$ 0.02 & 56.1 $\pm$ 0.2  & 0.41  $\pm$ 0.02 \\
24 & K3-50A            & 20:01:45.6 & +33:32:42 & $\leq$ 0.07     & -25.8 $\pm$ 0.6 & 0.17 $\pm$ 0.02 & -23.8 $\pm$ 0.5 & $\cdots$        & $\cdots$        & $\cdots$         \\
25 & G073.0633+01.7958 & 20:08:10.0 & +35:59:24 & 0.08 $\pm$ 0.01 & 0.7 $\pm$ 0.1   & 0.09 $\pm$ 0.01 & 0.8 $\pm$ 0.1   & $\cdots$        & $\cdots$        & $\cdots$         \\
26 & ON1               & 20:10:09.1 & +31:31:37 & 0.38 $\pm$ 0.02 & 11.3 $\pm$ 0.1  & 0.50 $\pm$ 0.02 & 11.4 $\pm$ 0.1  & 0.15 $\pm$ 0.02 & 10.8 $\pm$ 0.2  & 0.40  $\pm$ 0.05 \\
27 & 20126+4104        & 20:14:26.0 & +41:13:32 & $\leq$ 0.07 & -3.9 $\pm$ 0.2  & $\leq$ 0.05     & -3.9 $\pm$ 0.2  & $\leq$ 0.08     & -3.8 $\pm$ 0.2  & $\cdots$         \\
28 & ON2S              & 20:21:41.0 & +37:25:29 & $\leq$ 0.08 & -1.7 $\pm$ 0.3  & 0.12 $\pm$ 0.02 & -1.9 $\pm$ 0.3  & $\cdots$        & $\cdots$        & $\cdots$         \\
29 & ON2N              & 20:21:43.9 & +37:26:39 & 0.08 $\pm$ 0.01 & 1.2 $\pm$ 0.4   & 0.17 $\pm$ 0.01 & 0.8 $\pm$ 0.2   & $\cdots$        & $\cdots$        & $\cdots$         \\
30 & G083.7962+03.3058 & 20:33:48.0 & +45:40:54 & $\leq$ 0.14     & -4.4 $\pm$ 0.4  & $\leq$ 0.20     & -4.4 $\pm$ 0.1  & $\cdots$        & $\cdots$        & $\cdots$         \\
31 & G081.7133+00.5589 & 20:39:02.0 & +42:21:58 & 0.38 $\pm$ 0.02 & -3.5 $\pm$ 0.1  & 0.32 $\pm$ 0.02 & -2.9 $\pm$ 0.1  & $\leq$ 0.08 & -3.2 $\pm$ 0.3  & $\cdots$ \\
32 & G081.7624+00.5916 & 20:39:03.0 & +42:25:29 & 0.40 $\pm$ 0.02 & -4.0 $\pm$ 0.1  & 0.32 $\pm$ 0.02 & -4.2 $\pm$ 0.1  & $\cdots$        & $\cdots$        & $\cdots$         \\
33 & G094.4637-00.8043 & 21:35:09.0 & +50:53:09 & $\leq$ 0.07 & -44.7 $\pm$ 0.6 & $\leq$ 0.07 & -44.5 $\pm$ 0.3 & $\cdots$        & $\cdots$        & $\cdots$         \\
34 & 21391+5802        & 21:40:42.4 & +58:16:10 & 0.20 $\pm$ 0.01 & 0.5 $\pm$ 0.1   & 0.16 $\pm$ 0.01 & 0.8 $\pm$ 0.1   & 0.09 $\pm$ 0.01 & 0.7 $\pm$ 0.1   & 0.42  $\pm$ 0.05 \\
35 & NGC7538           & 23:13:44.8 & +61:26:50 & $\leq$ 0.27 & -57.4 $\pm$ 0.2 & 0.27 $\pm$ 0.03 & -57.3 $\pm$ 0.2 & $\cdots$        & $\cdots$        & $\cdots$         \\
36 & NGC7538A          & 23:13:45.6 & +61:28:18 & 0.11 $\pm$ 0.01 & -57.0 $\pm$ 0.2 & 0.18 $\pm$ 0.02 & -57.1 $\pm$ 0.2 & $\leq$ 0.08 & -57.4 $\pm$ 0.3 & $\cdots$ \\

\hline
\end{tabular}

    }   
\tablefoot{$\int T_{\rm mb}{\rm d}v$: velocity integrated intensity. $V_{\rm LSR}$: line central velocity. Line intensity ratio: ratio of $F$ = 3/2$^-$ - 1/2$^+$ to $F$ = 5/2$^+$ - 3/2$^-$.} 

    \end{minipage}
    }

\end{table*}


\subsection{Detection  of NO lines} \label{subsec:detection}


Under the condition that both strong lines ($F$ = 5/2$^+$ - 3/2$^-$ and $F$ = 5/2$^-$ - 3/2$^+$) satisfy $\int T_{\rm mb}{\rm d}v$ greater than 5$\sigma_{\text{area}}$, NO $J$ = 3/2 - 1/2 emission was detected in 28 out of 36 sources, which means an overall detection rate of 78\% for this sample and is shown in Table \ref{detective rate}. 
The detected intensities of two strong transitions are in close agreement, as expected under LTE assumption. The measured intensity ratios of $F$ = 3/2$^-$ - 1/2$^+$ to $F$ = 5/2$^+$ - 3/2$^-$ are about 0.38 on average, with a median of 0.40 (see Table \ref{parameters}), which is close to the theoretical value (0.372) under LTE and optically thin conditions.


As for 8 sources with NO lines undetected, the noise level of the velocity integrated intensity is estimated as $\sigma_{\text{area}} = rms \sqrt{\Delta v \delta v}$, where $\Delta v$ is the integrated range, $\delta v$ is channel separation in velocity and $rms$ is obtained for the spectrum under velocity resolution of $\delta v$.
Only $F$ = 5/2$^-$ - 3/2$^+$ transition was detected in G123.07-6.31, G133.7150+01.2155, K3-50A, ON2S and NGC7538. Neither of the two strong transitions was detected towards 20126+4104, G083.7962+03.3058 and G094.4637-00.8043.
Poor baseline causes difficulty for scientific analysis of NO molecule in W51D, even though there are clear detections.




\subsection{H$_2$CO and c-C$_3$H$_2$}
As one tracer of dense gas, H$_2$CO 2$_{11}$-1$_{10}$ was detected in all 36 sources. The central velocity derived from NO line agrees well with that obtained from H$_2$CO, confirming its reliability in identification.
Another optically thin $c$-C$_3$H$_2$ 4$_{14}$-3$_{03}$ line was detected in 33 sources, whose integrated intensity is stronger than NO and weaker than H$_2$CO. 

\subsection{Line profiles and widths}
\label{3.3}

The spectra with detectable signals regarding NO, H$_2$CO and $c$-C$_3$H$_2$ molecules can be well fitted with single component Gaussian profile.
Full widths at half maxima (FWHMs) of three NO transitions are quite close, as shown in Table \ref{NO_FWHM}, with median values of 4.0 km\,s$^{-1}$, 3.7 km\,s$^{-1}$ and 4.5 km\,s$^{-1}$, respectively.
As line profiles of SiO 5-4, with a median full width at zero power (FWZP) of $\sim$ 25 km\,s$^{-1}$ \citep{2019ApJ...878...29L}, are significantly broader than other molecules, line widths of NO, $c$-C$_3$H$_2$, H$_2$CO and SiO molecules in each source are found to follow the order: NO $\approx$ $c$-C$_3$H$_2$ \textless H$_2$CO $\ll$ SiO, as shown in Table \ref{H2CO&C3H2} and \ref{CDA of NO}.

\subsection{Column density of NO molecule} \label{sec:cd}
The small dipole moment ($\mu$ = 0.1587 D) and low column density ($N$ $\sim$ $10^{14}$ - $10^{15}$ cm$^{-2}$) of NO molecule make its emission lines normally optically thin \citep{2009A&A...493..557L}. 
Assuming LTE, the beam-averaged upper level column density can be calculated by the following equation:
\begin{equation}
N_u = \frac{8 \pi k \nu^2}{h c^3 A_{ul}} \int T_{\mathrm{mb}} \, dv
\end{equation}
where $k$ is the Boltzmann's constant, $h$ is the Planck constant and $c$ is the speed of light.
The total column density can then be calculated from below:
\begin{equation}
N_{\text{tot}} = \frac{N_u Q(T_{\mathrm{ex}})}{g_u} e^{E_u / k T_{\mathrm{ex}}}
\end{equation}
Since theoretical relative intensities of NO $F$ = 5/2$^+$ - 3/2$^-$ and $F$ = 5/2$^-$ - 3/2$^+$ lines are identical and their measured velocity integrated intensities are close to each other, we equivalently view this as two measurements of NO molecule and average a series of constant values for calculations to improve S/N ratio. In other words, according to Table \ref{NO lines}, parameters are adopted as below: line frequency $\nu$ = 150.3615 GHz, upper level energy $E_{\rm u}$/$k$ = 7.22662 K, Einstein coefficient $A_{\rm ul}$ = $10^{-6.478715}$ and upper state degeneracy $g_{\rm u}$ = 6. Kinetic temperatures of some sources have been estimated from ammonia molecules to calculate column densities, while excitation temperatures are adopted for remaining sources based on types, i.e., $T_{\rm ex}$ = 18 K for IRDCs, $T_{\rm ex}$ = 25 K for protostars and $T_{\rm ex}$ = 30 K for H II regions \citep{2019ApJ...878...29L}. As NO is a radical with fairly large fine structure 
splitting and with hyperfine splitting, the partition function $Q$ is determined by considering the contribution of hyperfine structure levels using data packages of upper state degeneracies and upper level energies in CDMS\footnote{\url{https://cdms.astro.uni-koeln.de/classic/predictions/pickett/beispiele/NO/main.v=0/c030517-only.egy}}, covering an energy range up to 6616 cm$^{-1}$, to ensure that it corresponds accurately to the excitation temperature $T_{\mathrm{ex}}$:
\begin{equation}
Q(T_{\mathrm{ex}}) = \sum_i g_i e^{-E_i / k T_{\mathrm{ex}}}
\end{equation}


The beam-averaged NO column densities are shown in Table \ref{CDA of NO}, which range from $1.5 \times 10^{14}$ cm$^{-2}$ to $3.2 \times 10^{15}$ cm$^{-2}$. The mean value is $8.0 \times 10^{14}$ cm$^{-2}$ and the median is $6.0 \times 10^{14}$ cm$^{-2}$. These results are similar to NO measurements in other massive star forming regions \citep{2006ApJS..164..450M,2018MNRAS.474.5694C}. 

\subsection{Relative abundance of NO molecule} \label{sec:ra}
To determine  NO abundance $X$(NO), H\textsubscript{2} column density derived by \cite{2019ApJ...878...29L} from dust emission was utilized. 
The results of $X$(NO), defined as $N$(NO)/$N$(H\textsubscript{2}), are shown in Table \ref{CDA of NO}. The maximum and minimum $X$(NO) for this sample are $3.2 \times 10^{-8}$ towards G022.35+00.41MM1 and $1.9 \times 10^{-9}$ towards G133.7150+01.2155, respectively, with median value of $6.5 \times 10^{-9}$. 
Note that as to the lowest abundance with rms values of 18 mK, the derived minimum is likely constrained by the sensitivity threshold rather than an intrinsic property of the source. 
In terms of evolutionary stages, NO abundances in IRDCs with a median of $2.4 \times 10^{-8}$ are significantly higher than those in protostars ($4.0 \times 10^{-9}$) and HII regions ($6.3 \times 10^{-9}$), which are shown in Table \ref{distribution}. 
Despite the limitation imposed by the sample size, no other clear relationships between NO molecule and stellar evolution have been identified according to this work.

For comparison, \cite{1991ApJ...373..535Z} reported $X$(NO) in star forming regions ranging from $4.2 \times 10^{-9}$ towards W51M to $1.1 \times 10^{-8}$ towards Sgr B2(N) and Orion-KL, while W51M in our results shows similar abundance of $(5.35 \pm 0.11) \times 10^{-9}$. 
In pre-protostellar cores, \cite{2007A&A...462..221A} inferred $X$(NO) approximately $8 \times 10^{-9}$ at the dust peak and in the range of $3 \times 10^{-8}$ to $5 \times 10^{-8}$ at offsets towards L1544, along with $1 \times 10^{-8}$ to $3 \times 10^{-8}$ towards L183.
\cite{1995A&A...303..541J} also derived $X$(NO) $\sim$ $0.2 \times 10^{-8}$ in Orion Bar. 
All abundances listed above are consistent with the derived values in this sample within an order of magnitude.


\section{Discussion} \label{sec:discussion}


\subsection{Can shock dominate NO enhancement in molecular clouds?}

\begin{figure}[!h]
\centering
\includegraphics[width=0.45\textwidth]{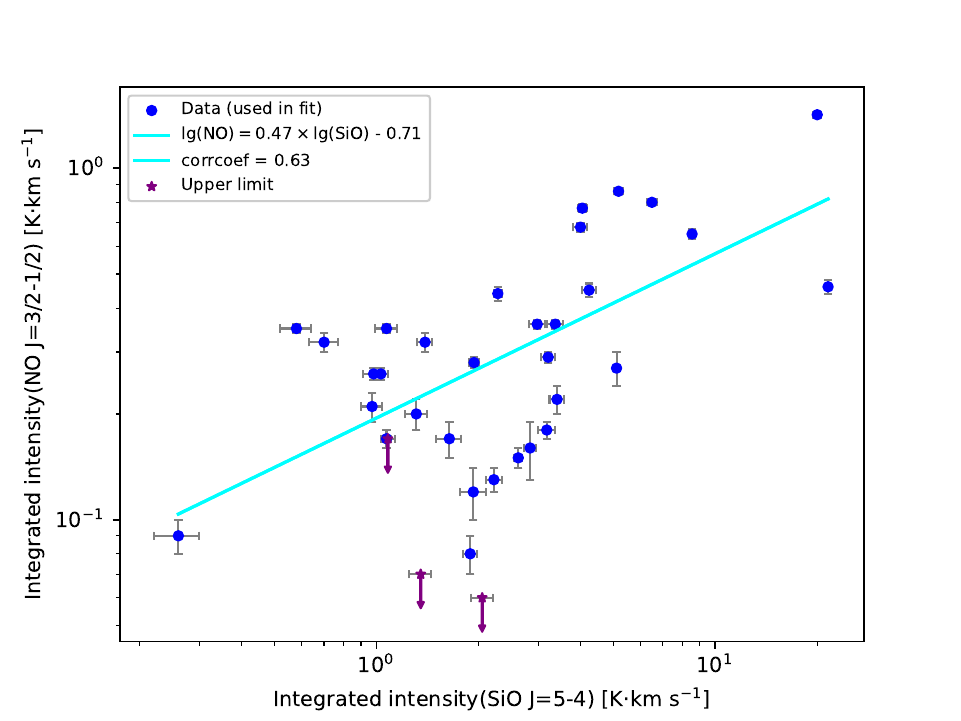}
\includegraphics[width=0.45\textwidth]{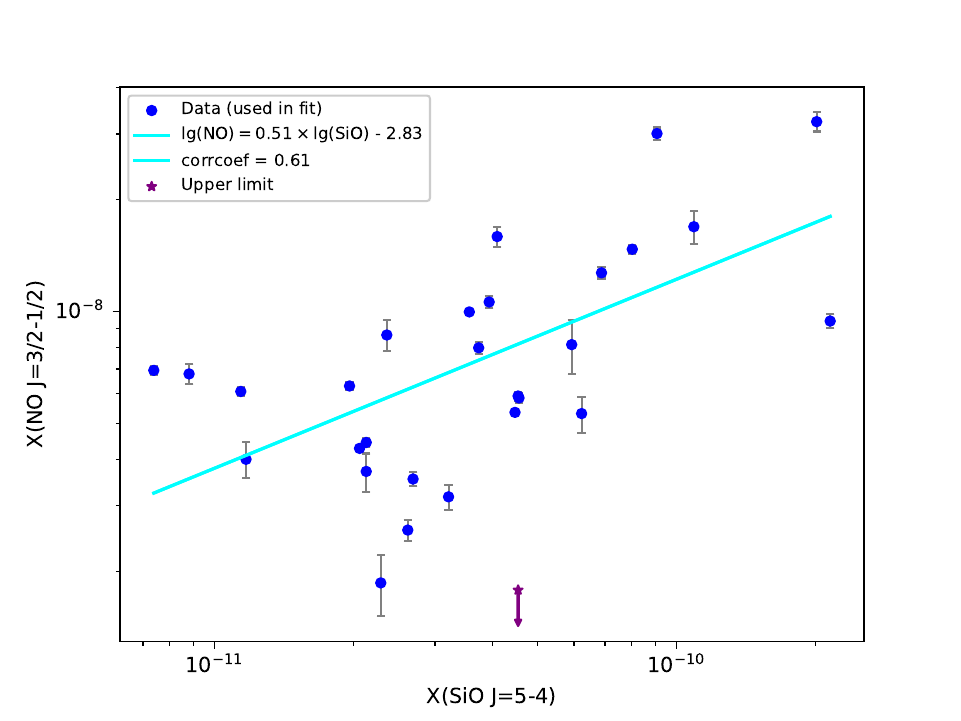}
\caption{Top: Integrated intensities between NO and SiO. Bottom: Relative abundances between NO and SiO.}
\label{flux&X}
\end{figure}

\begin{figure}[!h]
\centering
\includegraphics[width=0.45\textwidth]{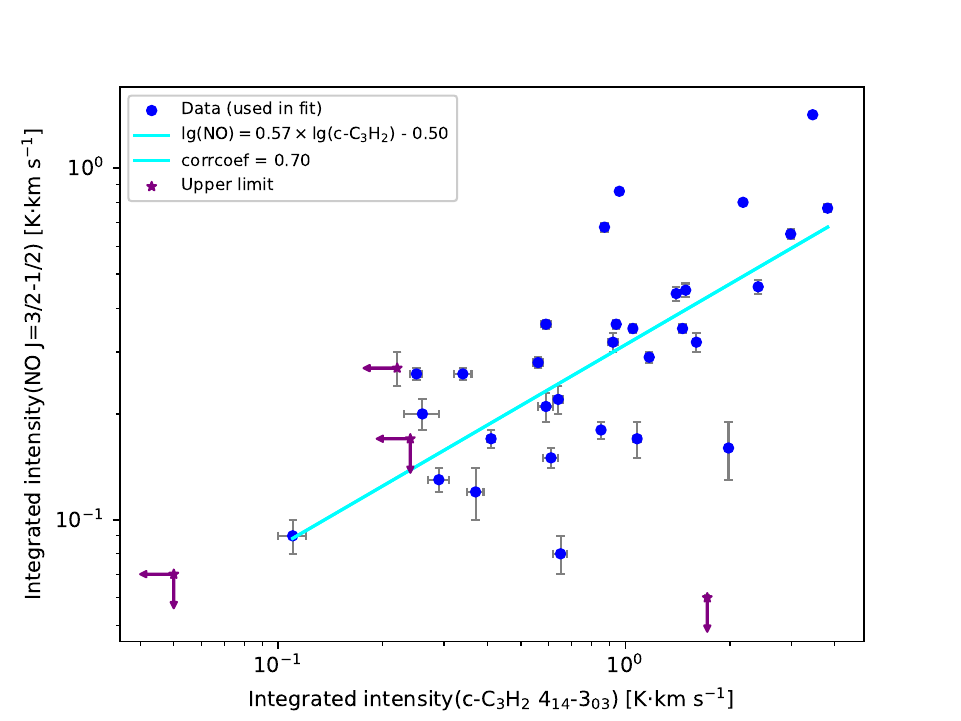}
\includegraphics[width=0.45\textwidth]{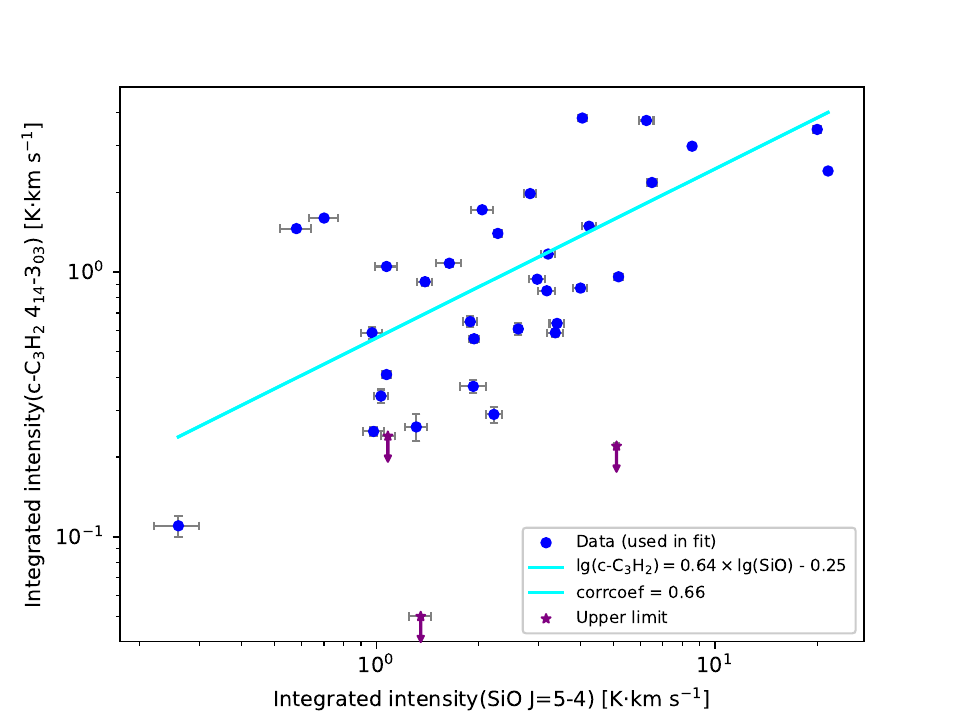}
\caption{Top: Integrated intensities between $c$-C$_3$H$_2$ and NO. Bottom: Integrated intensities between SiO and $c$-C$_3$H$_2$.}
\label{flux_C3H2}
\end{figure}

It was thought that NO molecule could be significantly enhanced in shocked regions \citep{2018MNRAS.474.5694C}, while SiO molecule serves as an excellent tracer for interstellar shocks. 
Thus, it is worth evaluating the connection between NO and SiO in molecular clouds. 
As shown at the top of Figure \ref{flux&X}, integrated intensity of NO $J$= 3/2 -1/2 exhibits a moderate relationship with SiO $J$ = 5-4, with a correlation coefficient of 0.63.  
A correlation coefficient of 0.61 is found in their relative abundances with respect to H$_2$, which is displayed at the bottom of Figure \ref{flux&X}. However, such relation may not be caused by the astro-chemical link between NO and SiO molecules, since similar trends can be found in integrated intensity for $c$-C$_3$H$_2$ with NO (r = 0.70) and with SiO (r = 0.66), shown in Figure \ref{flux_C3H2}.



The chemistry of NO behaves differently from that of SiO.
Under conditions where neutral–neutral reactions dominate the formation and destruction of NO molecules \citep{1992A&A...266..463G, 2007A&A...462..221A}, the efficiencies of these processes do not require elevated temperatures generated by shocks, since such reactions exhibit only a weak temperature dependence $k$ $\propto$ $T^{\rm 1/6}$ \citep{1973ApJ...185..505H}.
The essentially barrierless \citep{2011Sci...334.1538D, 2021MNRAS.508.1908B} NO-forming reaction, N + OH $\to$ NO + H, has a very low activation energy ($E_{A1}$) of 6 K according to KIDA datasheets\footnote{\url{https://kida.astrochem-tools.org}}, which implies that at typical cold cloud temperatures ($\sim$10 K), a sufficient number of reactant particles can readily acquire the necessary energy and surmount the minute barrier via thermal activation.
Indeed, N and OH can directly combine in the ground state to form a stable NOH intermediate with no activation energy, in which case the energy released in this step is enough to overcome the subsequent several kcal/mol barrier for the dissociation of NOH into NO + H \citep{1993CP....174...71P}. Furthermore, quantum tunneling may also contribute to enhancing the reaction rate in cold interstellar environments, as is expected for such reactions to bypass the classical activation barrier even in the absence of thermal energy. In contrast, the NO-consuming reaction N + NO $\to$ N$_2$ + O has a higher activation energy ($E_{A2}$) of 20 K, which is likely to be more strongly suppressed at low temperatures. Over the 10 - 300 K temperature range relevant to cold molecular clouds, the rate coefficient for N + OH remains consistently higher than that for N + NO, with the difference being most pronounced at the lowest temperatures and narrowing to near parity around 30 K, as given by the preferred values in KIDA. Taken together, these considerations suggest that NO can be produced and maintained even within quiescent cold envelopes, which does not necessarily involve shocks as a dominant mechanism, although shocks may still influence its abundance under certain conditions.

For most of the molecular clouds, such as massive SFRs detected in this work with a median abundance of $6.5 \times 10^{-9}$, observational results do not require  significant enrichment of NO molecule by shock activities, despite the presence of strong SiO emission. Theoretically, NO can also be efficiently formed in the normal warm gas phase without shocks, from the perspective of activation energy.

However, in particular astronomical environments, shock processes are able to significantly enrich NO molecule by certain means, such as reported in \cite{2018MNRAS.474.5694C}, with one order  higher abundance towards the low-mass L1157-B1 shock ($4 \text{-} 7 \times 10^{-6}$) as well as that tentatively derived for the SVS13-A outflow ($0.1 \text{-} 5 \times 10^{-7}$), suggesting the enhancement of NO production possibly via direct release from dust mantles in shock-dominated regions.


\subsection{Emission distribution from line profiles and constraints on shock enhancement}

In this sample, the emission lines of NO, $c$-C$_3$H$_2$ and H$_2$CO exhibit Gaussian profiles, whereas SiO does not. These four molecules can be ranked by line width as follows: NO $\approx$ $c$-C$_3$H$_2$ \textless H$_2$CO $\ll$ SiO, where SiO stands out with a significantly broader velocity range.
These clear differences suggest that NO, $c$-C$_3$H$_2$ and H$_2$CO (FWHM $\leq$ 10 km\,s$^{-1}$) probably originate from the same emission region within the molecular cloud, whereas SiO (FWZP $\geq$ 20 km\,s$^{-1}$) has roots in strong shocks or turbulence driven by outflows, which is distinct from the other three molecules.
Spectra in Figure \ref{spectra of 3 molecules} also demonstrate that the central velocities of NO, $c$-C$_3$H$_2$ and H$_2$CO emissions are in close agreement.
Therefore, most of NO molecule should  be produced under normal molecular cloud conditions, unlike SiO which requires shocks.

An enhancement of approximately two orders of magnitude in NO abundance attributed to shock activity was reported based on IRAM 30 m observations of L1157-B1 \citep{2018MNRAS.474.5694C}. To assess whether the absence of such enhancement in the present sample could be attributed to beam dilution, we consider the following estimate. Assuming the shock region is compact relative to both beams, the enhancement factor scales with the beam filling factor, with the ratio of beam areas between ARO 12 m and IRAM 30 m being $(12/30)^2 \approx 0.16$. Under the assumption that the intrinsic enhancement is $f = 100$, the same shock region observed with the ARO 12 m telescope would yield an apparent enhancement of only $100 \times (12/30)^2 = 16$, while NO abundances in our sample are close to that in the SVS13-A envelope as a normal molecular cloud environment \citep{2018MNRAS.474.5694C}. Even though NO abundance can be enhanced particularly in shocked regions, most of the NO molecules are not enhanced by shock chemistry in the sampled sources.

\section{Summary} \label{sec:Summary}

Considering the possibility that NO enrichment is linked to shock activities, we observed three NO lines around 150 GHz together with $c$-C$_3$H$_2$ and H$_2$CO via the ARO 12 m telescope towards a sample of 36 massive star forming regions with SiO 5-4 lines stronger than 0.1 K. Main results are summarized as follows:

1. NO $J$ = 3/2 - 1/2 was detected in 28 out of 36 sources, giving a detection rate of 78\%, where detection necessitates both strong hyperfine components $F$ = 5/2$^+$ - 3/2$^-$ and $F$ = 5/2$^-$ - 3/2$^+$ reach $\int T_{\rm mb}{\rm d}v$ $>$ 5$\sigma_{\text{area}}$. 
$c$-C$_3$H$_2$ 4$_{14}$-3$_{03}$ and H$_2$CO 2$_{11}$-1$_{10}$ were also detected in 33 and 36 sources with rates of 92\% and 100\%, respectively.

2. All detective lines can be well fitted with a single-Gaussian component, with median FWHM values of 4.0 km\,s$^{-1}$, 3.7 km\,s$^{-1}$ and 4.5 km\,s$^{-1}$ for $F$ = 5/2$^+$ - 3/2$^-$, $F$ = 5/2$^-$ - 3/2$^+$ and $F$ = 3/2$^-$ - 1/2$^+$, respectively. Line widths of three molecules increase in the order of NO $\approx$ $c$-C$_3$H$_2$ \textless H$_2$CO.

3. Beam-averaged NO column densities, as well as abundances with respect to H$_2$, are derived with medians of $6.0\times 10^{14}$ cm$^{-2}$ and $6.5 \times 10^{-9}$, respectively. 
Restricted by the sample size, there is insufficient evidence for a clear relationship between NO molecule and evolutionary stages. Only higher abundances in IRDCs ($2.4 \times 10^{-8}$) relative to protostars ($4.0 \times 10^{-9}$) and HII regions ($6.3 \times 10^{-9}$) are found in this work.

4. Based on the relation between line profiles of NO, SiO and $c$-C$_3$H$_2$, as well as derived NO abundances, shock chemistry does not significantly enhance NO molecule in the sampled sources.


\begin{acknowledgements}
      We would like to appreciate the staff at the Arizona Radio Observatory for assistance during the observations. This work is supported by National Key R\&D Program of China under grant 2023YFA1608204, the Guangxi Science and Technology Innovation Platform Program (Leitai Action Plan, Grant No. Guike LT2600640026), Guangxi Key R\&D Program (Guangxi Funeng Action Plan, Grant No. FN2504240040), and the "Guangxi Highland of Innovation Talents" Program.
\end{acknowledgements}


\bibliographystyle{aa}
\bibliography{references}



\clearpage

\appendix
\onecolumn
\section{Tables}

\begin{table}[h!]\tiny
\centering
\caption{Detection of NO molecule in 36 sources.}
\begin{tabular}{ccccccc}
\hline\hline

$F$ = 5/2$^+$ - 3/2$^-$ &  $F$ = 5/2$^-$ - 3/2$^+$ &  $F$ = 3/2$^-$ - 1/2$^+$ &  Number of sources \\
\hline

YES                     & YES                     &  NO                      &  11                \\
YES                     & YES                     & YES                     & 17                 \\
YES                     &  NO                      & NO                      &  0                \\
NO                      & YES                    &  NO                     &  5                 \\
NO                      &  NO                      &  NO                    &  3                 \\

\hline
\end{tabular}
\label{detective rate}
\end{table}

\begin{table*}[h!]\tiny
\caption{Line widths of three NO transitions in km\,s$^{-1}$.}
\centering
\begin{tabular}{ccccc}
\hline\hline
Transition        & FWHM$_{min}$ & FWHM$_{max}$ & FWHM$_{average}$ & FWHM$_{median}$ \\
\hline
$F$ = 5/2$^+$ - 3/2$^-$ & 1.4          & 9.7          & 4.3              & 4.0             \\
$F$ = 5/2$^-$ - 3/2$^+$ & 1.7          & 11.3         & 4.2              & 3.7             \\
$F$ = 3/2$^-$ - 1/2$^+$ & 2.2          & 9.4          & 4.7              & 4.5             \\   
\hline
\label{NO_FWHM}
\end{tabular}
\end{table*}

\begin{table*}[h!]\tiny
\caption{Sources with H$_2$CO and $c$-C$_3$H$_2$ detections.}
\centering
\begin{tabular}{cccccccc}
\hline\hline
 &            & \multicolumn{3}{c}{H$_2$CO}             & \multicolumn{3}{c}{$c$-C$_3$H$_2$}             \\
\cmidrule(lr){3-5} \cmidrule(lr){6-8}
No &    Source         &    $\int T_{\rm mb}{\rm d}v$         & $V_{\rm LSR}$              & FWHM             & $\int T_{\rm mb}{\rm d}v$            & $V_{\rm LSR}$              & FWHM             \\
 &              & (K·km\,s$^{-1}$)             & (km\,s$^{-1}$)              & (km\,s$^{-1}$)             & (K·km\,s$^{-1}$)            & (km\,s$^{-1}$)              & (km\,s$^{-1}$)             \\
\hline
1  & G121.30+0.66      & 9.68 $\pm$ 0.07  & -17.82 $\pm$ 0.01 & 4.17 $\pm$ 0.04  & 0.92 $\pm$ 0.03 & -17.55 $\pm$ 0.04 & 2.49 $\pm$ 0.08  \\
2  & G123.07-6.31      & 9.86 $\pm$ 0.07  & -30.91 $\pm$ 0.01 & 4.71 $\pm$ 0.04  & 0.65 $\pm$ 0.03 & -30.08 $\pm$ 0.09 & 3.79 $\pm$ 0.24  \\
3  & G133.6945+01.2166 & 14.02 $\pm$ 0.08 & -42.96 $\pm$ 0.01 & 4.04 $\pm$ 0.03  & 1.60 $\pm$ 0.03 & -42.27 $\pm$ 0.05 & 5.30 $\pm$ 0.13  \\
4  & G133.7150+01.2155 & 8.77 $\pm$ 0.04  & -38.80 $\pm$ 0.02 & 8.00 $\pm$ 0.04  & 1.98 $\pm$ 0.04 & -39.00 $\pm$ 0.06 & 6.66 $\pm$ 0.16  \\
5  & S231              & 11.60 $\pm$ 0.06 & -16.21 $\pm$ 0.01 & 4.79 $\pm$ 0.03  & 0.56 $\pm$ 0.02 & -16.14 $\pm$ 0.05 & 2.66 $\pm$ 0.13  \\
6  & S255              & 14.20 $\pm$ 0.09 & 7.38 $\pm$ 0.01   & 3.02 $\pm$ 0.02  & 0.59 $\pm$ 0.03 & 7.35 $\pm$ 0.06   & 2.91 $\pm$ 0.15  \\
7  & G5.89-0.39        & 19.13 $\pm$ 0.21 & 9.47 $\pm$ 0.03   & 5.50 $\pm$ 0.08  & 2.41 $\pm$ 0.03 & 9.14 $\pm$ 0.03   & 4.25 $\pm$ 0.06  \\
8  & G10.6-0.4         & 32.02 $\pm$ 0.08 & -3.56 $\pm$ 0.00  & 7.48 $\pm$ 0.02  & 2.99 $\pm$ 0.03 & -2.91 $\pm$ 0.03  & 6.28 $\pm$ 0.08  \\
9  & W33IRS3           & 20.93 $\pm$ 0.05 & 36.37 $\pm$ 0.01  & 7.84 $\pm$ 0.02  & 3.82 $\pm$ 0.09 & 35.30 $\pm$ 0.06  & 5.84 $\pm$ 0.16  \\
10 & I18182-1433MM1    & 5.43 $\pm$ 0.03  & 59.26 $\pm$ 0.01  & 4.21 $\pm$ 0.03  & 0.34 $\pm$ 0.02 & 58.81 $\pm$ 0.07  & 2.55 $\pm$ 0.18  \\
11 & G019.27+00.07MM1  & 3.51 $\pm$ 0.07  & 27.10 $\pm$ 0.05  & 5.65 $\pm$ 0.14  & 0.26 $\pm$ 0.03 & 26.37 $\pm$ 0.12  & 2.13 $\pm$ 0.25  \\
12 & G19.61-0.23       & 7.45 $\pm$ 0.05  & 42.06 $\pm$ 0.03  & 9.72 $\pm$ 0.08  & 1.49 $\pm$ 0.02 & 43.07 $\pm$ 0.05  & 8.24 $\pm$ 0.12  \\
13 & G20.08-0.13       & 6.11 $\pm$ 0.04  & 41.87 $\pm$ 0.03  & 9.01 $\pm$ 0.07  & 0.59 $\pm$ 0.02 & 42.77 $\pm$ 0.11  & 6.87 $\pm$ 0.25  \\
14 & G022.35+00.41MM1  & 3.69 $\pm$ 0.09  & 52.65 $\pm$ 0.06  & 5.64 $\pm$ 0.18  & 0.41 $\pm$ 0.01 & 52.84 $\pm$ 0.03  & 3.15 $\pm$ 0.06  \\
15 & G023.60+00.00MM1  & 4.69 $\pm$ 0.07  & 107.06 $\pm$ 0.04 & 6.51 $\pm$ 0.12  & 0.25 $\pm$ 0.01 & 106.16 $\pm$ 0.10 & 3.42 $\pm$ 0.22  \\
16 & G24.49-0.04       & 7.41 $\pm$ 0.09  & 110.05 $\pm$ 0.03 & 5.15 $\pm$ 0.08  & 0.64 $\pm$ 0.02 & 110.38 $\pm$ 0.07 & 4.68 $\pm$ 0.18  \\
17 & W43S              & 9.80 $\pm$ 0.10  & 97.19 $\pm$ 0.02  & 4.43 $\pm$ 0.06  & 1.17 $\pm$ 0.02 & 97.29 $\pm$ 0.03  & 4.12 $\pm$ 0.07  \\
18 & G31.41+0.31       & 5.34 $\pm$ 0.03  & 97.57 $\pm$ 0.04  & 12.52 $\pm$ 0.08 & 0.87 $\pm$ 0.02 & 97.28 $\pm$ 0.06  & 6.00 $\pm$ 0.15  \\
19 & W43Main3          & 12.16 $\pm$ 0.12 & 96.34 $\pm$ 0.05  & 9.81 $\pm$ 0.14  & 0.96 $\pm$ 0.01 & 98.05 $\pm$ 0.03  & 6.71 $\pm$ 0.07  \\
20 & G34.26+0.15       & 20.74 $\pm$ 0.43 & 56.81 $\pm$ 0.04  & 4.13 $\pm$ 0.10  & 2.18 $\pm$ 0.02 & 58.15 $\pm$ 0.03  & 5.78 $\pm$ 0.08  \\
21 & G35.20-0.74       & 16.09 $\pm$ 0.07 & 35.23 $\pm$ 0.02  & 8.31 $\pm$ 0.04  & 1.46 $\pm$ 0.01 & 33.91 $\pm$ 0.02  & 3.98 $\pm$ 0.04  \\
22 & W51D              & 28.60 $\pm$ 0.38 & 60.55 $\pm$ 0.05  & 8.14 $\pm$ 0.14  & 3.74 $\pm$ 0.09 & 60.90 $\pm$ 0.08  & 6.58 $\pm$ 0.20  \\
23 & W51M              & 52.27 $\pm$ 0.09 & 56.09 $\pm$ 0.01  & 9.86 $\pm$ 0.02  & 3.46 $\pm$ 0.02 & 56.77 $\pm$ 0.02  & 11.03 $\pm$ 0.07 \\
24 & K3-50A            & 8.95 $\pm$ 0.04  & -23.97 $\pm$ 0.02 & 8.81 $\pm$ 0.05  & 1.08 $\pm$ 0.03 & -23.74 $\pm$ 0.12 & 8.91 $\pm$ 0.28  \\
25 & G073.0633+01.7958 & 1.82 $\pm$ 0.05  & 0.73 $\pm$ 0.03   & 2.24 $\pm$ 0.07  & 0.11 $\pm$ 0.01 & 0.75 $\pm$ 0.07   & 1.49 $\pm$ 0.15  \\
26 & ON1               & 7.29 $\pm$ 0.07  & 12.47 $\pm$ 0.02  & 4.80 $\pm$ 0.06  & 1.40 $\pm$ 0.02 & 11.32 $\pm$ 0.03  & 4.34 $\pm$ 0.07  \\
27 & 20126+4104        & 5.43 $\pm$ 0.05  & -3.47 $\pm$ 0.02  & 3.54 $\pm$ 0.04  & 1.72 $\pm$ 0.02 & -3.57 $\pm$ 0.01  & 2.31 $\pm$ 0.03  \\
28 & ON2S              & 7.60 $\pm$ 0.17  & -1.15 $\pm$ 0.07  & 6.54 $\pm$ 0.19  & 0.37 $\pm$ 0.02 & -1.62 $\pm$ 0.07  & 3.50 $\pm$ 0.17  \\
29 & ON2N              & 5.23 $\pm$ 0.08  & 0.19 $\pm$ 0.05   & 7.43 $\pm$ 0.17  & 0.29 $\pm$ 0.02 & 0.13 $\pm$ 0.18   & 4.11 $\pm$ 0.38  \\
30 & G083.7962+03.3058 & 5.27 $\pm$ 0.10  & -4.17 $\pm$ 0.03  & 3.26 $\pm$ 0.08  & $\leq$ 0.24     & -4.49 $\pm$ 0.29  & $\cdots$         \\
31 & G081.7133+00.5589 & 10.43 $\pm$ 0.05 & -4.25 $\pm$ 0.00  & 4.36 $\pm$ 0.02  & 1.05 $\pm$ 0.02 & -3.40 $\pm$ 0.03  & 2.89 $\pm$ 0.07  \\
32 & G081.7624+00.5916 & 8.48 $\pm$ 0.16  & -4.01 $\pm$ 0.04  & 4.75 $\pm$ 0.11  & 0.94 $\pm$ 0.02 & -4.44 $\pm$ 0.02  & 2.07 $\pm$ 0.05  \\
33 & G094.4637-00.8043 & 1.03 $\pm$ 0.07  & -44.57 $\pm$ 0.09 & 2.90 $\pm$ 0.23  & $\leq$ 0.05     & -44.93 $\pm$ 0.29 & $\cdots$         \\
34 & 21391+5802        & 10.78 $\pm$ 0.22 & 0.37 $\pm$ 0.04   & 4.34 $\pm$ 0.12  & 0.85 $\pm$ 0.01 & 0.55 $\pm$ 0.02   & 2.35 $\pm$ 0.05  \\
35 & NGC7538           & 19.89 $\pm$ 0.11 & -56.40 $\pm$ 0.02 & 5.94 $\pm$ 0.04  & $\leq$ 0.22     & -57.57 $\pm$ 0.11 & $\cdots$         \\
36 & NGC7538A          & 12.02 $\pm$ 0.29 & -57.16 $\pm$ 0.05 & 4.49 $\pm$ 0.14  & 0.61 $\pm$ 0.03 & -56.68 $\pm$ 0.08 & 4.06 $\pm$ 0.24  \\ 
\hline
\end{tabular}
\label{H2CO&C3H2}
\end{table*}

\begin{table*}[h!]\tiny
\caption{Beam-averaged column densities and relative abundances.}
\centering
\resizebox{\textwidth}{!}{%
\begin{tabular}{ccccccccccc}
\hline\hline
No & Source            & $T_{\mathrm{ex}}$  & $Q$      & \multicolumn{3}{c}{FWHM (km\,s$^{-1}$)}                              &    $\int T_{\rm mb} {\rm d}v$     & $N$(NO)                 & $N$(H\textsubscript{2})  & $X$(NO)                            \\
\cline{5-7}
   &                   &   (K)    &        & $F$ = 5/2$^+$ - 3/2$^-$ &  $F$ = 5/2$^-$ - 3/2$^+$ &  $F$ = 3/2$^-$ - 1/2$^+$             &      (K·km\,s$^{-1}$)           &         (cm$^{-2}$)                         &         (cm$^{-2}$)                                             &                                  \\
\hline
1  & G121.30+0.66      & 30    & 80.17  & 2.06 $\pm$ 0.17  & 2.16 $\pm$ 0.16  & 3.40 $\pm$ 0.62 & 0.32 $\pm$ 0.02 & $(7.20 \pm 0.45) \times 10^{14}$ & $4.52\times 10^{22}$                                 & $(1.59 \pm 0.10) \times 10^{-8}$ \\
2  & G123.07-6.31      & 30    & 80.17  & $\cdots$         & 1.74 $\pm$ 0.31  & $\cdots$        & 0.08 $\pm$ 0.01 & $(1.80 \pm 0.22) \times 10^{14}$ & $\cdots$                                             & $\cdots$                         \\
3  & G133.6945+01.2166 & 30    & 80.17  & 4.23 $\pm$ 0.54  & 3.39 $\pm$ 0.31  & $\cdots$        & 0.32 $\pm$ 0.02 & $(7.20 \pm 0.45) \times 10^{14}$ & $1.06\times 10^{23}$                                 & $(6.79 \pm 0.42) \times 10^{-9}$ \\
4  & G133.7150+01.2155 & 25    & 67.57  & $\cdots$         & 3.40 $\pm$ 0.79  & $\cdots$        & 0.16 $\pm$ 0.03 & $(3.18 \pm 0.60) \times 10^{14}$ & $1.71\times 10^{23}$                                 & $(1.86 \pm 0.35) \times 10^{-9}$ \\
5  & S231              & 25.7  & 69.33  & 3.96 $\pm$ 0.29  & 3.46 $\pm$ 0.27  & $\cdots$        & 0.28 $\pm$ 0.01 & $(5.67 \pm 0.20) \times 10^{14}$ & $7.11\times 10^{22}$                                 & $(7.98 \pm 0.28) \times 10^{-9}$ \\
6  & S255              & 30    & 80.17  & 2.17 $\pm$ 0.35  & 3.08 $\pm$ 0.37  & $\cdots$        & 0.21 $\pm$ 0.02 & $(4.72 \pm 0.45) \times 10^{14}$ & $5.47\times 10^{22}$                                 & $(8.64 \pm 0.82) \times 10^{-9}$ \\
7  & G5.89-0.39        & 39.7  & 105.28 & 3.96 $\pm$ 0.22  & 4.19 $\pm$ 0.26  & 2.67 $\pm$ 0.62 & 0.46 $\pm$ 0.02 & $(1.28 \pm 0.06) \times 10^{15}$ & $1.36\times 10^{23}$                                 & $(9.42 \pm 0.41) \times 10^{-9}$ \\
8  & G10.6-0.4         & 30    & 80.17  & 7.08 $\pm$ 0.36  & 6.31 $\pm$ 0.36  & 8.42 $\pm$ 1.44 & 0.65 $\pm$ 0.02 & $(1.46 \pm 0.04) \times 10^{15}$ & $2.50\times 10^{23}$                                 & $(5.85 \pm 0.18) \times 10^{-9}$ \\
9  & W33IRS3           & 30    & 80.17  & 6.40 $\pm$ 0.21  & 5.59 $\pm$ 0.20  & 5.46 $\pm$ 0.58 & 0.77 $\pm$ 0.02 & $(1.73 \pm 0.04) \times 10^{15}$ & $2.75\times 10^{23}$                                 & $(6.30 \pm 0.16) \times 10^{-9}$ \\
10 & I18182-1433MM1    & 22.67 & 61.75  & 2.94 $\pm$ 0.17  & 3.09 $\pm$ 0.29  & 3.47 $\pm$ 0.69 & 0.26 $\pm$ 0.01 & $(4.87 \pm 0.19) \times 10^{14}$ & $1.62\times 10^{22}$                                 & $(3.01 \pm 0.12) \times 10^{-8}$ \\
11 & G019.27+00.07MM1  & 18    & 50.13  & 3.90 $\pm$ 0.69  & 2.89 $\pm$ 0.45  & $\cdots$        & 0.20 $\pm$ 0.02 & $(3.30 \pm 0.33) \times 10^{14}$ & $1.96\times 10^{22}$                                 & $(1.69 \pm 0.17) \times 10^{-8}$ \\
12 & G19.61-0.23       & 21.4  & 58.58  & 9.74 $\pm$ 0.75  & 6.59 $\pm$ 0.34  & 5.24 $\pm$ 0.88 & 0.45 $\pm$ 0.02 & $(8.15 \pm 0.36) \times 10^{14}$ & $2.30\times 10^{23}$                                 & $(3.54 \pm 0.16) \times 10^{-9}$ \\
13 & G20.08-0.13       & 30.8  & 82.20  & 4.95 $\pm$ 0.32  & 5.19 $\pm$ 0.27  & 4.45 $\pm$ 0.66 & 0.36 $\pm$ 0.01 & $(8.25 \pm 0.23) \times 10^{14}$ & $5.60\times 10^{22}$                                 & $(1.47 \pm 0.04) \times 10^{-8}$ \\
14 & G022.35+00.41MM1  & 18    & 50.13  & 2.67 $\pm$ 0.29  & 2.44 $\pm$ 0.15  & 2.31 $\pm$ 0.24 & 0.17 $\pm$ 0.01 & $(2.81 \pm 0.17) \times 10^{14}$ & $8.66\times 10^{21}$                                 & $(3.24 \pm 0.19) \times 10^{-8}$ \\
15 & G023.60+00.00MM1  & 18    & 50.13  & 2.93 $\pm$ 0.19  & 3.34 $\pm$ 0.24  & 2.87 $\pm$ 0.40 & 0.26 $\pm$ 0.01 & $(4.29 \pm 0.17) \times 10^{14}$ & $4.07\times 10^{22}$                                 & $(1.06 \pm 0.04) \times 10^{-8}$ \\
16 & G24.49-0.04       & 30    & 80.17  & 4.71 $\pm$ 0.72  & 3.95 $\pm$ 0.43  & 7.20 $\pm$ 0.95 & 0.22 $\pm$ 0.02 & $(4.95 \pm 0.45) \times 10^{14}$ & $\cdots$                                             & $\cdots$                         \\
17 & W43S              & 39.8  & 105.54 & 4.88 $\pm$ 0.35  & 5.33 $\pm$ 0.37  & $\cdots$        & 0.29 $\pm$ 0.01 & $(8.09 \pm 0.28) \times 10^{14}$ & $6.35\times 10^{22}$                                 & $(1.27 \pm 0.04) \times 10^{-8}$ \\
18 & G31.41+0.31       & 21.2  & 58.09  & 6.01 $\pm$ 0.29  & 4.00 $\pm$ 0.18  & 5.32 $\pm$ 0.44 & 0.68 $\pm$ 0.02 & $(1.22 \pm 0.04) \times 10^{15}$ & $2.76\times 10^{23}$                                 & $(4.44 \pm 0.13) \times 10^{-9}$ \\
19 & W43Main3          & 30    & 80.17  & 6.65 $\pm$ 0.22  & 6.92 $\pm$ 0.18  & 5.18 $\pm$ 0.46 & 0.86 $\pm$ 0.01 & $(1.93 \pm 0.02) \times 10^{15}$ & $1.94\times 10^{23}$                                 & $(9.97 \pm 0.12) \times 10^{-9}$ \\
20 & G34.26+0.15       & 30    & 80.17  & 7.24 $\pm$ 0.13  & 6.87 $\pm$ 0.16  & 5.88 $\pm$ 0.43 & 0.80 $\pm$ 0.01 & $(1.80 \pm 0.02) \times 10^{15}$ & $4.20\times 10^{23}$                                 & $(4.28 \pm 0.05) \times 10^{-9}$ \\
21 & G35.20-0.74       & 27.2  & 73.09  & 3.32 $\pm$ 0.22  & 2.99 $\pm$ 0.19  & 3.58 $\pm$ 0.36 & 0.35 $\pm$ 0.01 & $(7.36 \pm 0.21) \times 10^{14}$ & $1.06\times 10^{23}$                                 & $(6.94 \pm 0.20) \times 10^{-9}$ \\
22 & W51D              & 51.4  & 137.39 & 10.20 $\pm$ 1.49 & 10.05 $\pm$ 1.47 & $\cdots$        & 0.78 $\pm$ 0.06 & $(2.72 \pm 0.21) \times 10^{15}$ & $1.62\times 10^{23}$                                 & $(1.68 \pm 0.13) \times 10^{-8}$ \\
23 & W51M              & 30    & 80.17  & 7.92 $\pm$ 0.23  & 11.25 $\pm$ 0.45 & 9.36 $\pm$ 0.42 & 1.42 $\pm$ 0.03 & $(3.19 \pm 0.07) \times 10^{15}$ & $5.97\times 10^{23}$                                 & $(5.35 \pm 0.11) \times 10^{-9}$ \\
24 & K3-50A            & 30    & 80.17  & $\cdots$         & 5.84 $\pm$ 1.10  & $\cdots$        & 0.17 $\pm$ 0.02 & $(3.82 \pm 0.45) \times 10^{14}$ & $1.03\times 10^{23}$                                 & $(3.71 \pm 0.44) \times 10^{-9}$ \\
25 & G073.0633+01.7958 & 17.6  & 49.14  & 1.44 $\pm$ 0.17  & 1.74 $\pm$ 0.22  & $\cdots$        & 0.09 $\pm$ 0.01 & $(1.47 \pm 0.16) \times 10^{14}$ & $3.68\times 10^{22}$                                 & $(4.00 \pm 0.44) \times 10^{-9}$ \\
26 & ON1               & 30    & 80.17  & 3.55 $\pm$ 0.24  & 4.76 $\pm$ 0.25  & 2.83 $\pm$ 0.36 & 0.44 $\pm$ 0.02 & $(9.90 \pm 0.45) \times 10^{14}$ & $\cdots$                                             & $\cdots$                         \\
27 & 20126+4104        & 23.1  & 62.82  & $\cdots$         & $\cdots$         & $\cdots$        & $\leq$ 0.06     & $\leq$ $1.14 \times 10^{14}$     & $6.37\times 10^{22}$                                 & $\leq$ $1.78 \times 10^{-9}$     \\
28 & ON2S              & 45.1  & 119.81 & $\cdots$         & 3.88 $\pm$ 0.57  & $\cdots$        & 0.12 $\pm$ 0.02 & $(3.72 \pm 0.62) \times 10^{14}$ & $4.57\times 10^{22}$                                 & $(8.14 \pm 1.36) \times 10^{-9}$ \\
29 & ON2N              & 30    & 80.17  & 4.12 $\pm$ 0.70  & 5.83 $\pm$ 0.50  & $\cdots$        & 0.13 $\pm$ 0.01 & $(2.92 \pm 0.22) \times 10^{14}$ & $9.23\times 10^{22}$                                 & $(3.17 \pm 0.24) \times 10^{-9}$ \\
30 & G083.7962+03.3058 & 30    & 80.17  & $\cdots$         & $\cdots$         & $\cdots$        & $\leq$ 0.17     & $\leq$ $3.82 \times 10^{14}$     & $\cdots$                                             & $\cdots$                         \\
31 & G081.7133+00.5589 & 28.6  & 76.62  & 2.97 $\pm$ 0.21  & 2.16 $\pm$ 0.13  & $\cdots$        & 0.35 $\pm$ 0.01 & $(7.61 \pm 0.22) \times 10^{14}$ & $1.25\times 10^{23}$                                 & $(6.09 \pm 0.17) \times 10^{-9}$ \\
32 & G081.7624+00.5916 & 18.9  & 52.37  & 2.43 $\pm$ 0.15  & 1.86 $\pm$ 0.15  & $\cdots$        & 0.36 $\pm$ 0.01 & $(6.09 \pm 0.17) \times 10^{14}$ & $1.03\times 10^{23}$                                 & $(5.92 \pm 0.16) \times 10^{-9}$ \\
33 & G094.4637-00.8043 & 25    & 67.57  & $\cdots$         & $\cdots$         & $\cdots$        & $\leq$ 0.07     & $\leq$ $1.39 \times 10^{14}$     & $\cdots$                                             & $\cdots$                         \\
34 & 21391+5802        & 25    & 67.57  & 2.37 $\pm$ 0.17  & 2.06 $\pm$ 0.13  & 2.18 $\pm$ 0.23 & 0.18 $\pm$ 0.01 & $(3.58 \pm 0.20) \times 10^{14}$ & $\cdots$                                             & $\cdots$                         \\
35 & NGC7538           & 28.4  & 76.12  & $\cdots$         & 2.37 $\pm$ 0.37  & $\cdots$        & 0.27 $\pm$ 0.03 & $(5.85 \pm 0.65) \times 10^{14}$ & $1.10\times 10^{23}$                                 & $(5.31 \pm 0.59) \times 10^{-9}$ \\
36 & NGC7538A          & 30.8  & 82.20  & 2.29 $\pm$ 0.36  & 4.76 $\pm$ 0.52  & $\cdots$        & 0.15 $\pm$ 0.01 & $(3.44 \pm 0.23) \times 10^{14}$ & $1.33\times 10^{23}$                                 & $(2.58 \pm 0.17) \times 10^{-9}$ \\
\hline
\end{tabular}
}
\tablefoot{$\int T_{\rm mb}{\rm d}v$: Average of two NO lines ($F$ = 5/2$^+$ - 3/2$^-$ and $F$ = 5/2$^-$ - 3/2$^+$) when both detected above 5$\sigma$; otherwise, value of the single detected one.}
\label{CDA of NO}
\end{table*}

\begin{table*}[h!]\tiny
\caption{Distributions of NO column densities and abundances.}
\centering
\begin{tabular}{ccccccccc}
\hline\hline
Source                           & $N($NO$)_{\rm min}$                        & $N($NO$)_{\rm max}$                        & $N($NO$)_{\rm average}$ & $N($NO$)_{\rm median}$ & $X($NO$)_{\rm min}$ & $X($NO$)_{\rm max}$ & $X($NO$)_{\rm average}$ & $X($NO$)_{\rm median}$ \\
types                            & (cm$^{-2}$)                                & (cm$^{-2}$)                                & (cm$^{-2}$)             & (cm$^{-2}$)            &                     &                     &                         &                        \\
\hline
H II                             & $1.8\times 10^{14}$                        & $3.2\times 10^{15}$                        & $9.5\times 10^{14}$     & $7.5\times 10^{14}$    & $2.6\times 10^{-9}$ & $1.6\times 10^{-8}$ & $7.2\times 10^{-9}$     & $6.3\times 10^{-9}$    \\
protostar &  $1.5\times 10^{14}$ &  $6.1\times 10^{14}$ & $3.6\times 10^{14}$     & $3.4\times 10^{14}$    & $1.9\times 10^{-9}$ & $5.9\times 10^{-9}$ & $3.9\times 10^{-9}$     & $4.0\times 10^{-9}$    \\
IRDC      &  $2.8\times 10^{14}$ &  $4.9\times 10^{14}$ & $3.8\times 10^{14}$     & $3.8\times 10^{14}$    & $1.1\times 10^{-8}$ & $3.2\times 10^{-8}$ & $2.3\times 10^{-8}$     & $2.4\times 10^{-8}$    \\
Total                            & $1.5\times 10^{14}$                        & $3.2\times 10^{15}$                        & $8.0\times 10^{14}$     & $6.0\times 10^{14}$    & $1.9\times 10^{-9}$ & $3.2\times 10^{-8}$ & $9.1\times 10^{-9}$     & $6.5\times 10^{-9}$    \\
\hline
\end{tabular}
\label{distribution}
\end{table*}

\clearpage
\section{Spectra}

Spectra of NO lines in all sources are shown, together with $c$-C$_3$H$_2$ and H$_2$CO.

\begin{figure}[!h]
\centering
\includegraphics[width=0.33\textwidth]{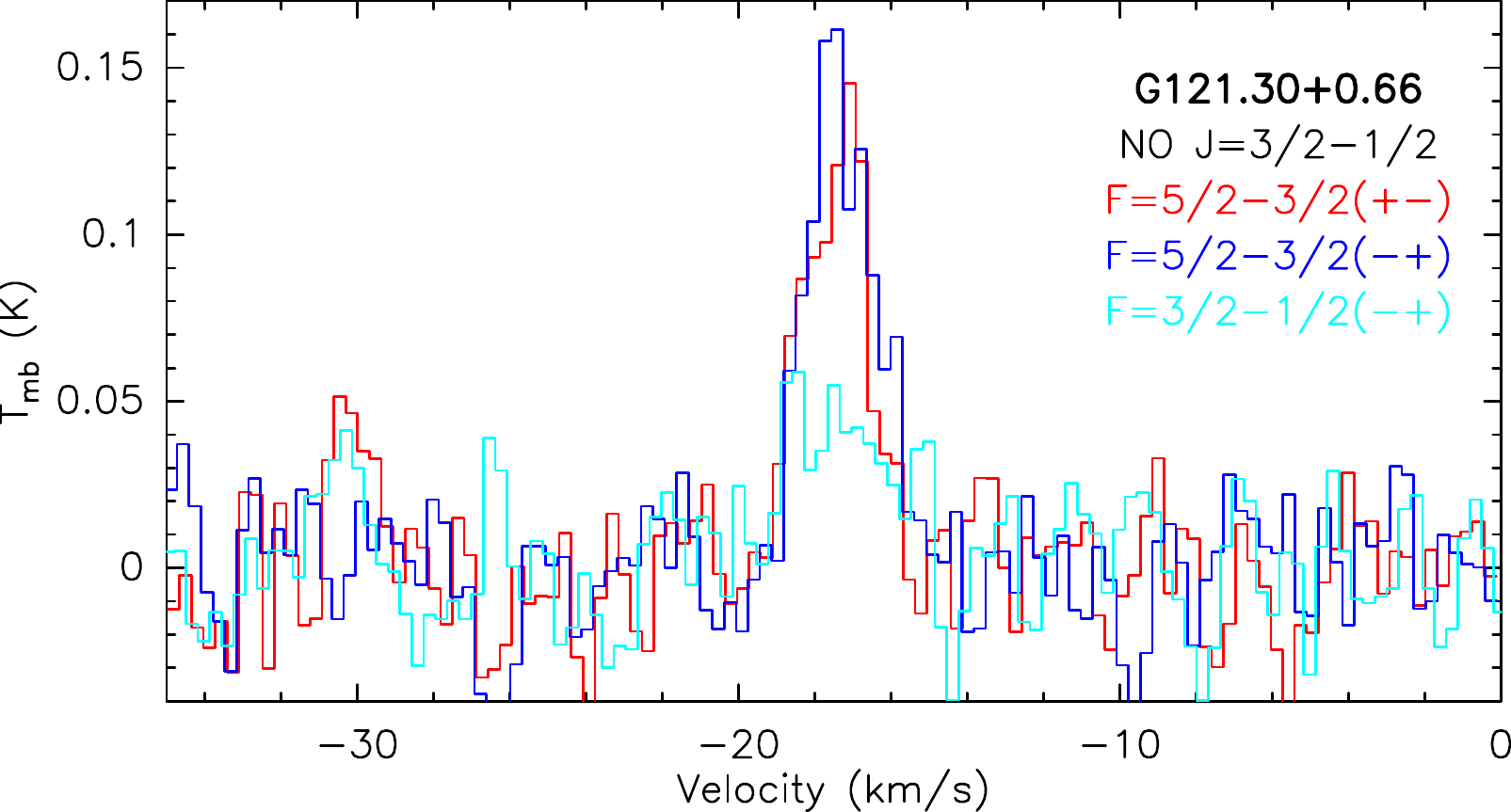}
\includegraphics[width=0.33\textwidth]{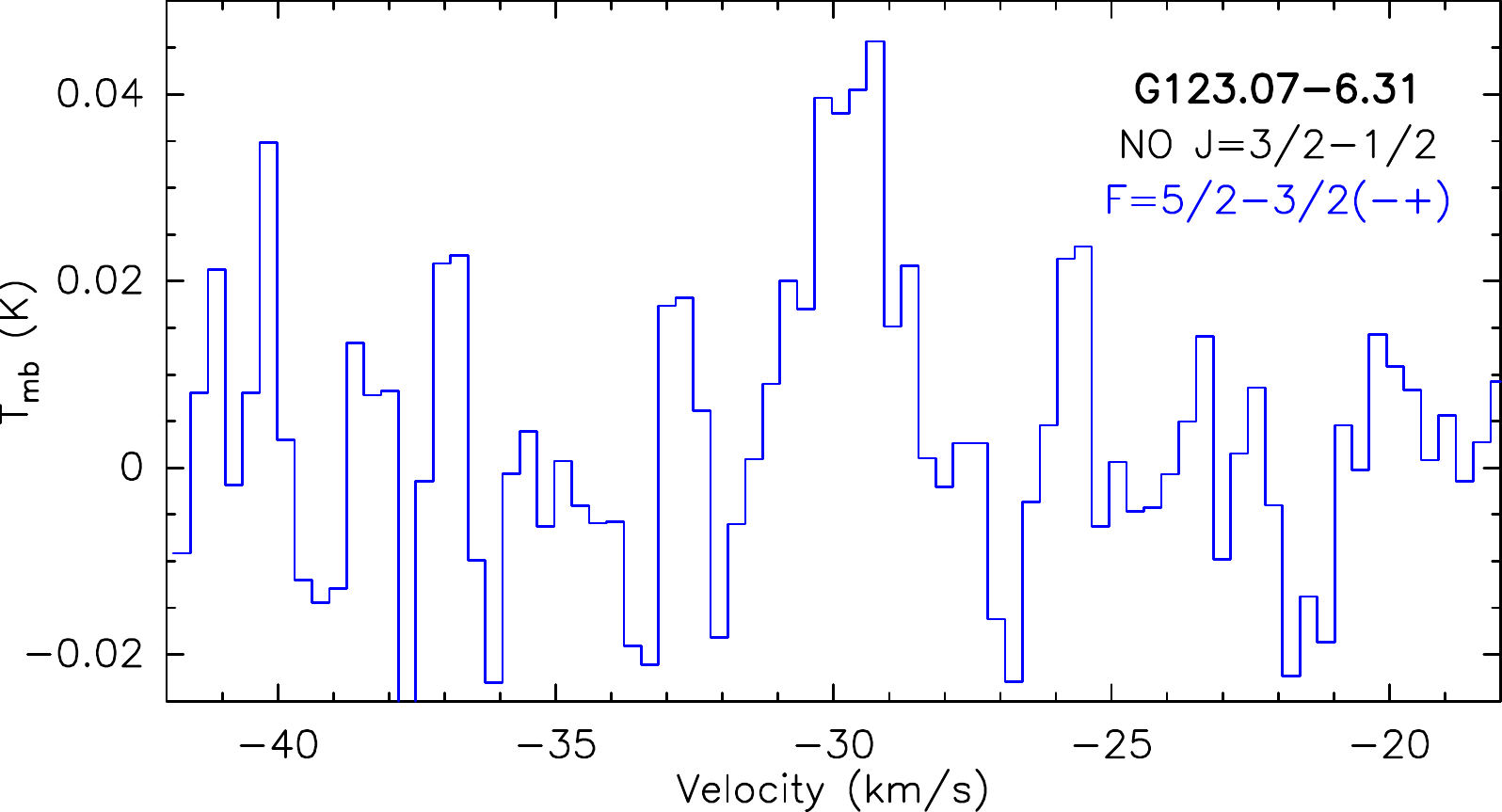}
\includegraphics[width=0.33\textwidth]{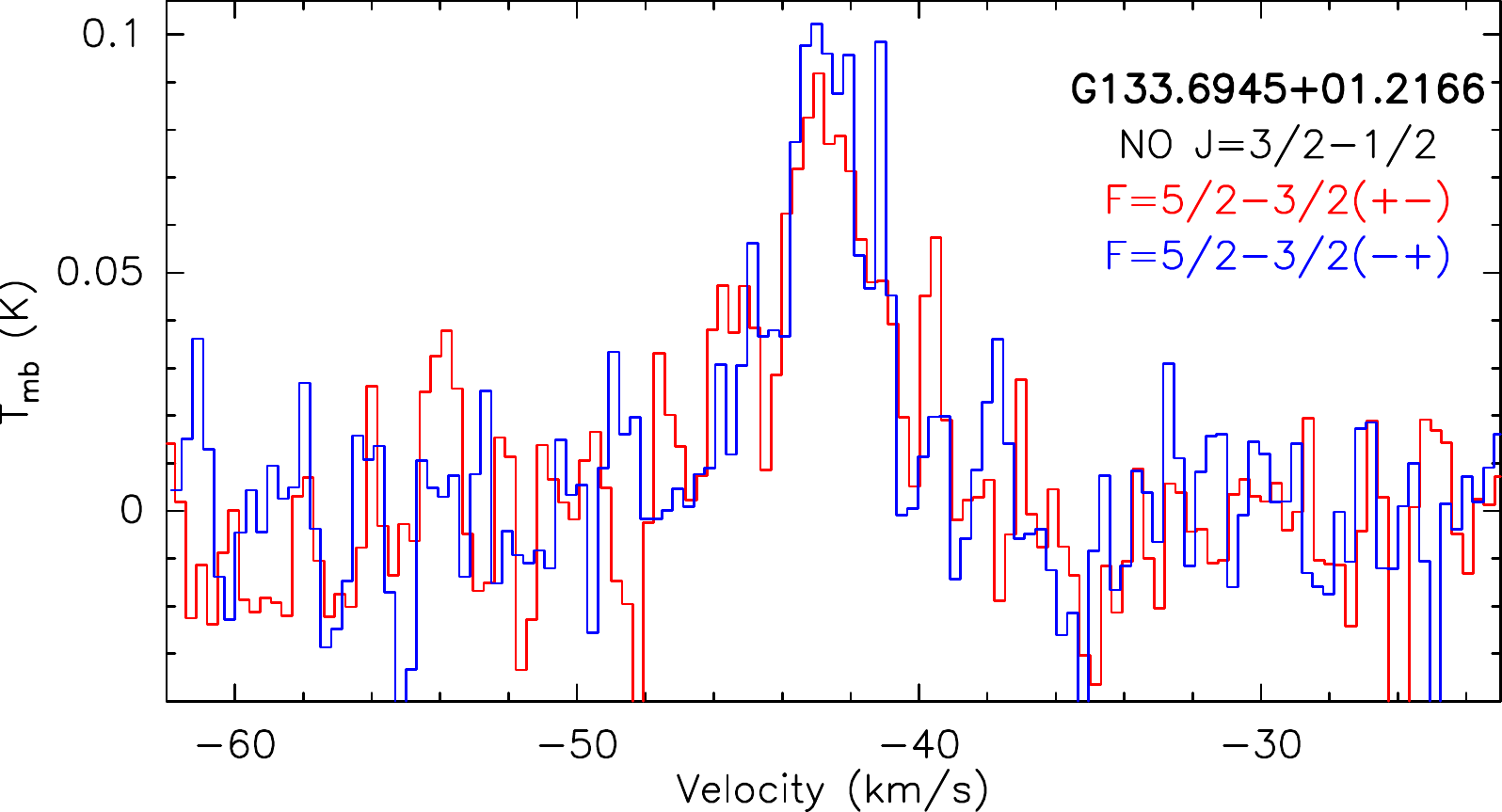}
\includegraphics[width=0.33\textwidth]{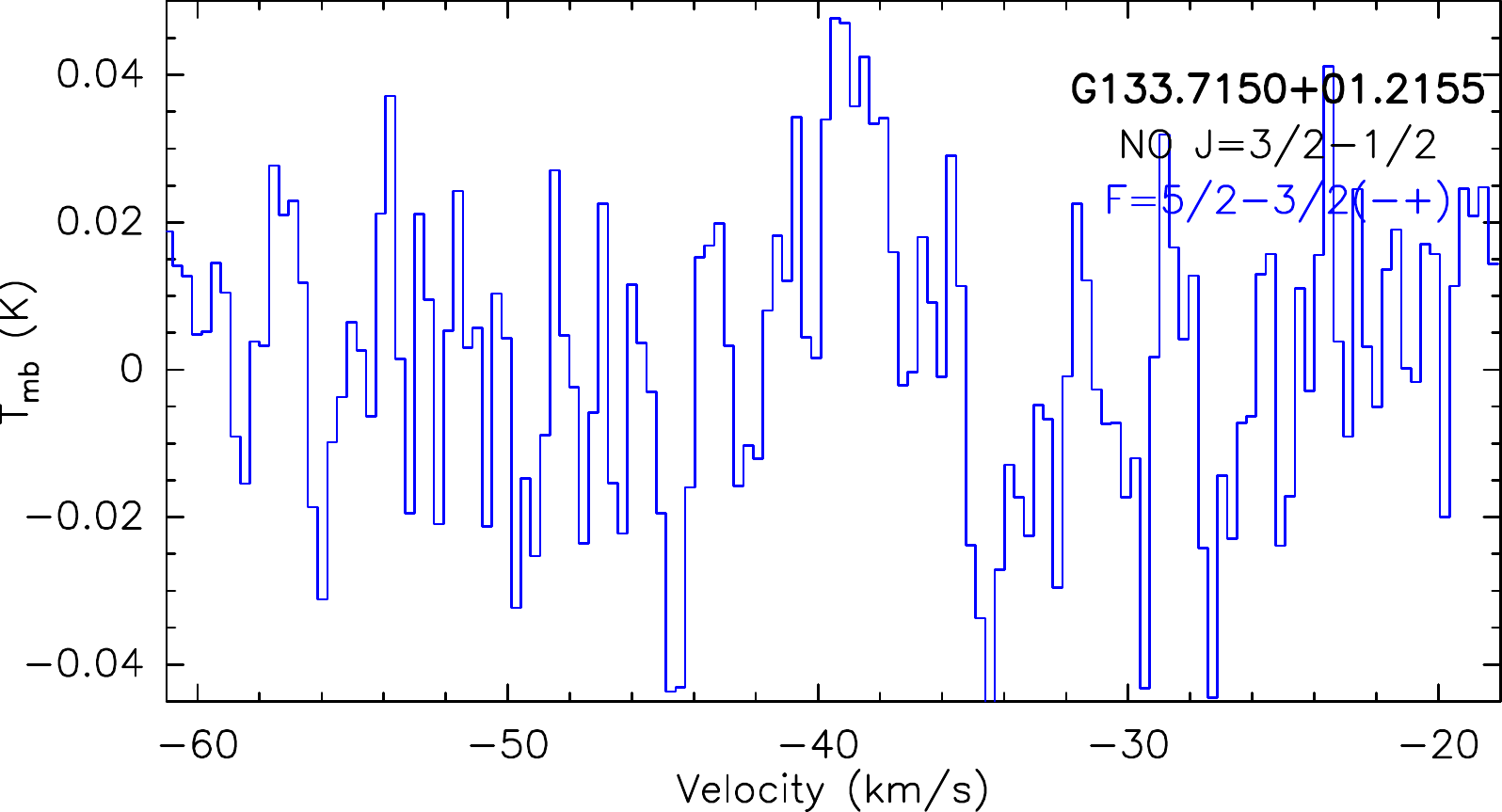}
\includegraphics[width=0.33\textwidth]{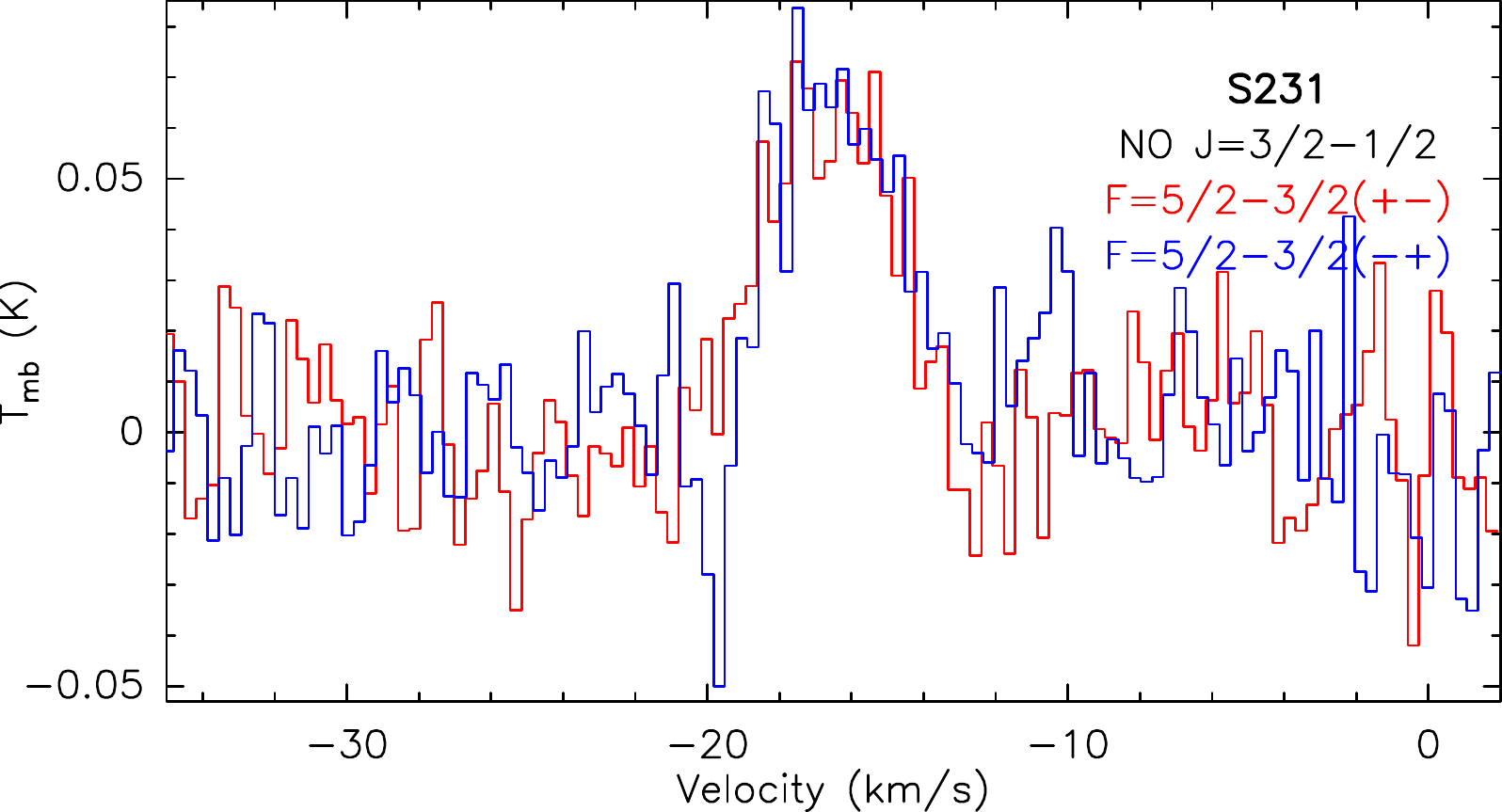}
\includegraphics[width=0.33\textwidth]{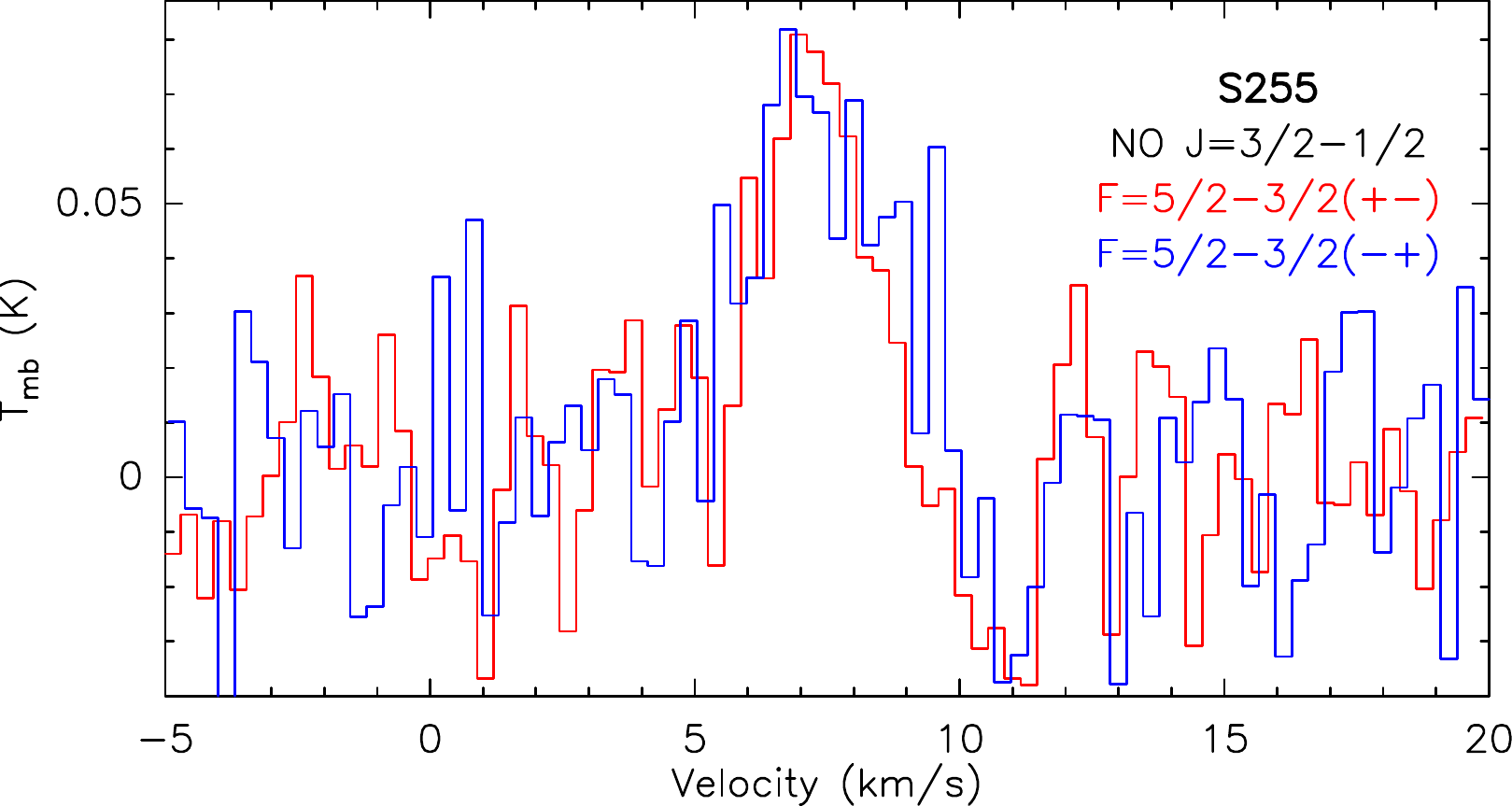}
\includegraphics[width=0.33\textwidth]{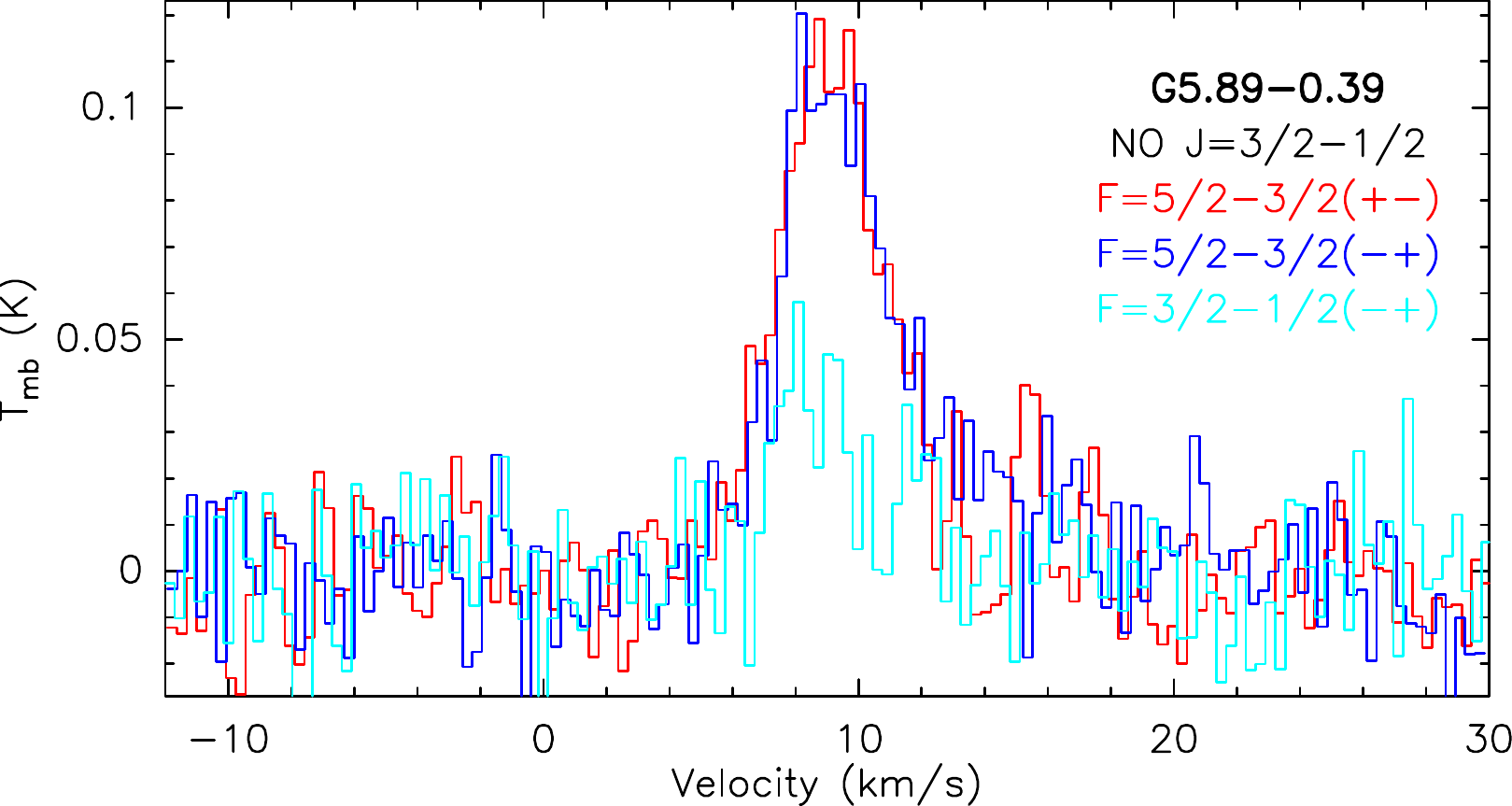}
\includegraphics[width=0.33\textwidth]{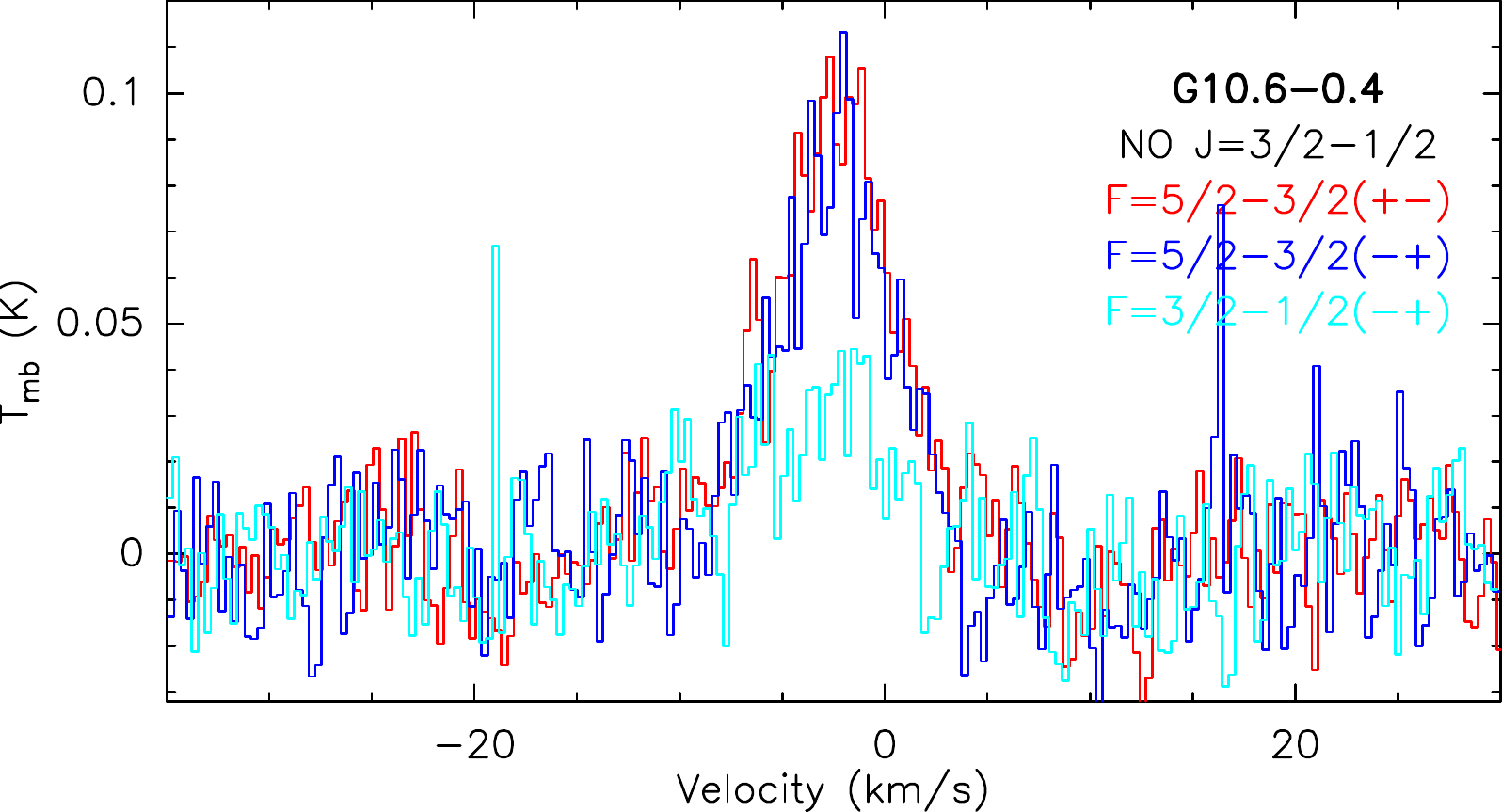}
\includegraphics[width=0.33\textwidth]{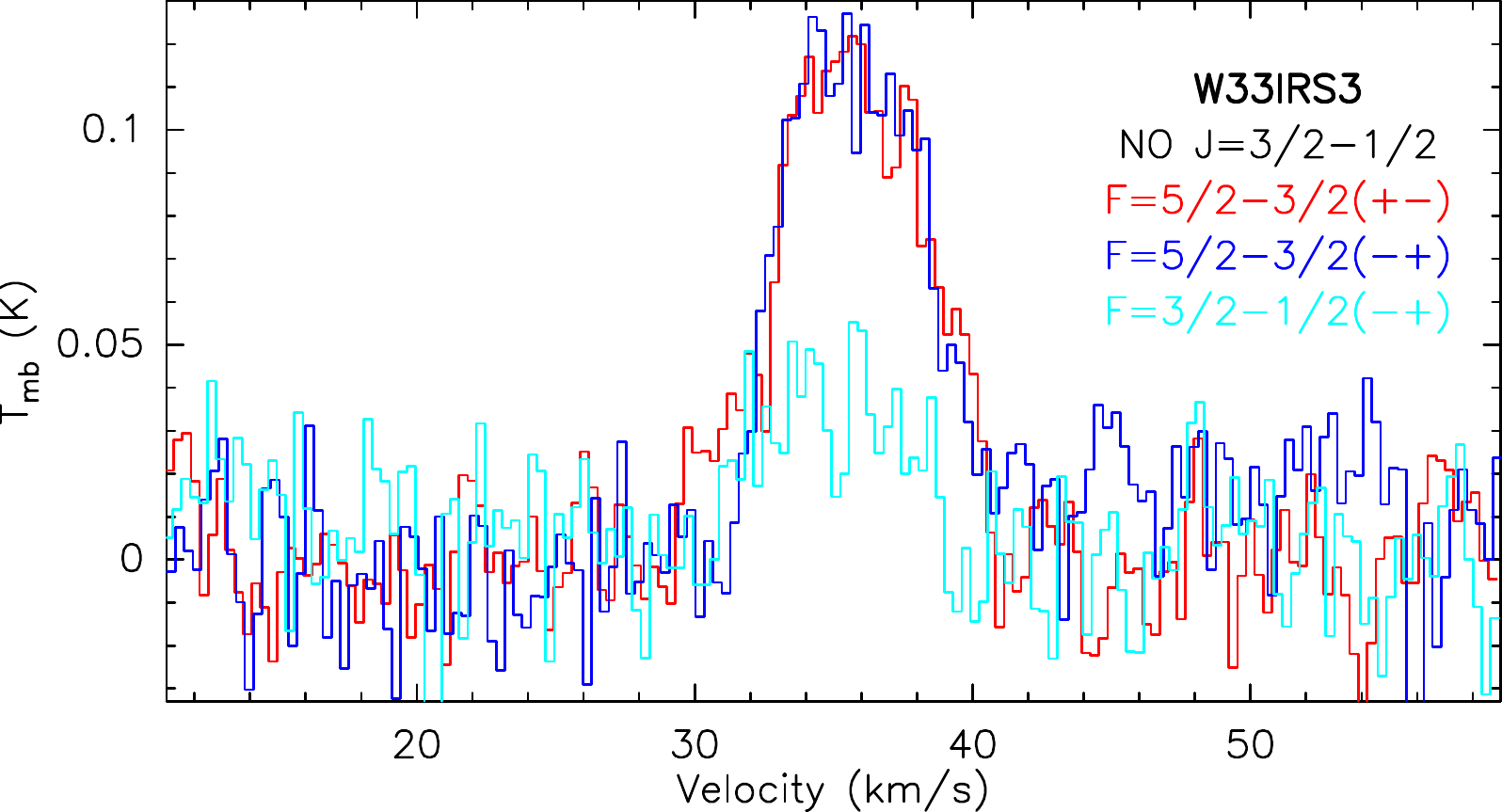}
\includegraphics[width=0.33\textwidth]{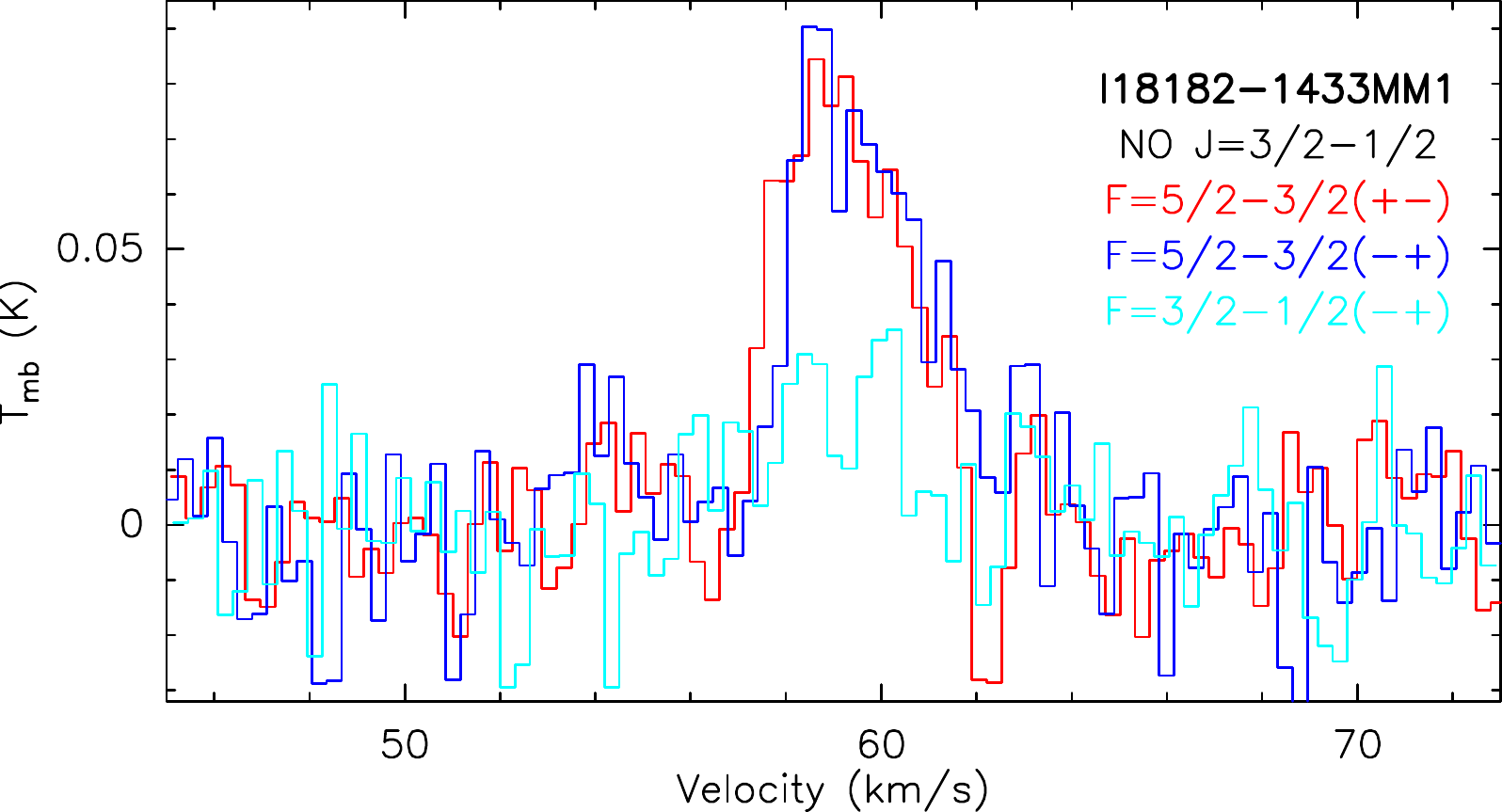}
\includegraphics[width=0.33\textwidth]{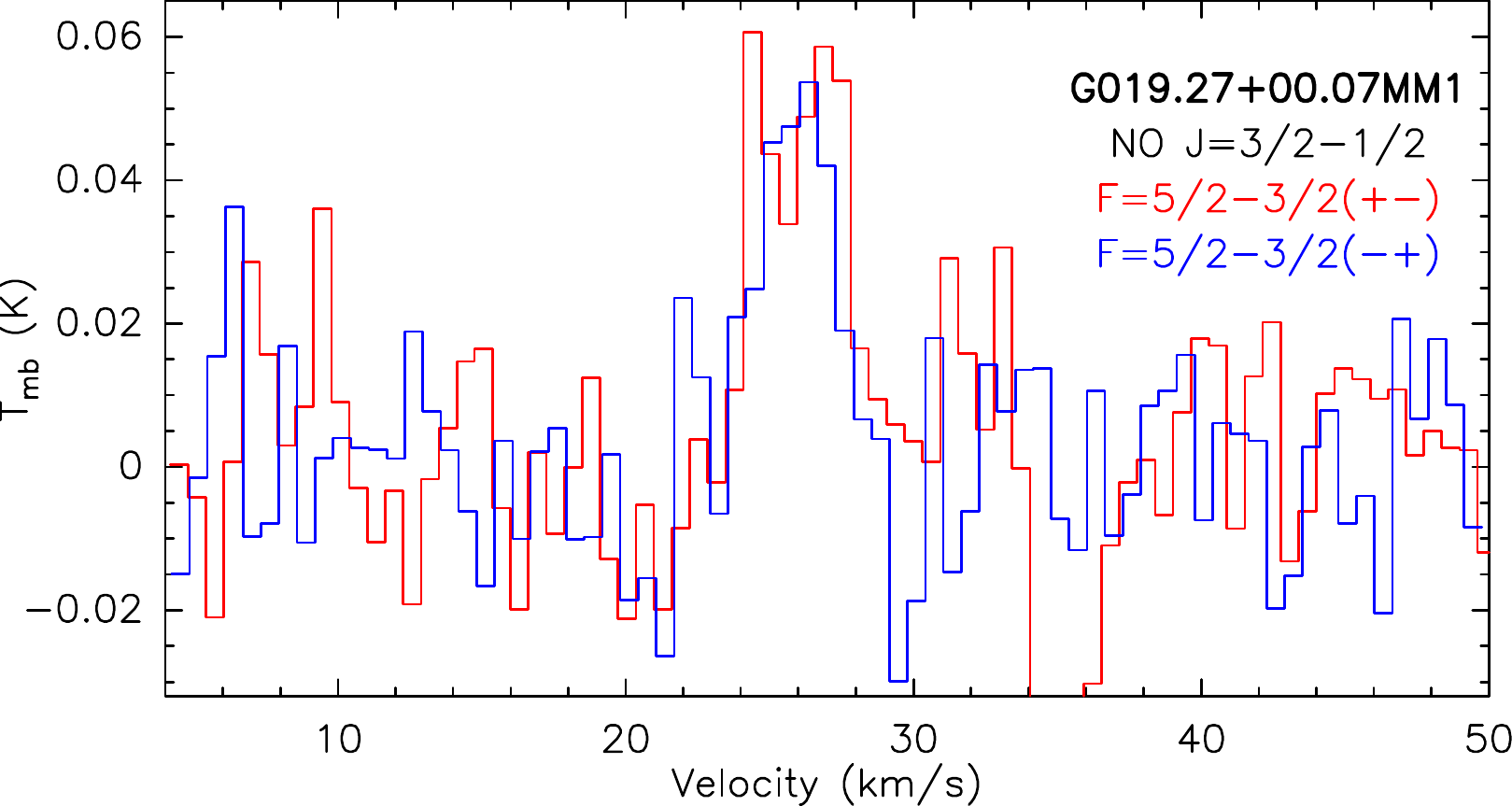}
\includegraphics[width=0.33\textwidth]{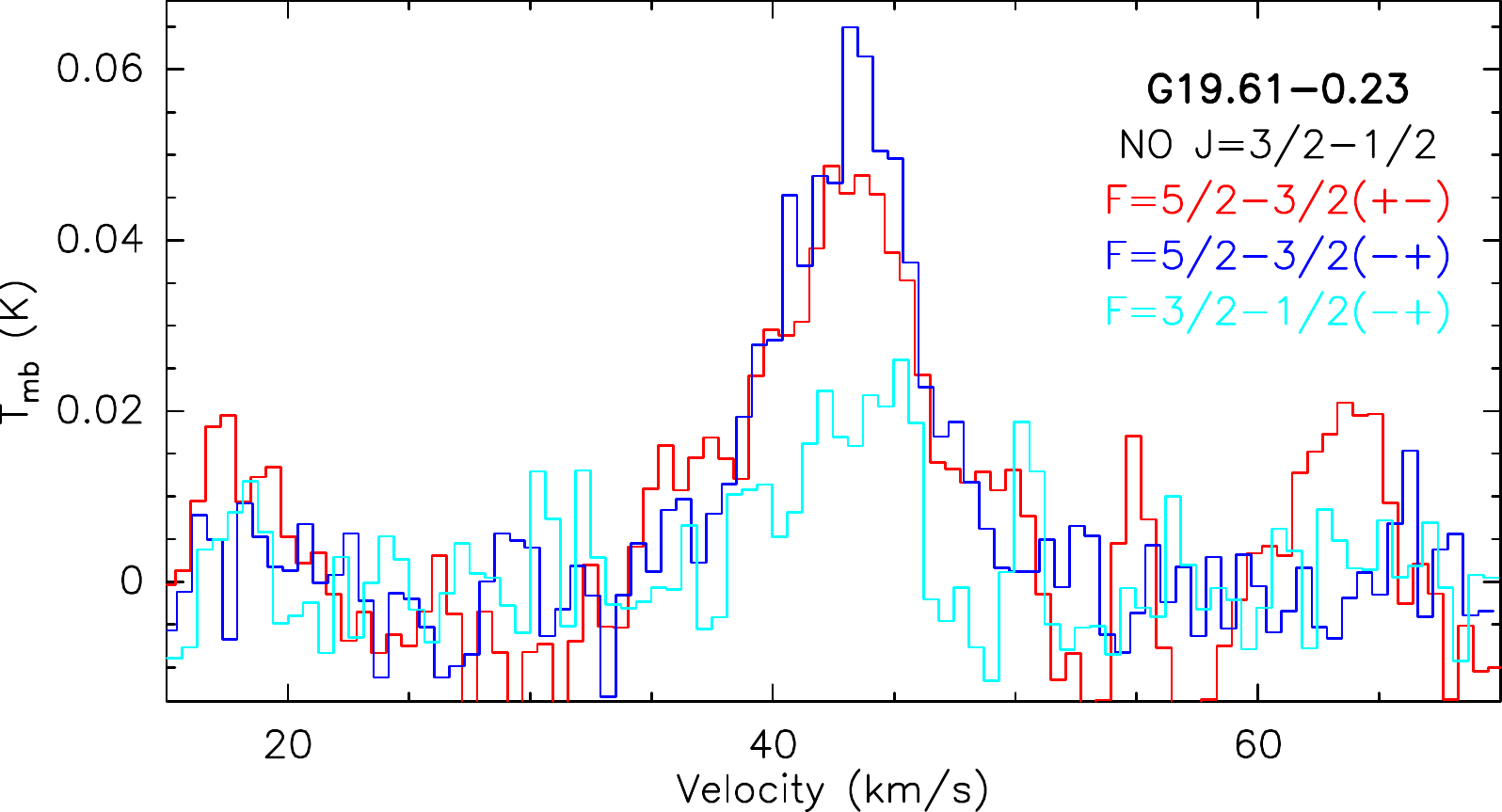}
\includegraphics[width=0.33\textwidth]{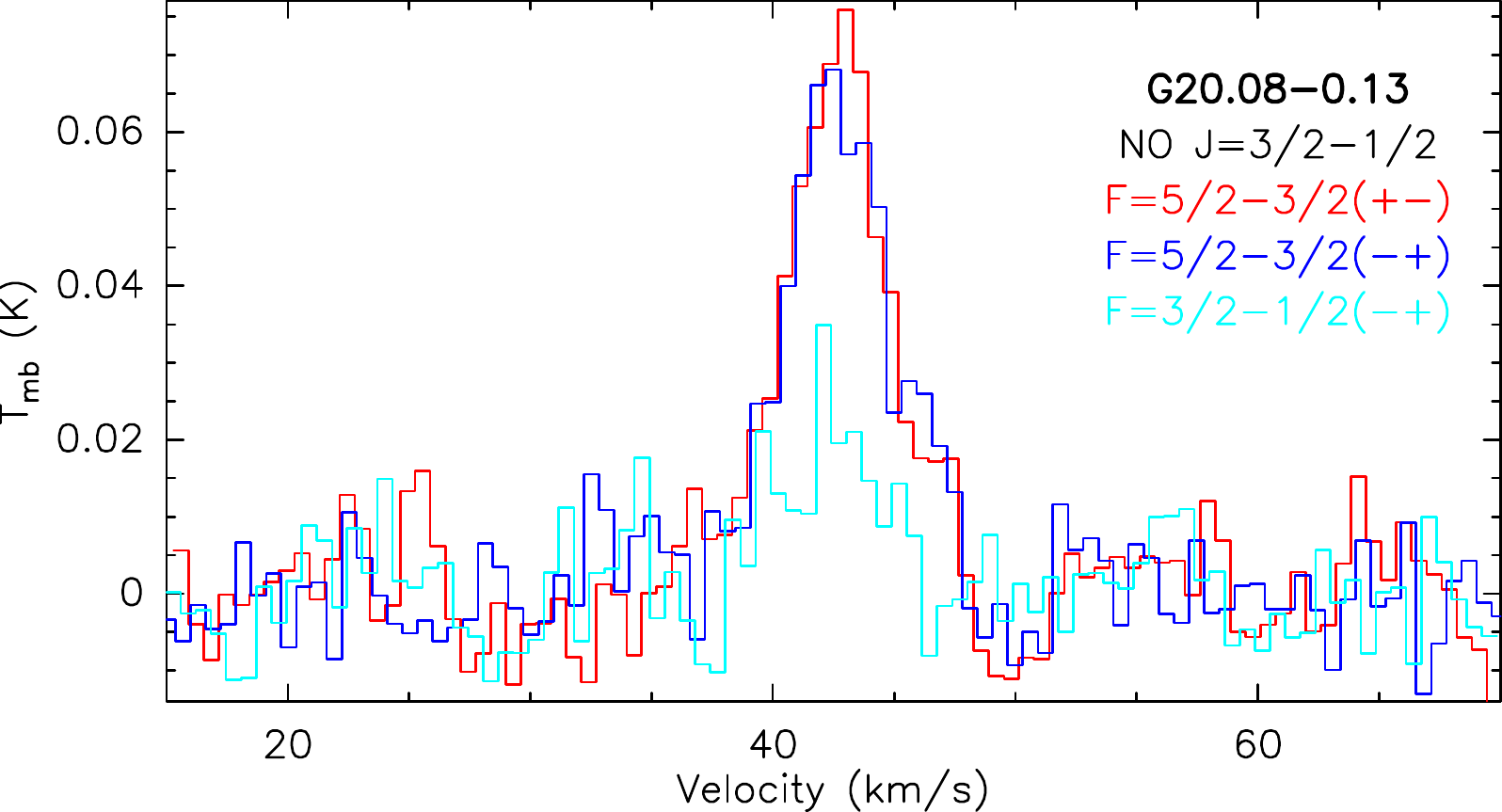}
\includegraphics[width=0.33\textwidth]{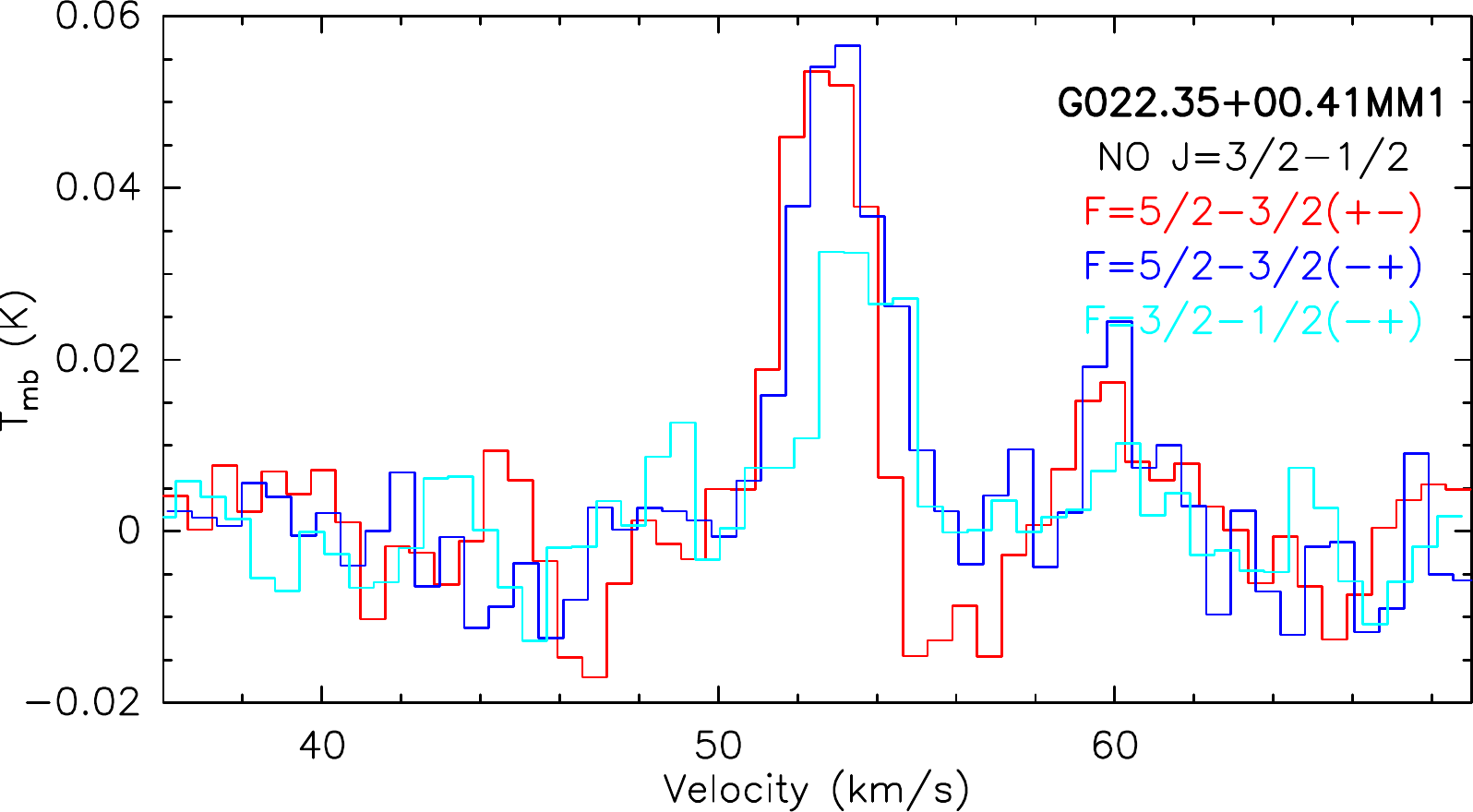}
\includegraphics[width=0.33\textwidth]{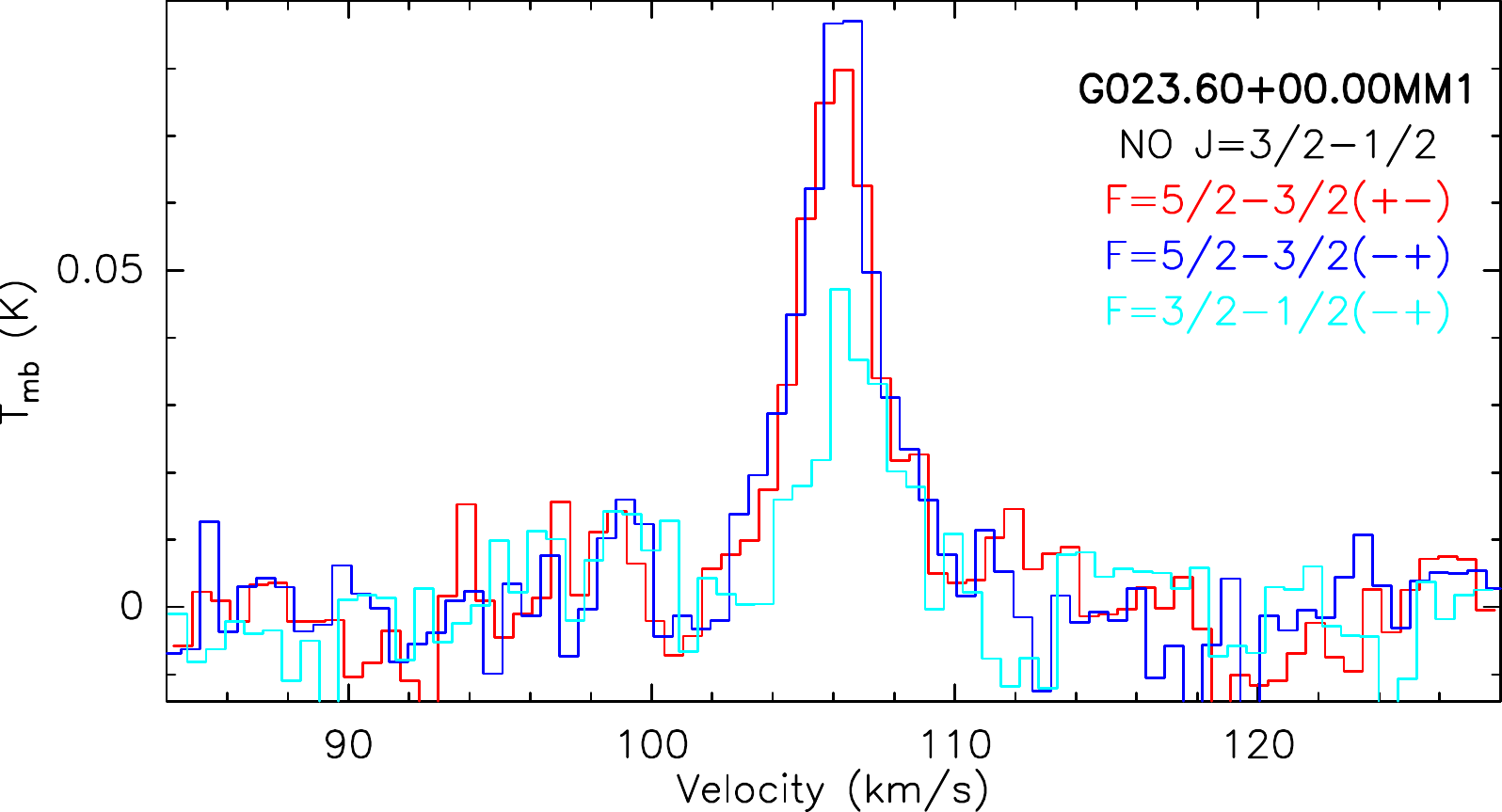}
\includegraphics[width=0.33\textwidth]{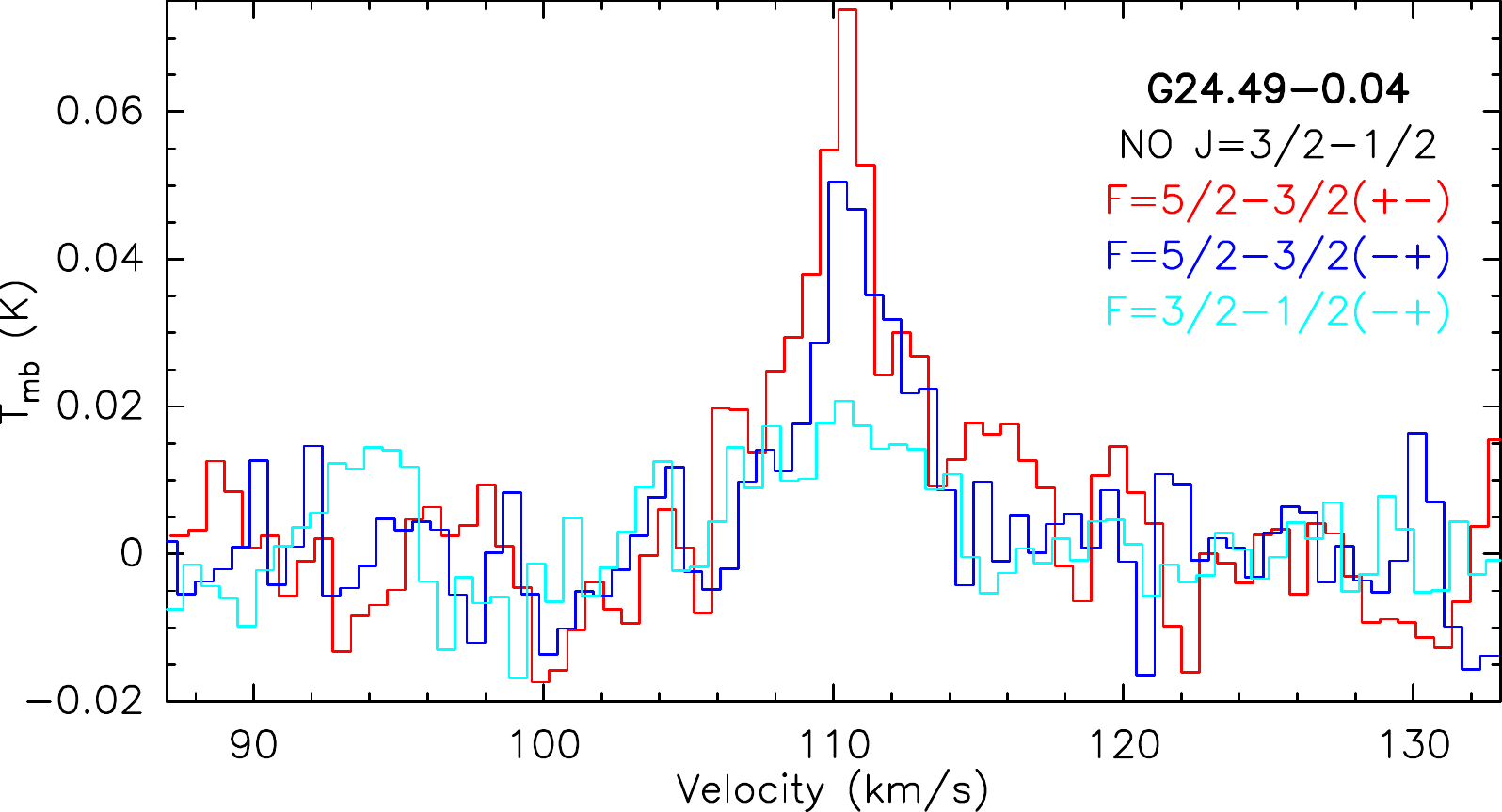}
\includegraphics[width=0.33\textwidth]{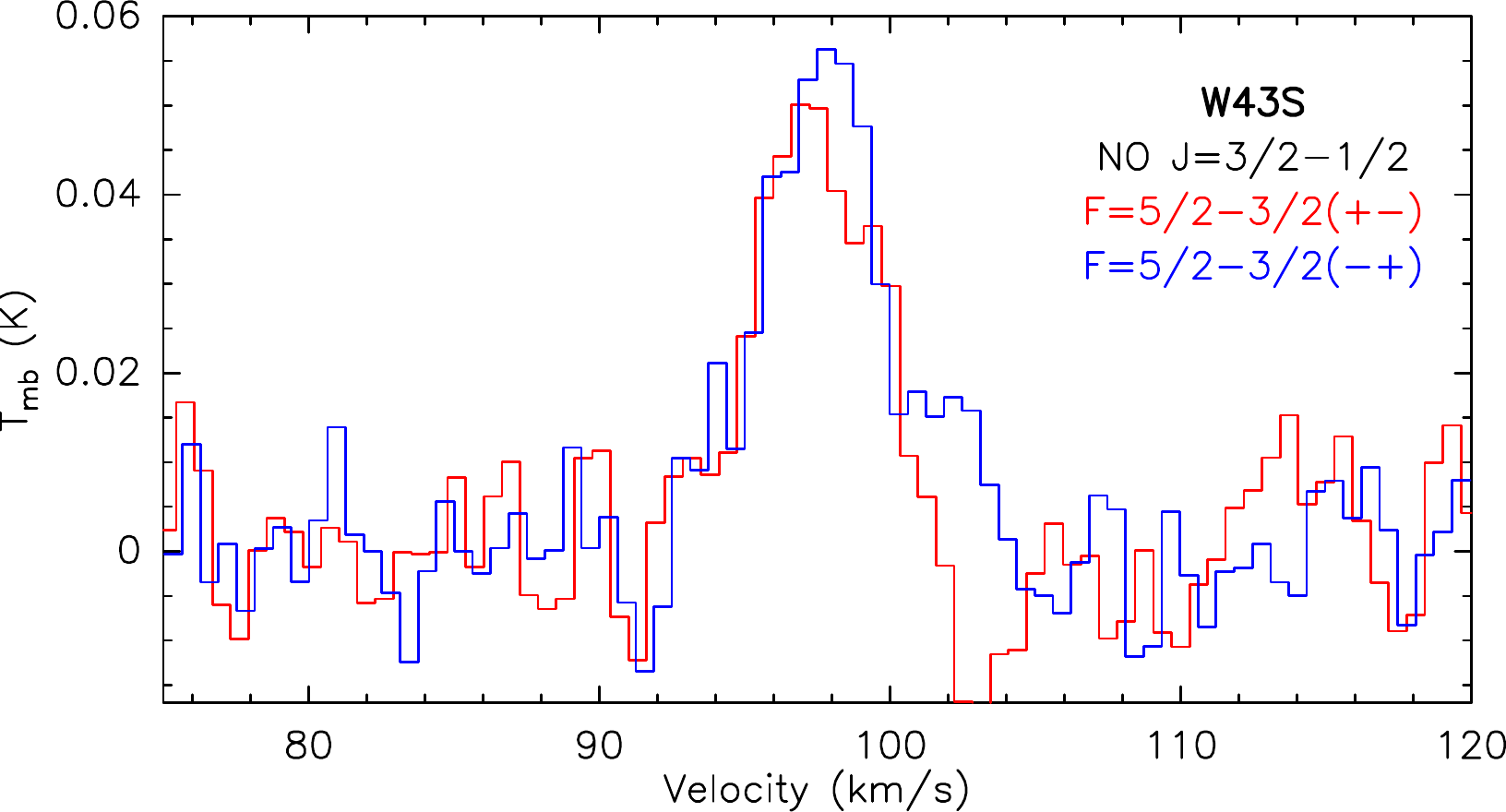}
\includegraphics[width=0.33\textwidth]{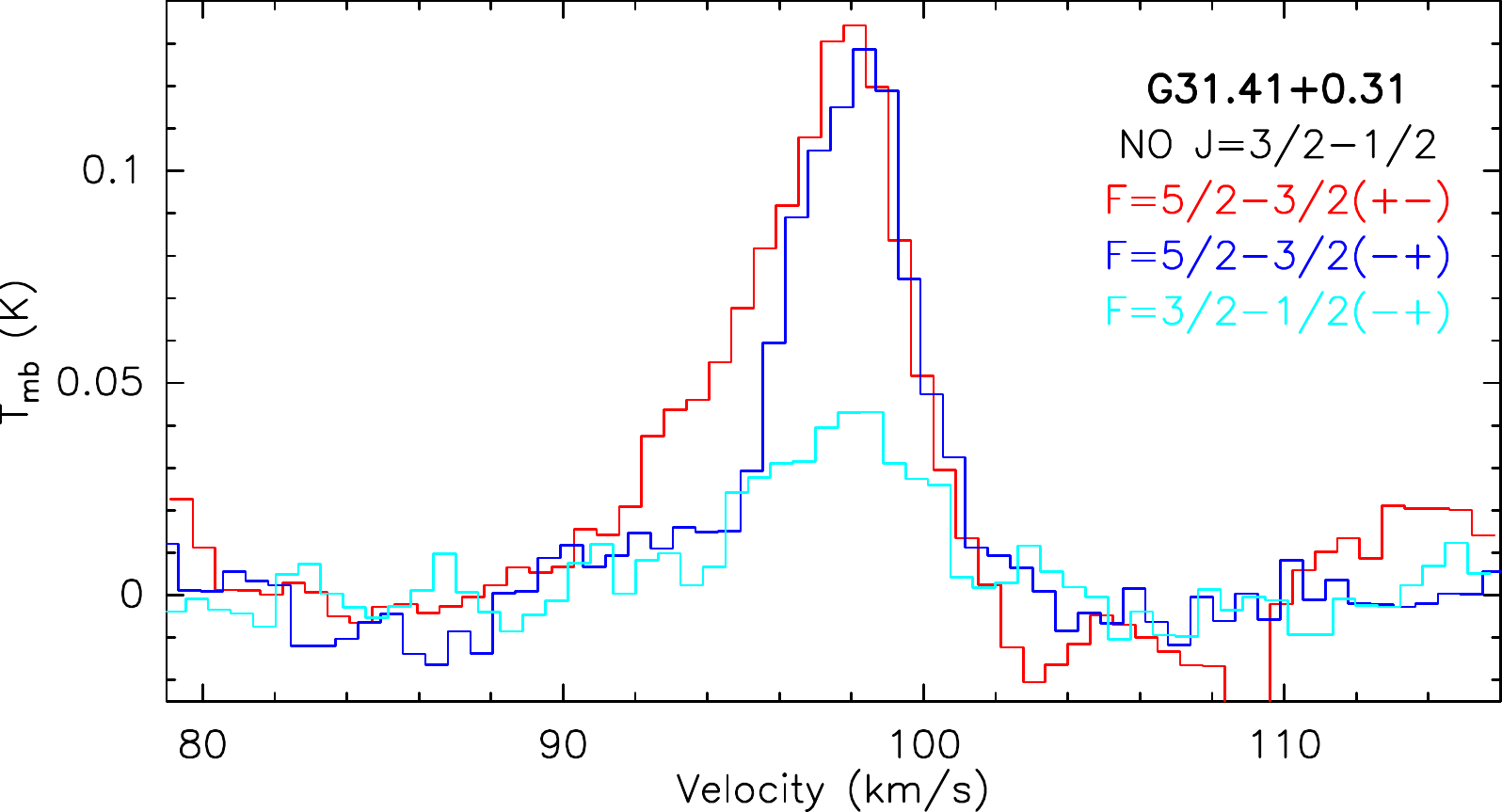}
\includegraphics[width=0.33\textwidth]{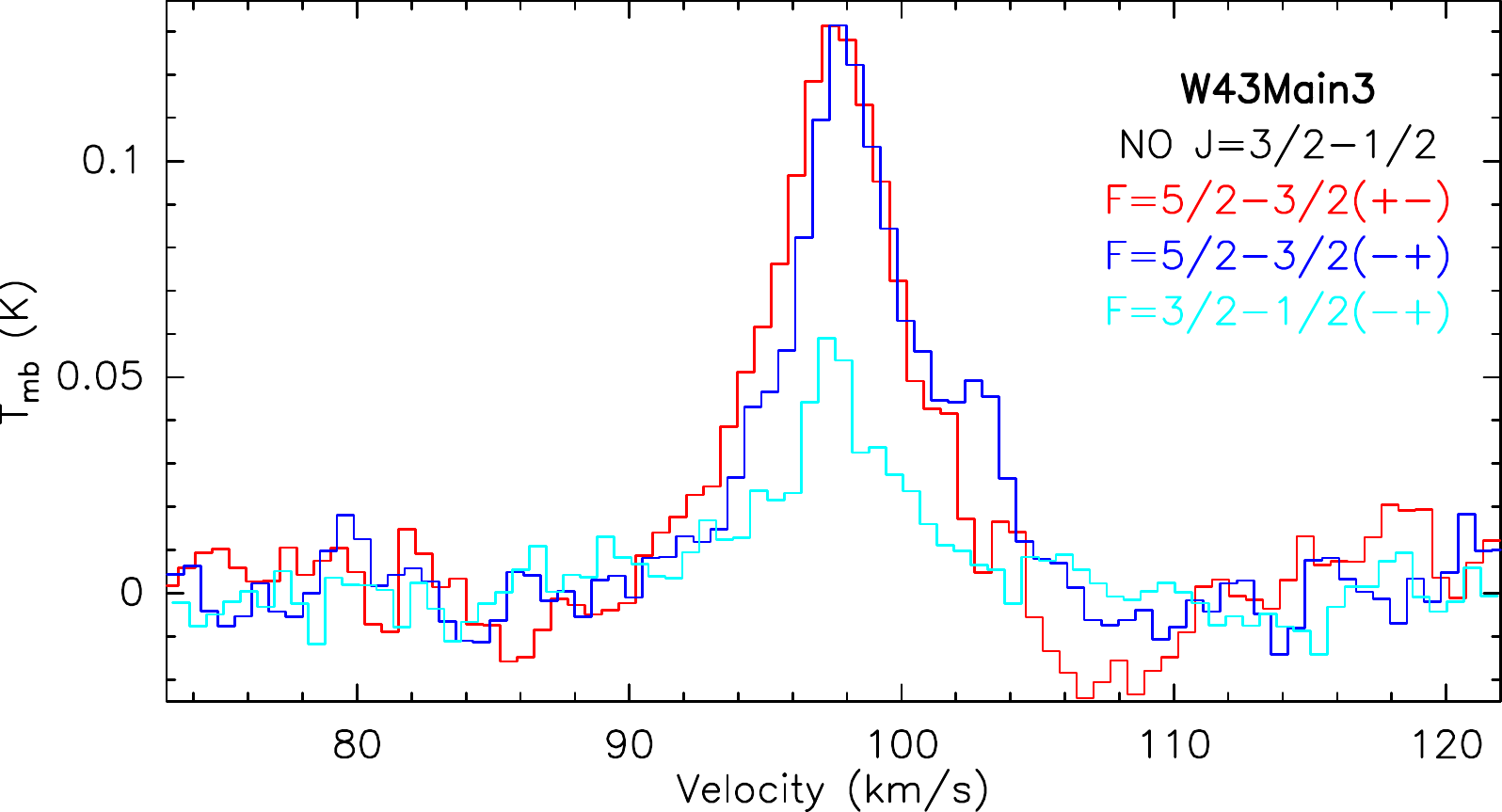}
\includegraphics[width=0.33\textwidth]{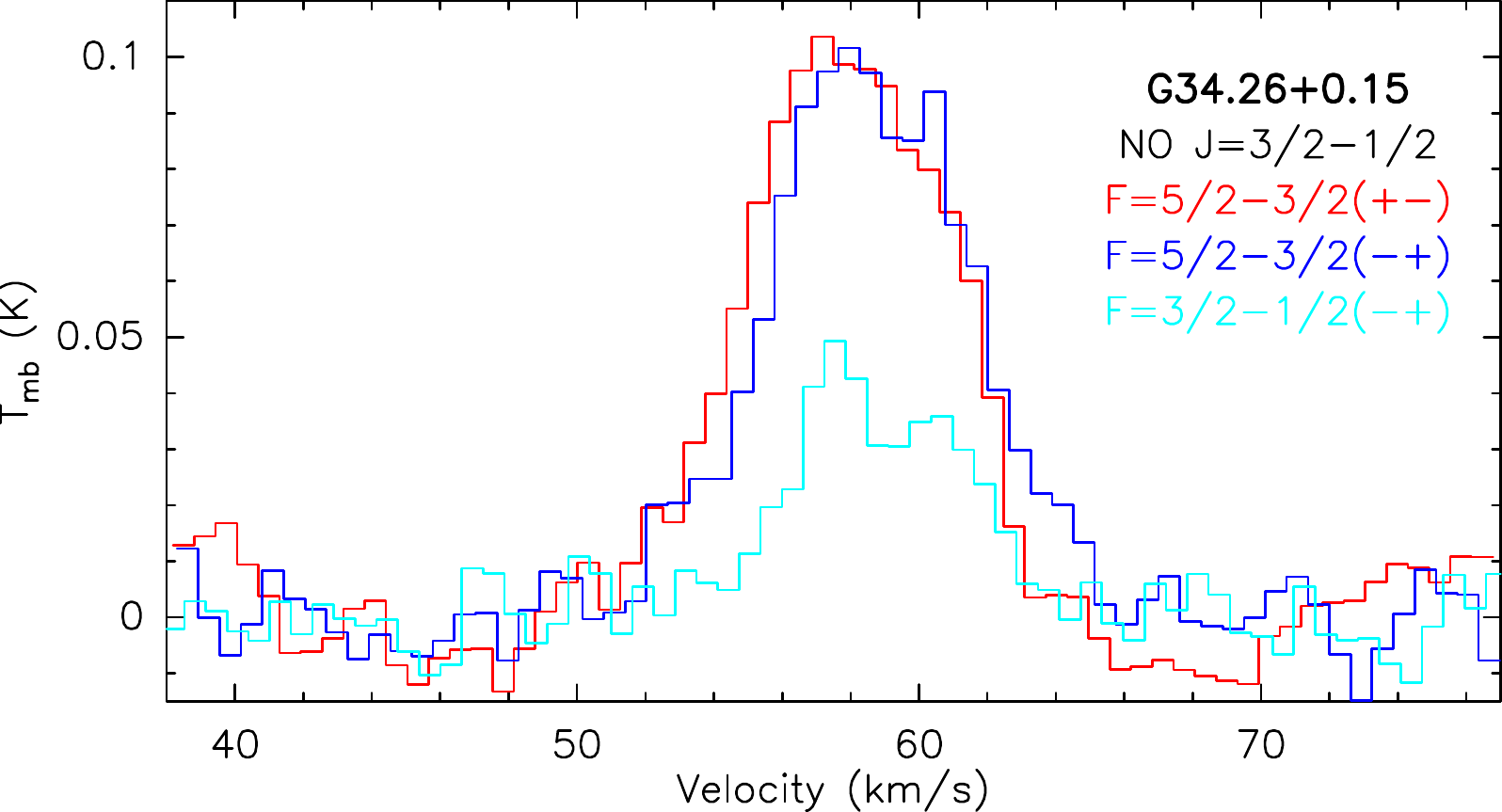}
\includegraphics[width=0.33\textwidth]{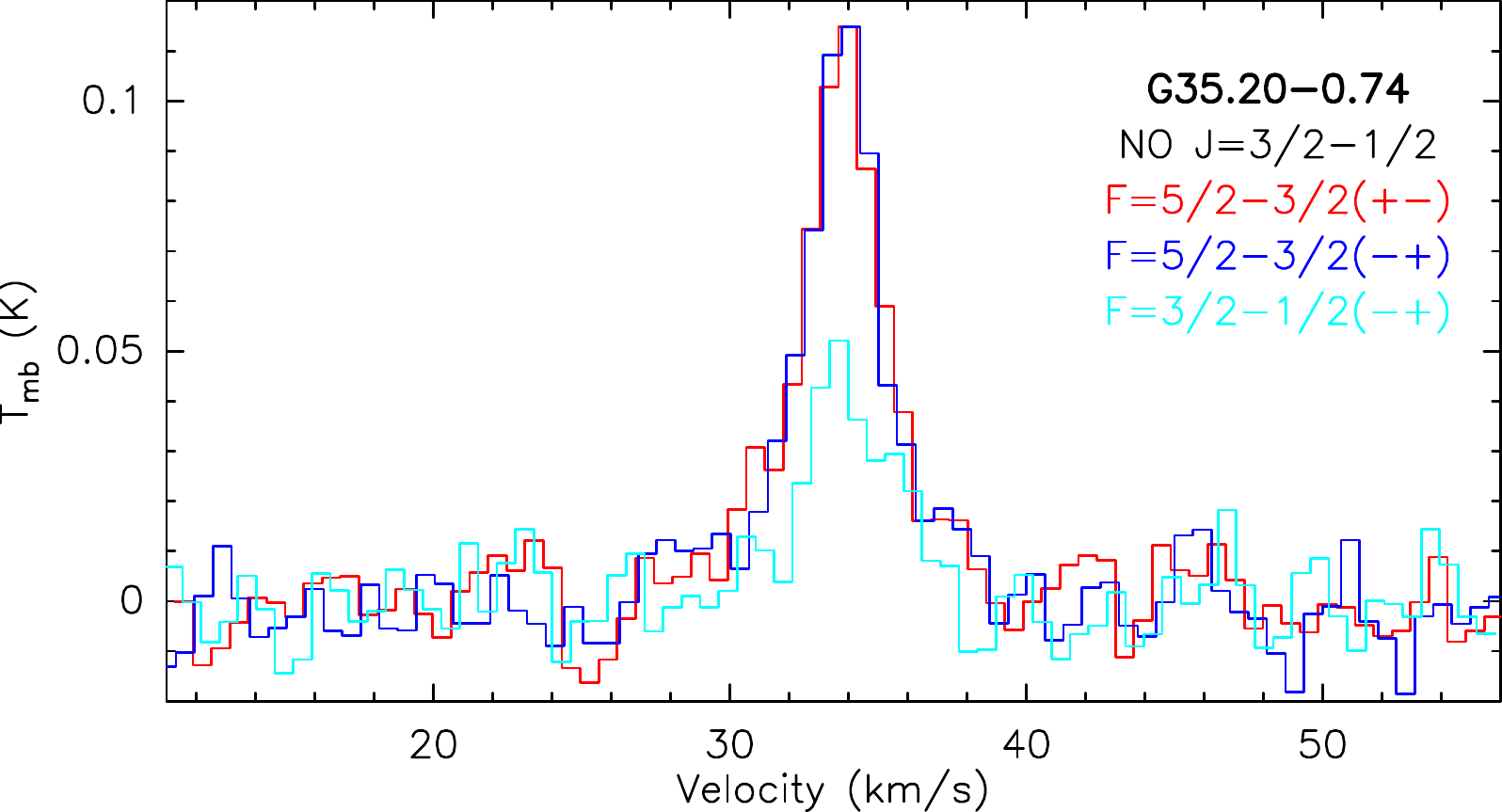}
\end{figure}

\onecolumn
\begin{figure}[!h]
\centering
\includegraphics[width=0.33\textwidth]{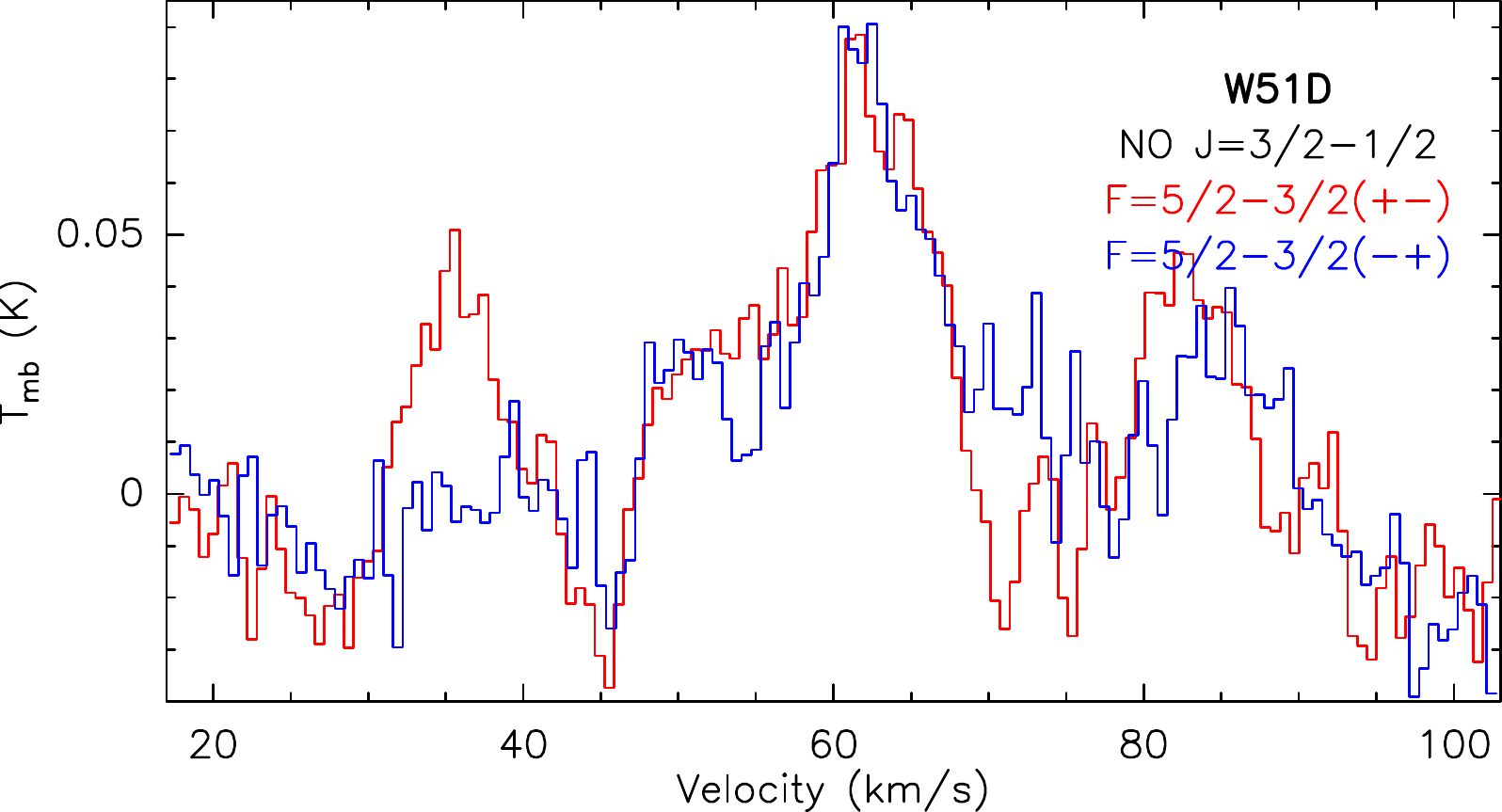}
\includegraphics[width=0.33\textwidth]{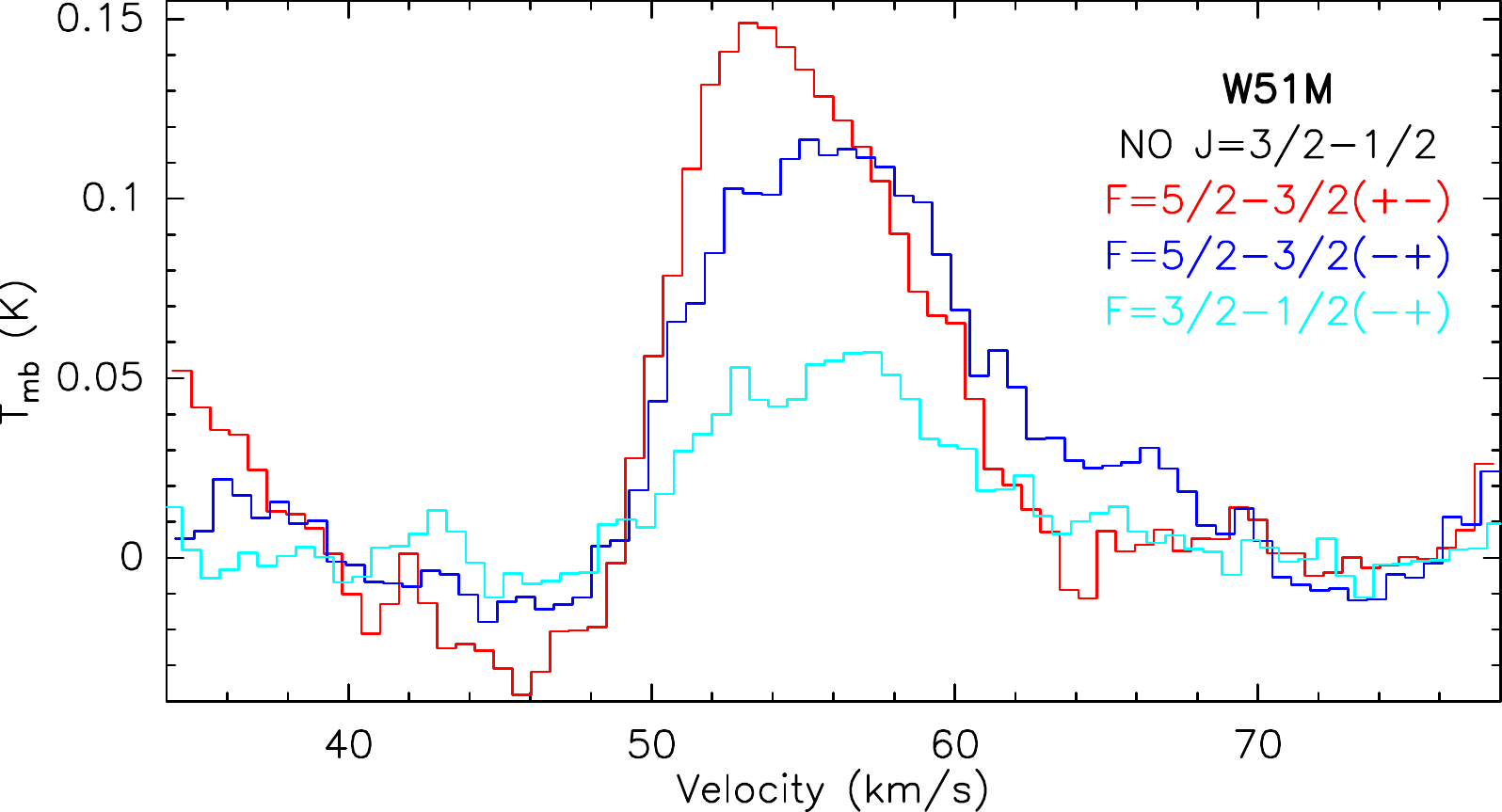}
\includegraphics[width=0.33\textwidth]{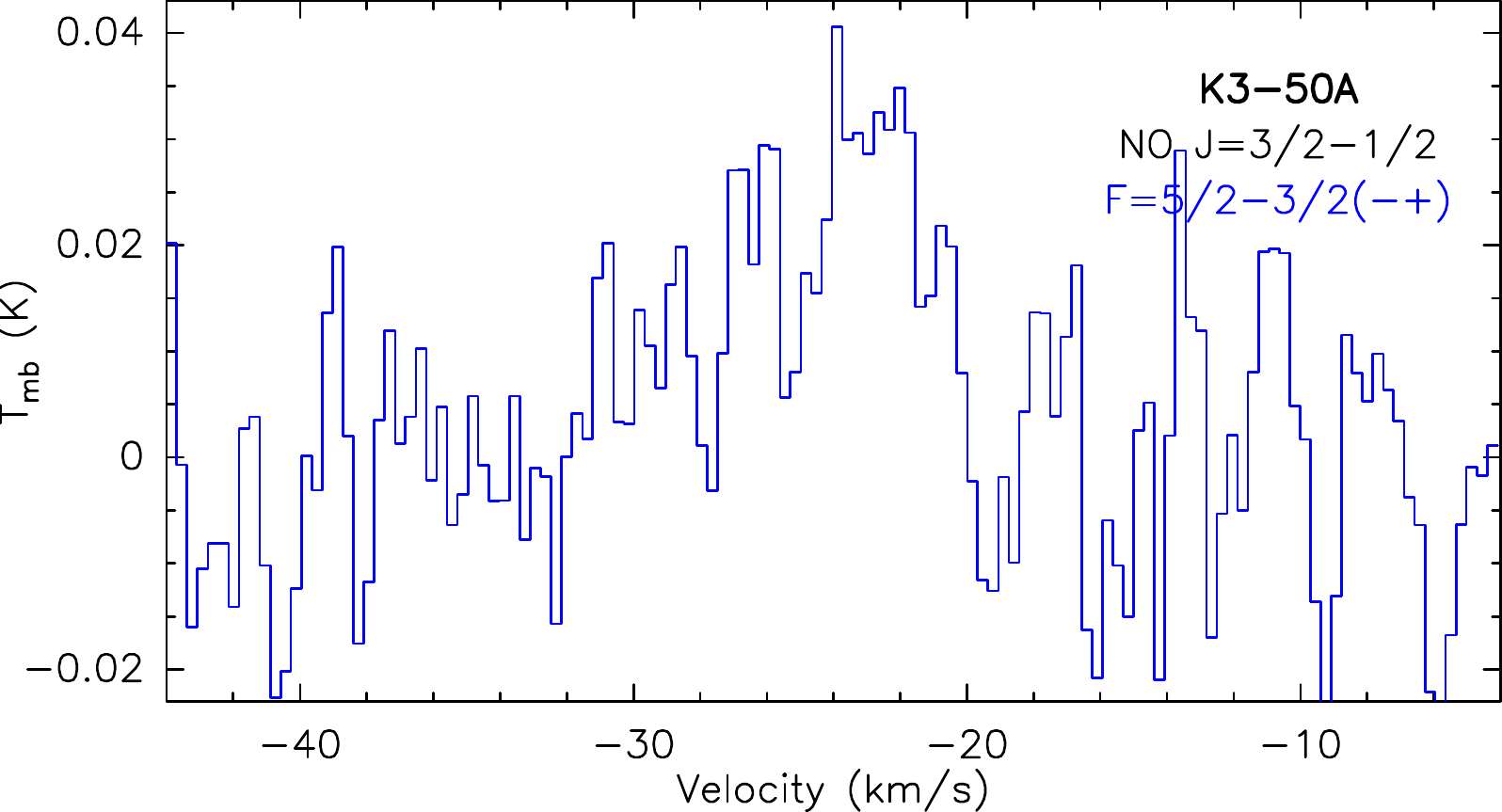}
\includegraphics[width=0.33\textwidth]{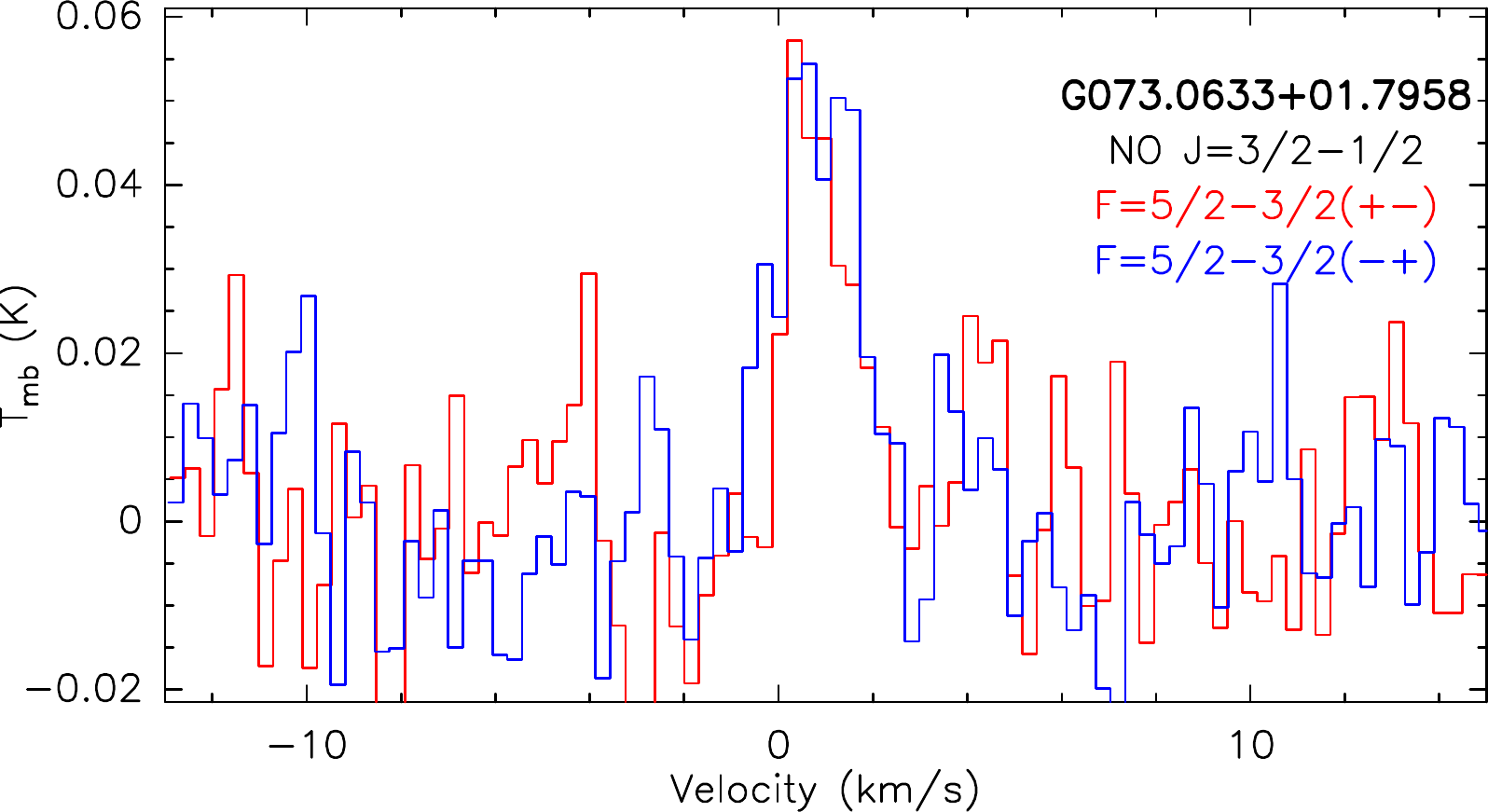}
\includegraphics[width=0.33\textwidth]{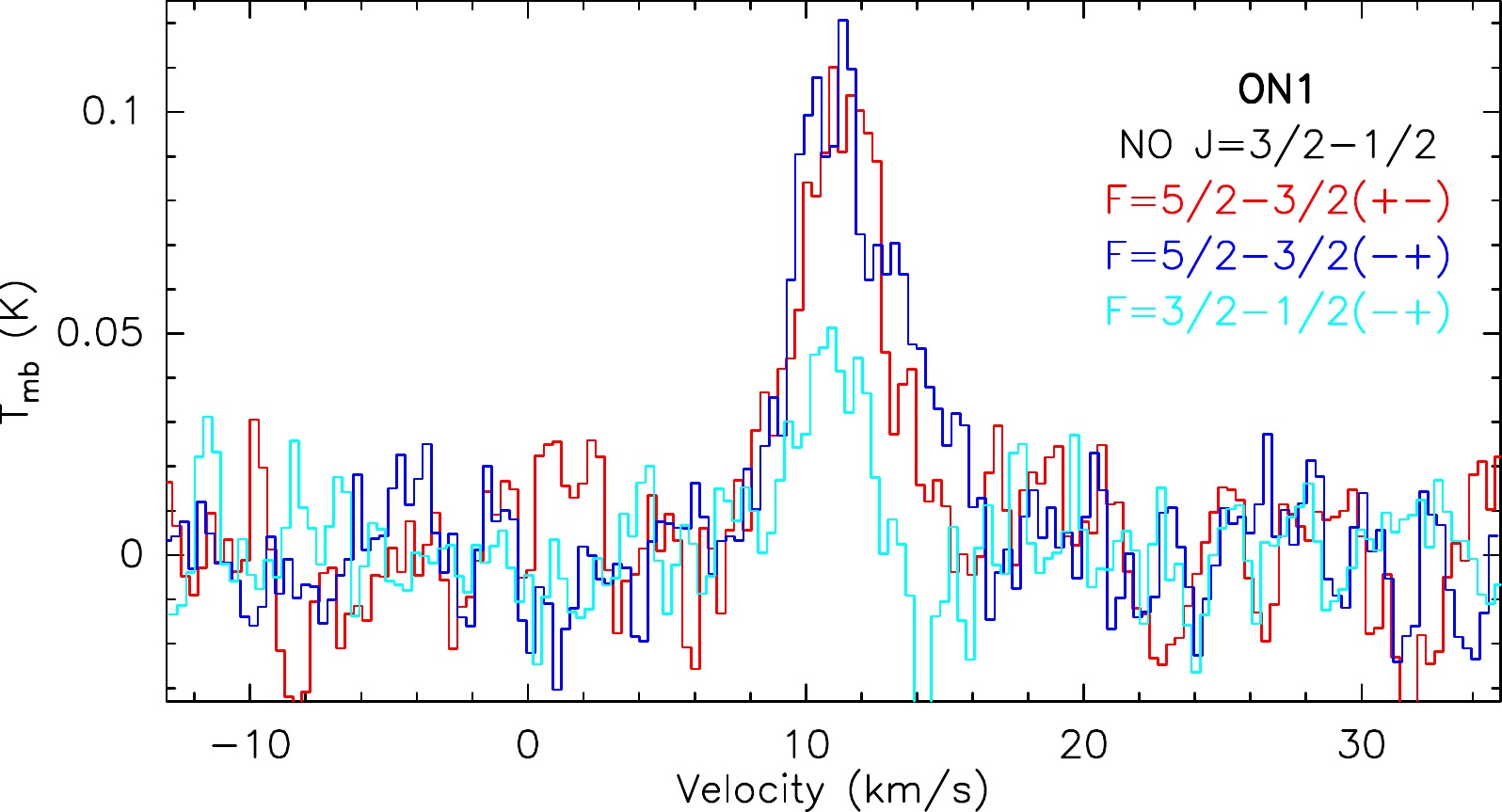}
\includegraphics[width=0.33\textwidth]{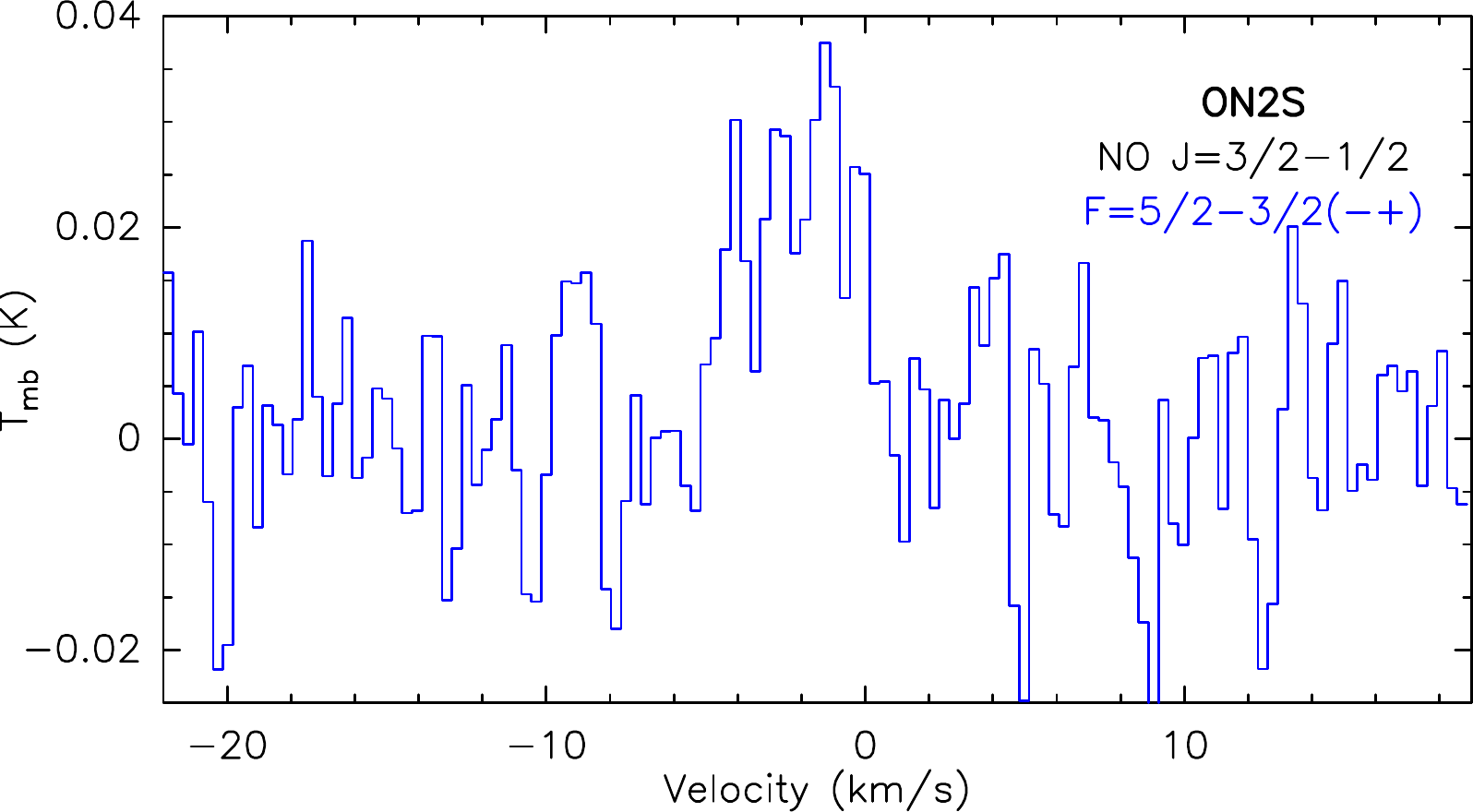}
\includegraphics[width=0.33\textwidth]{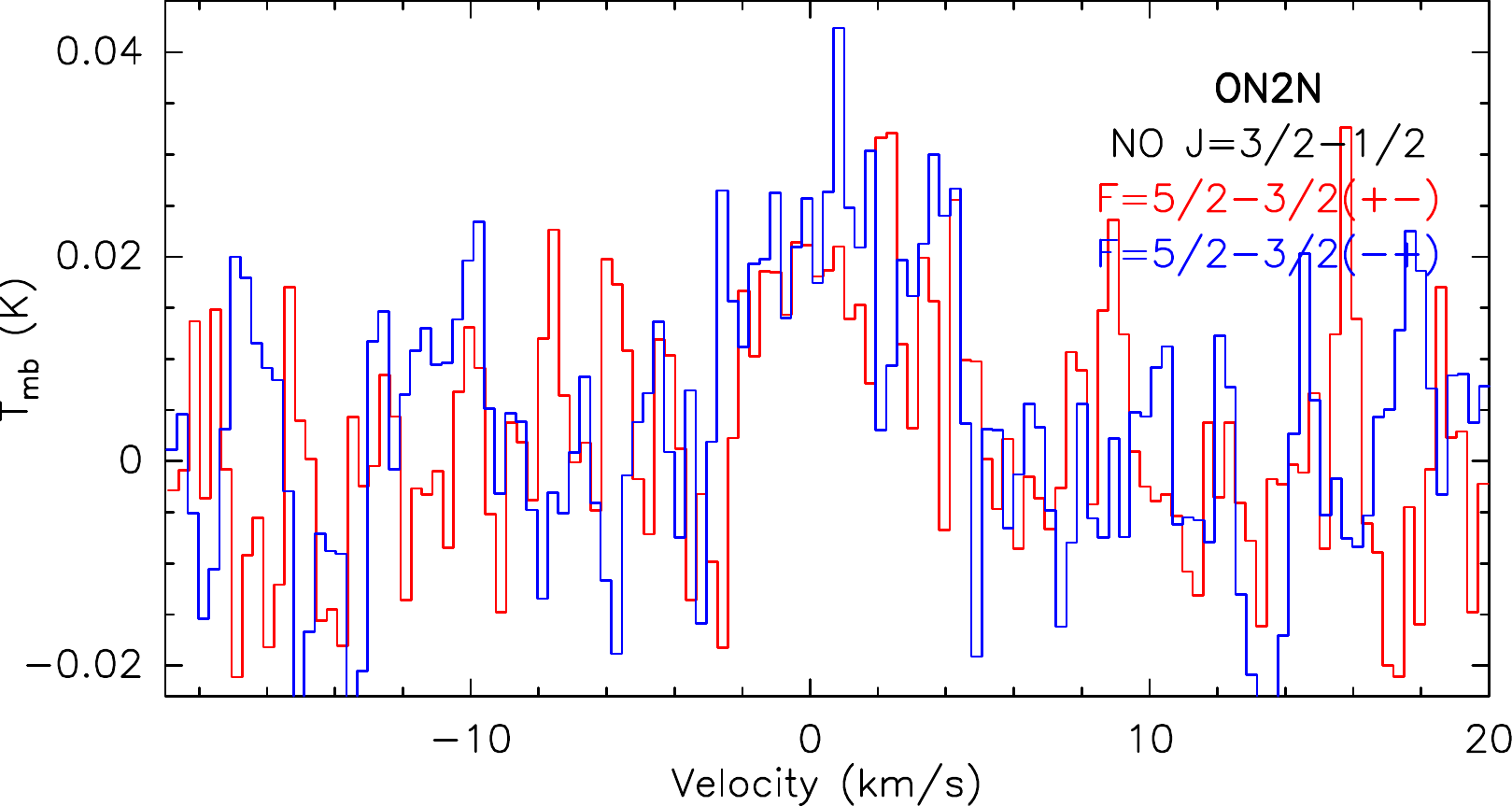}
\includegraphics[width=0.33\textwidth]{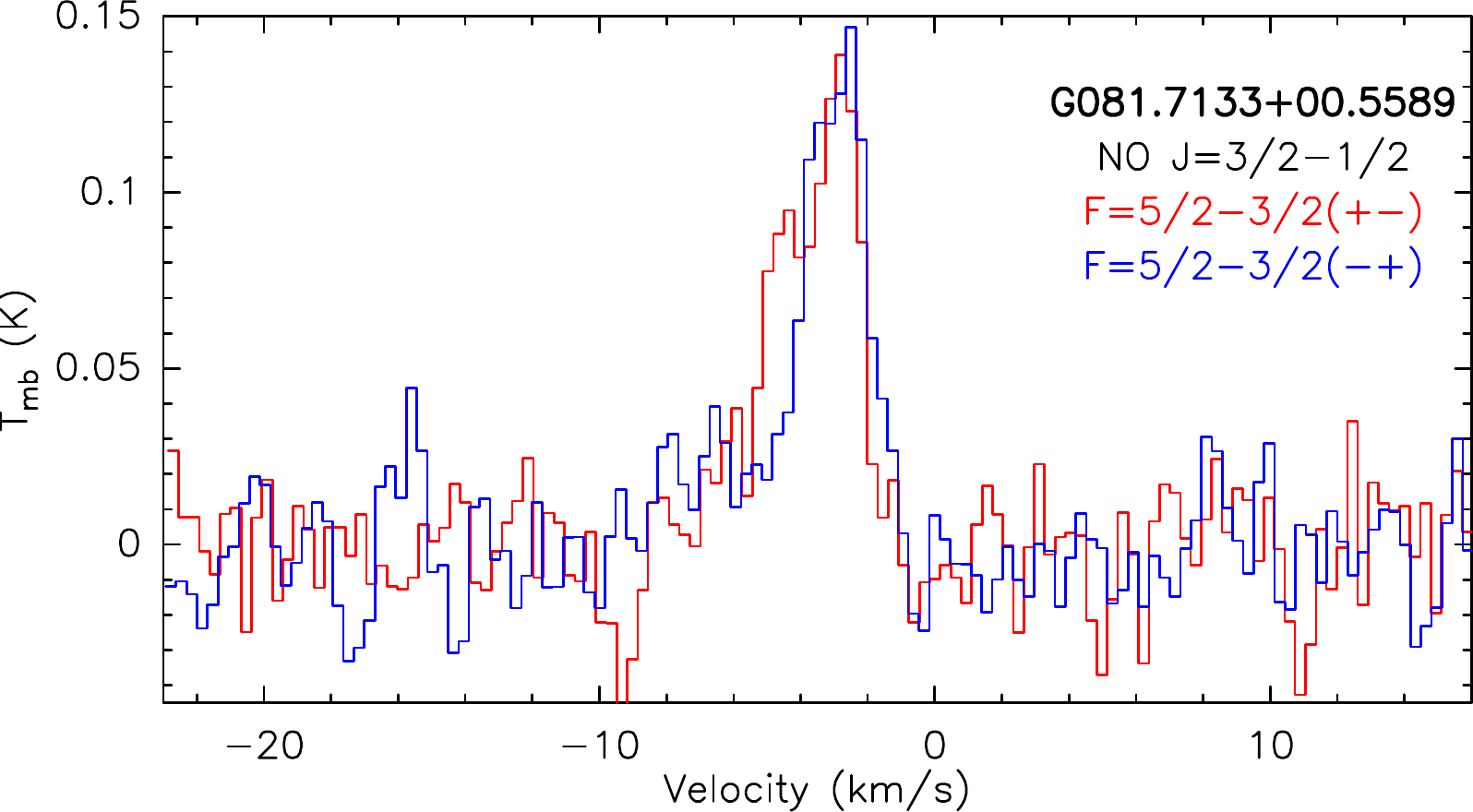}
\includegraphics[width=0.33\textwidth]{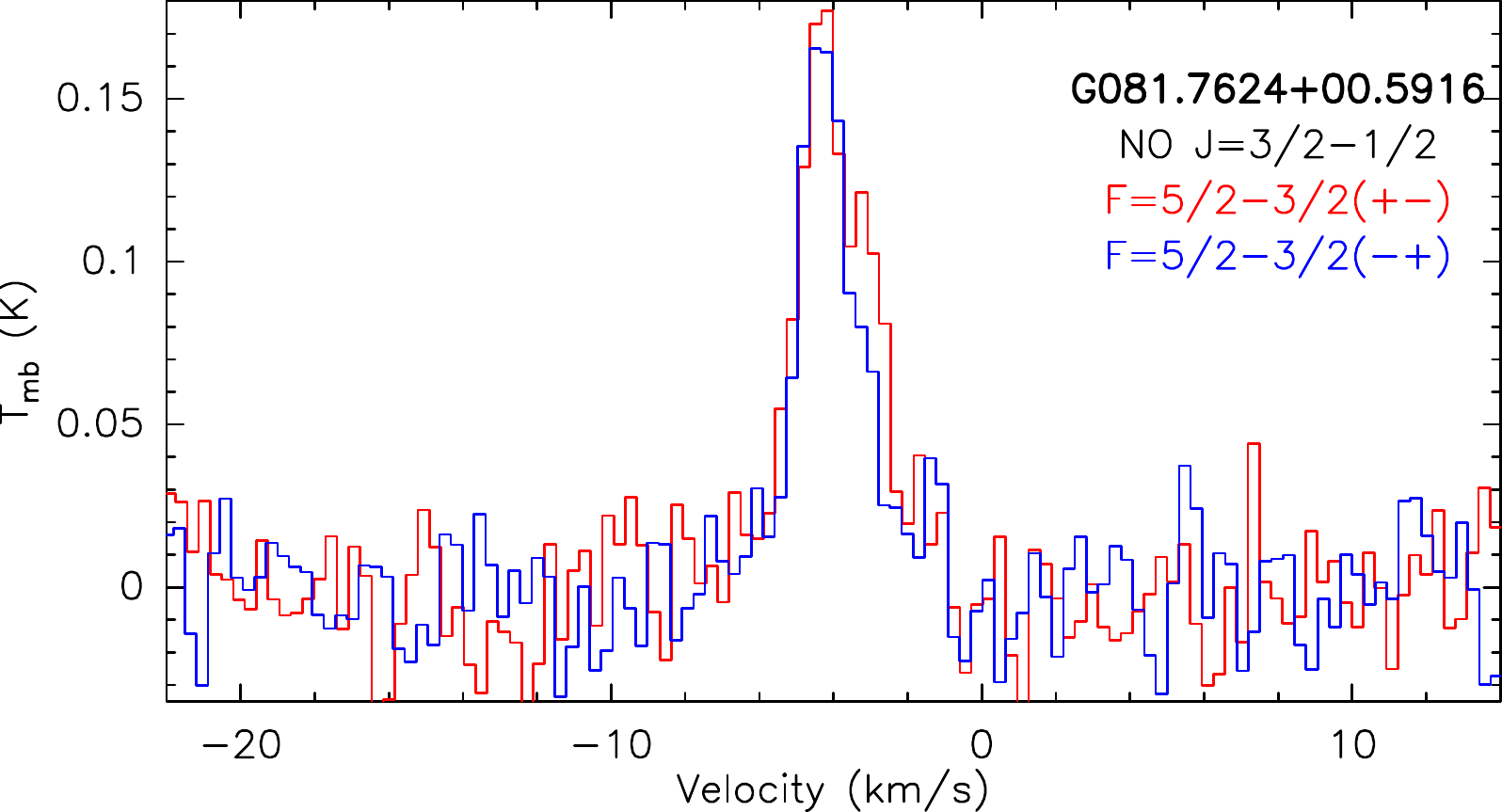}
\includegraphics[width=0.33\textwidth]{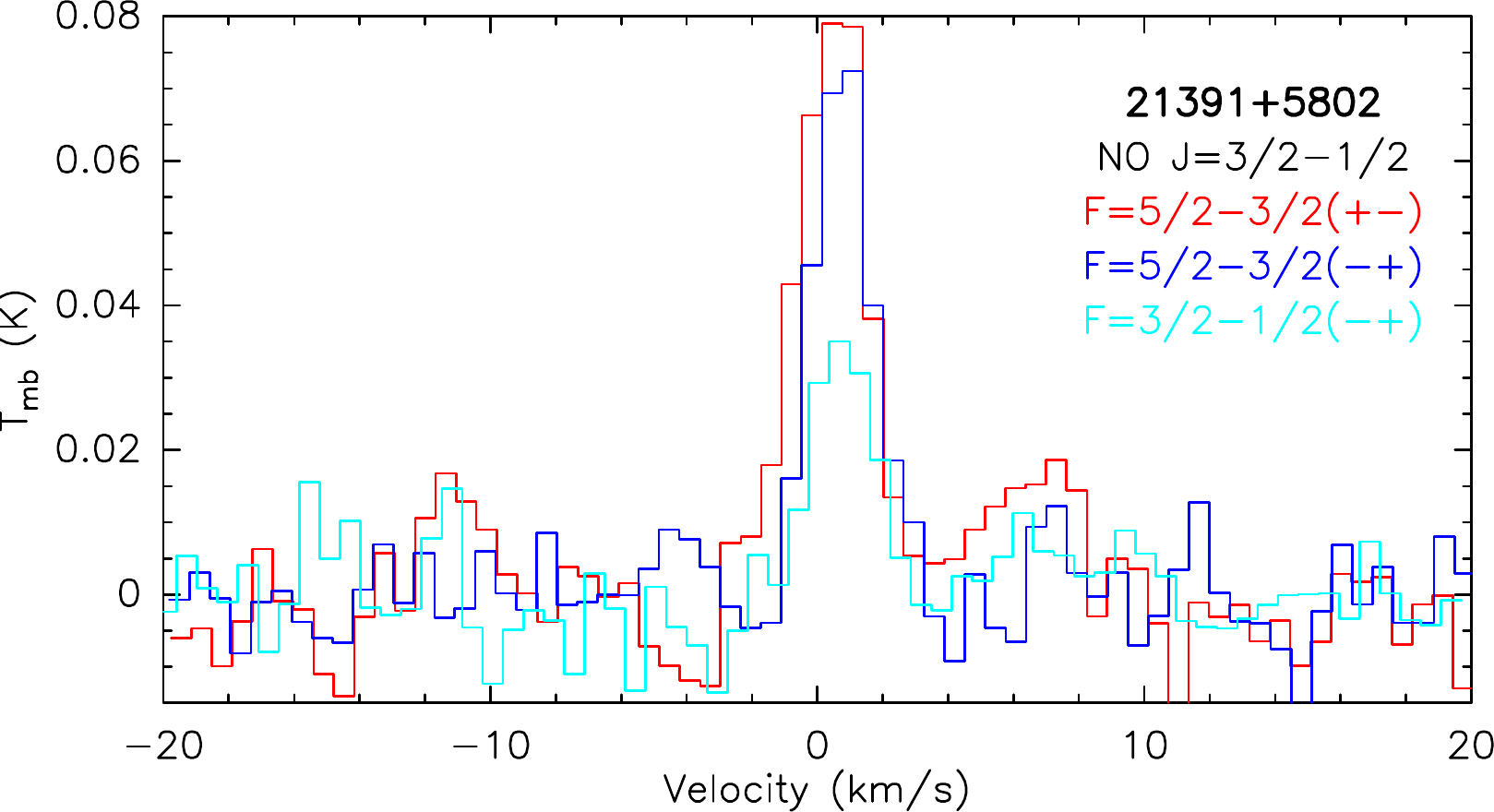}
\includegraphics[width=0.33\textwidth]{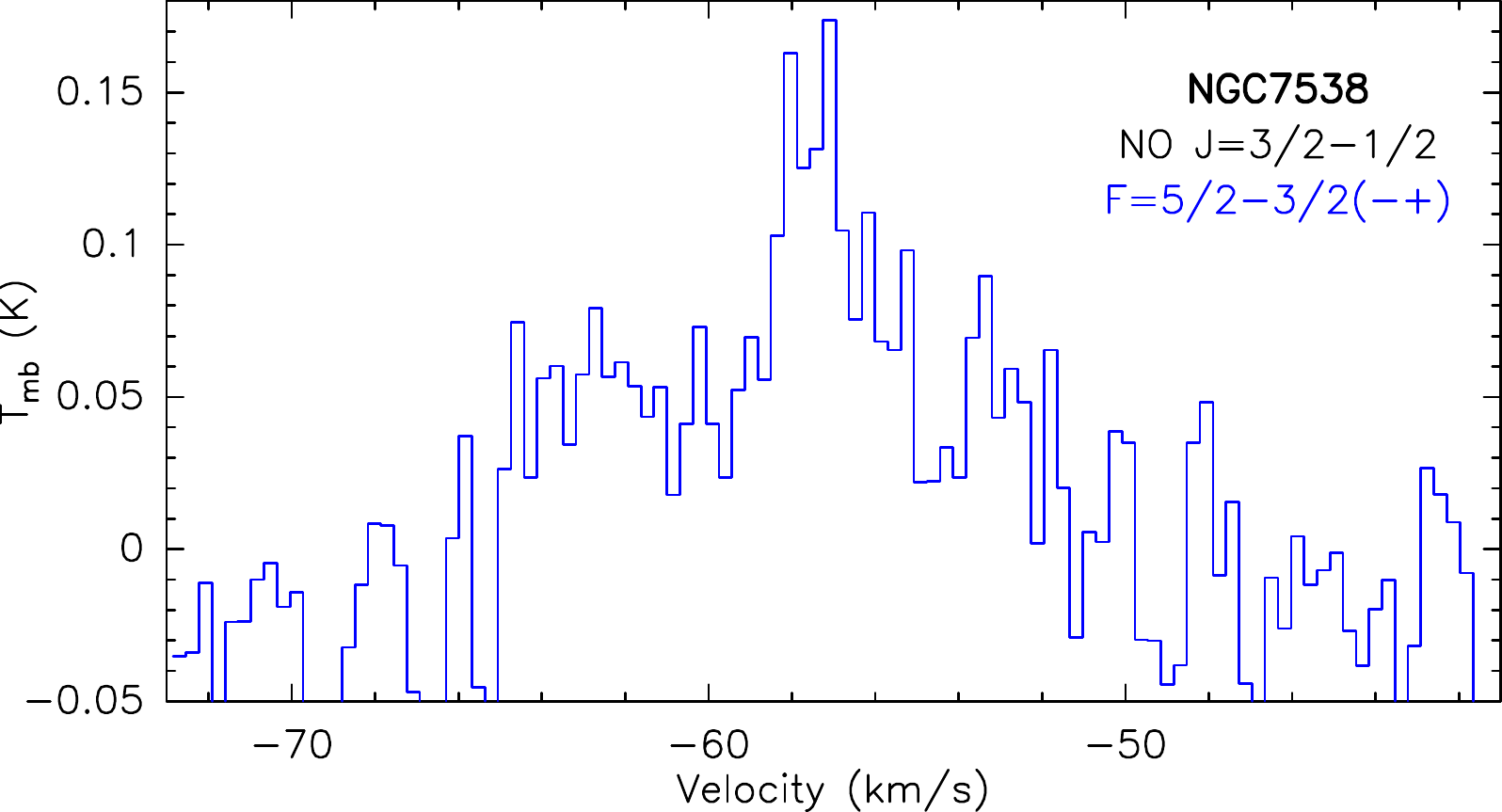}
\includegraphics[width=0.33\textwidth]{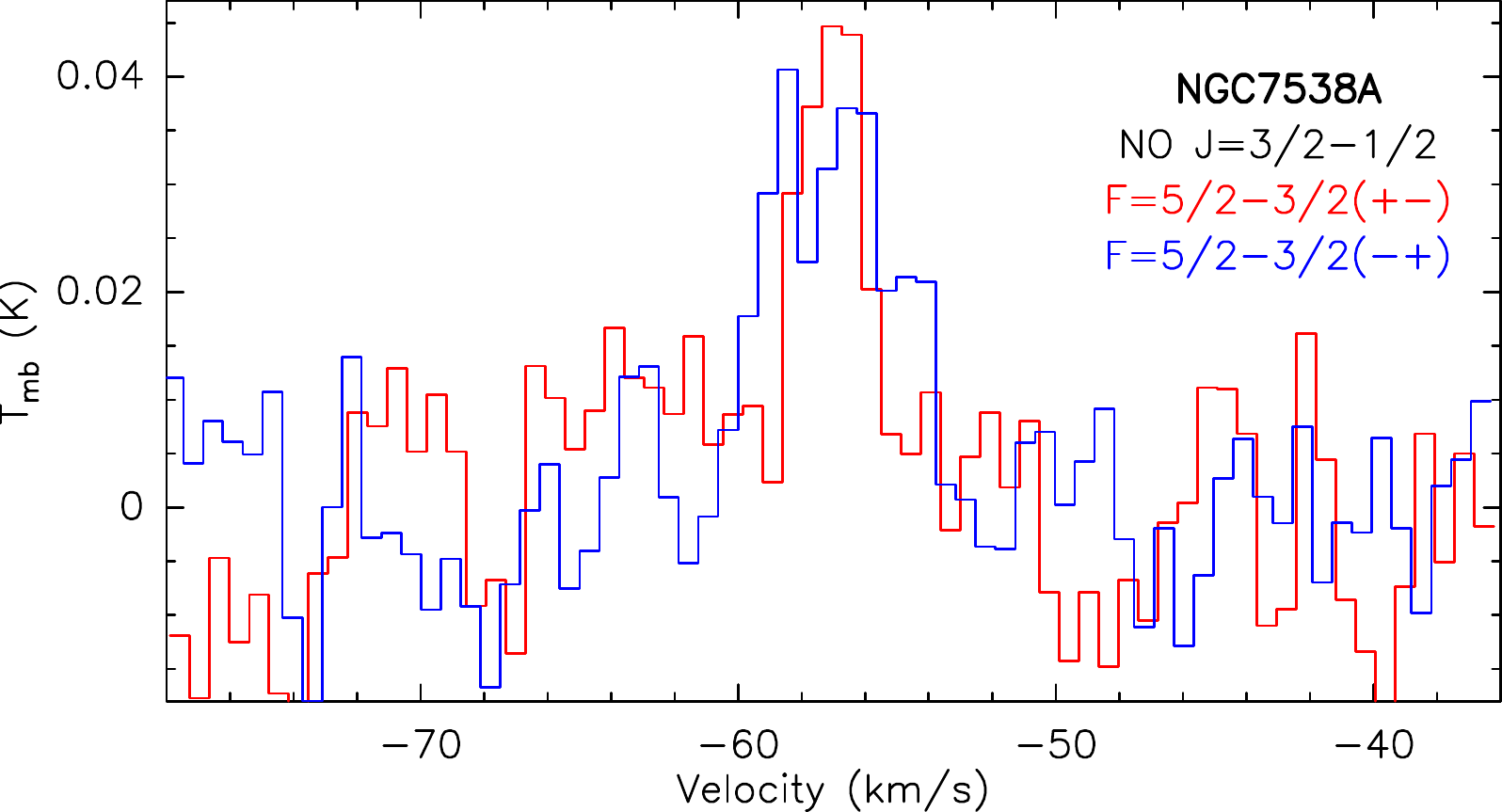}
\caption{Detected NO lines of three transitions.}
\label{spectra of NO}
\end{figure}


\onecolumn
\begin{figure}[!h]
\centering
\includegraphics[width=0.33\textwidth]{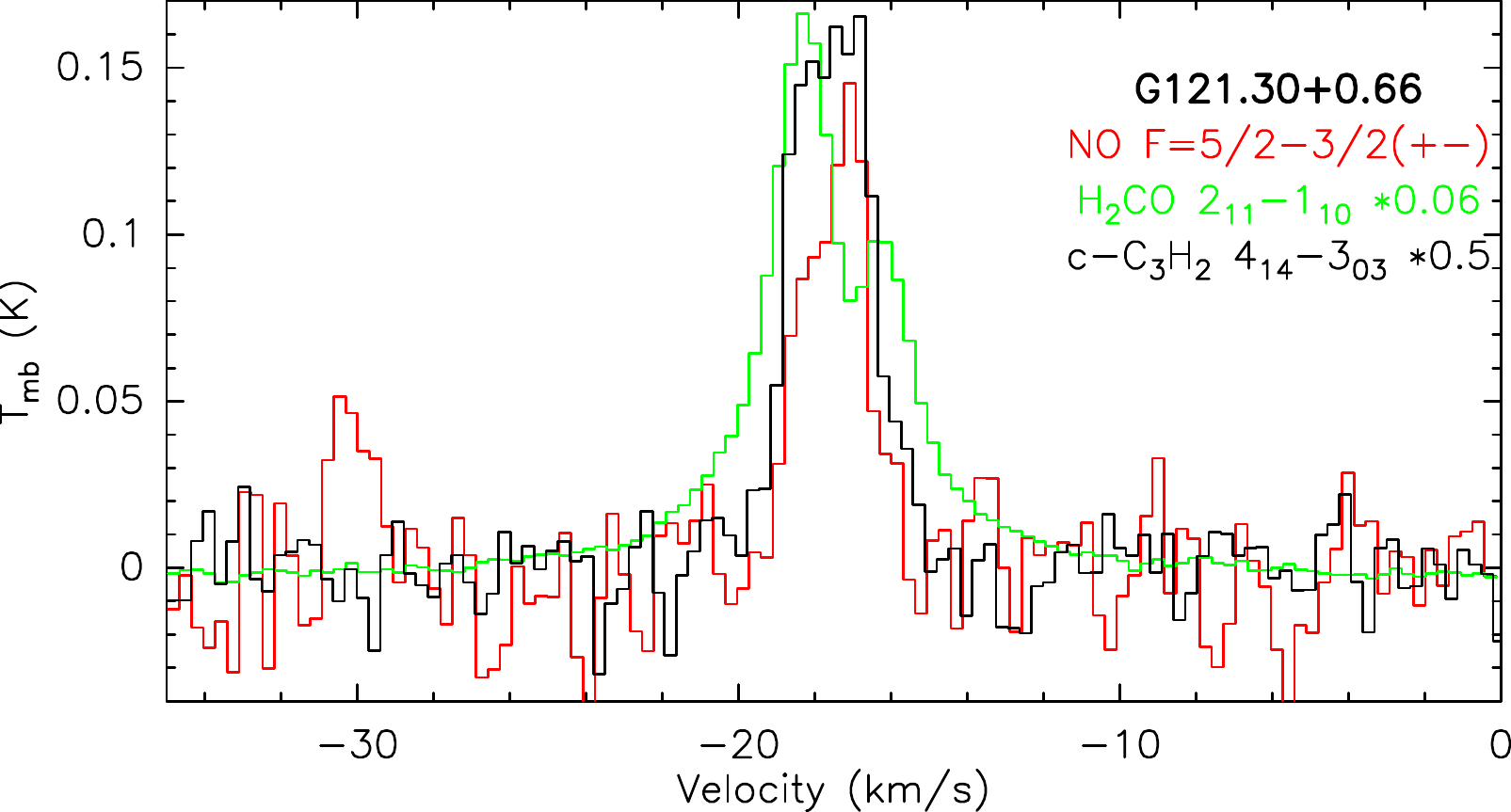}
\includegraphics[width=0.33\textwidth]{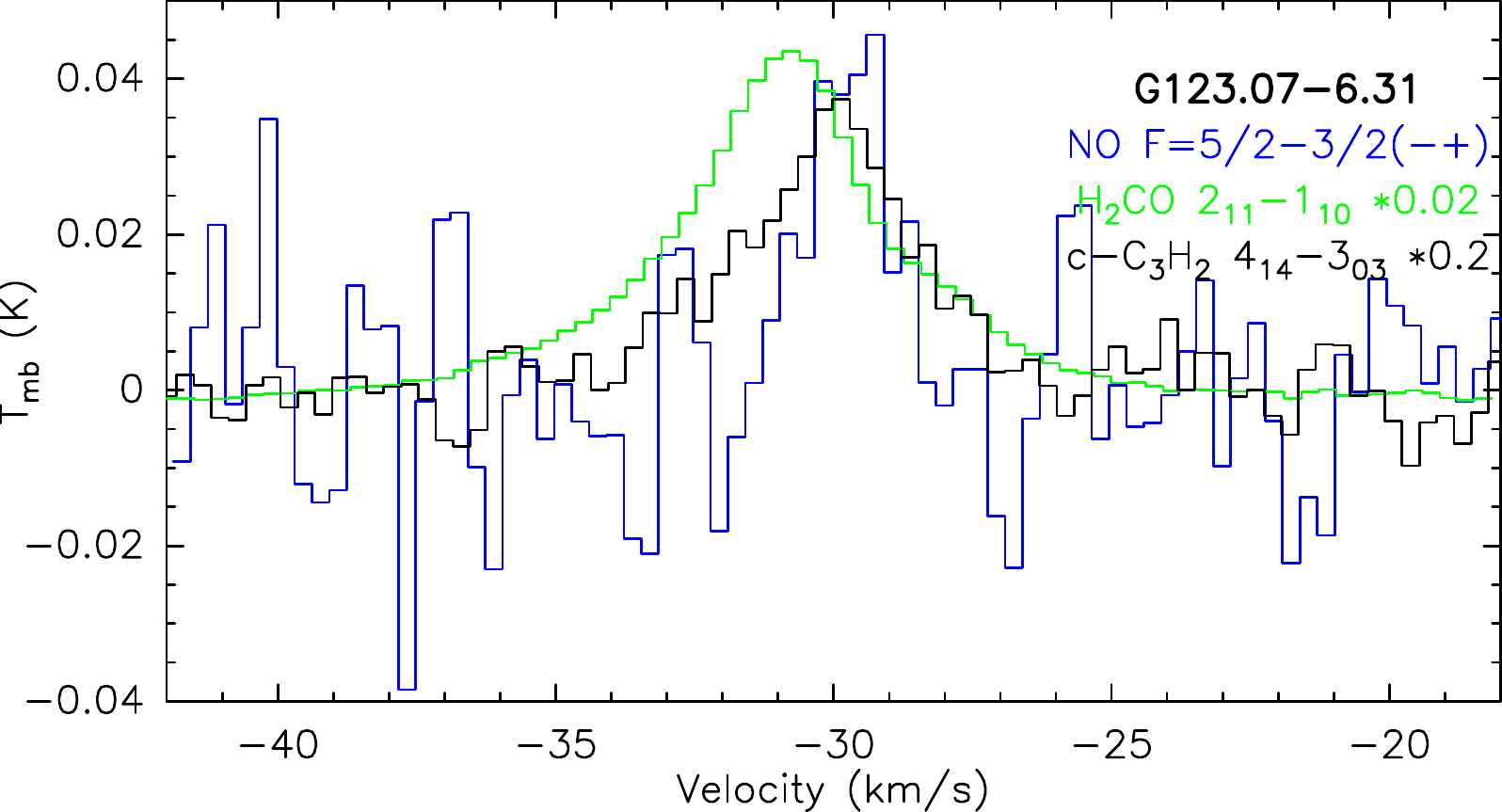}
\includegraphics[width=0.33\textwidth]{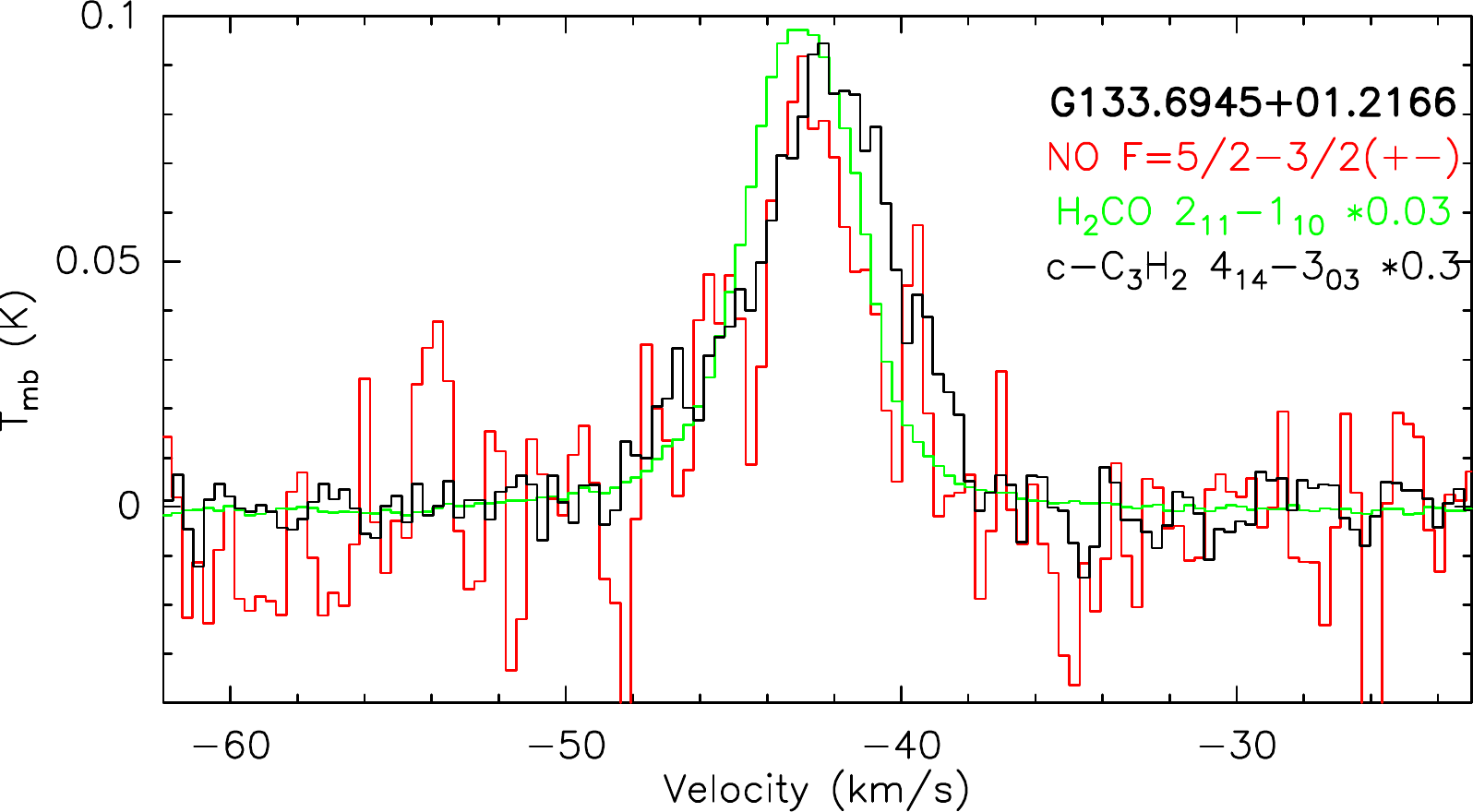}
\includegraphics[width=0.33\textwidth]{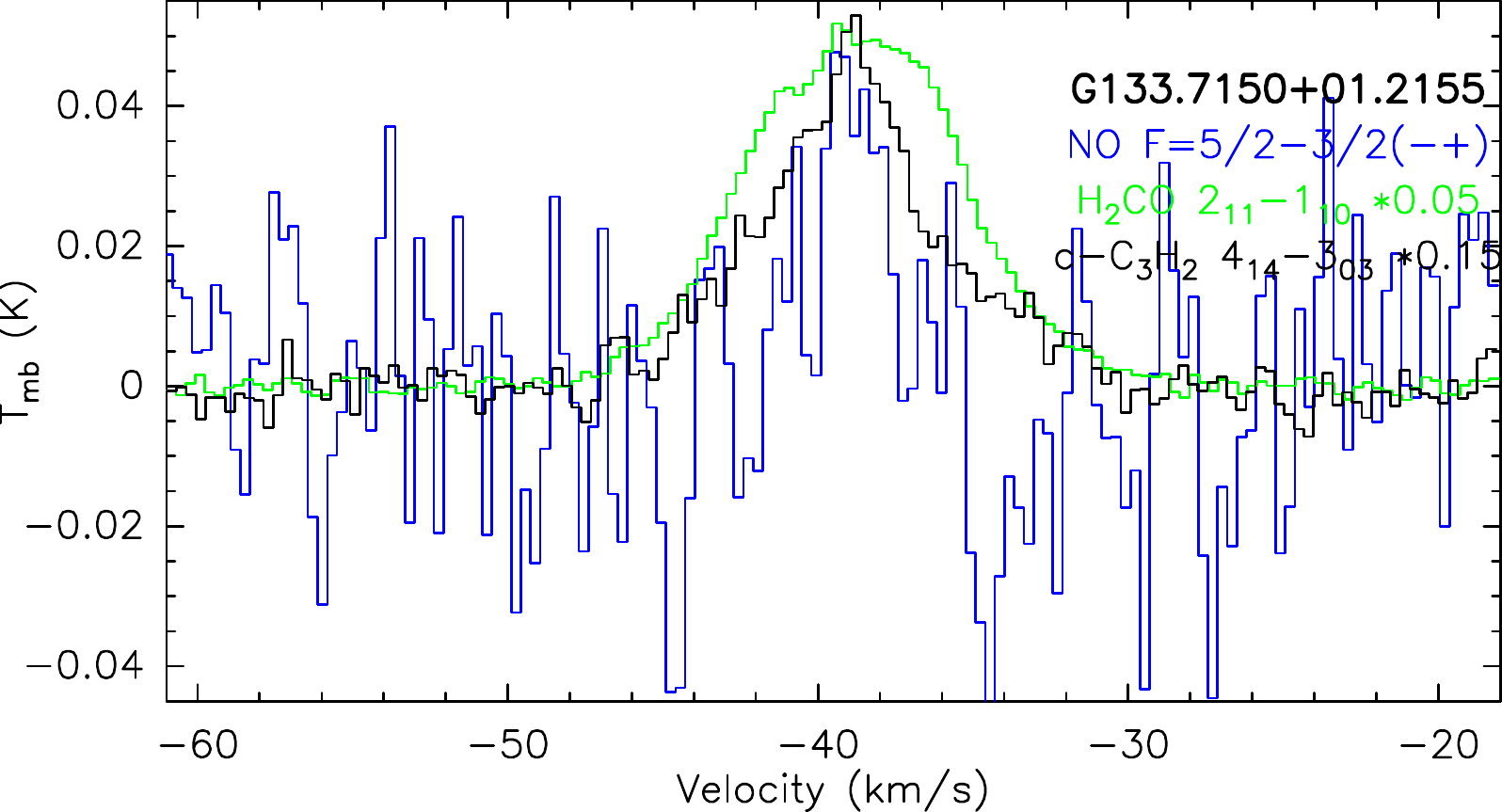}
\includegraphics[width=0.33\textwidth]{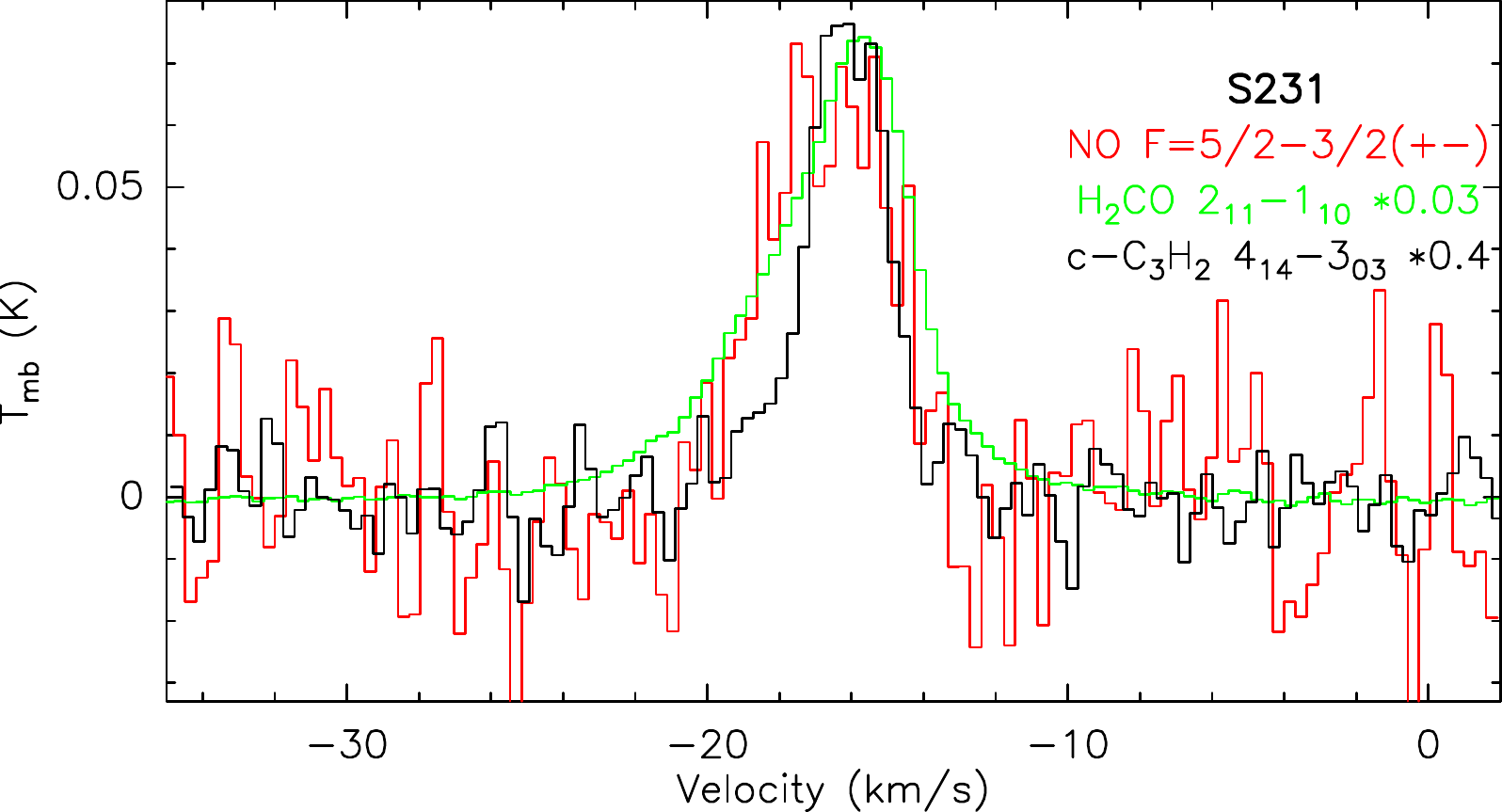}
\includegraphics[width=0.33\textwidth]{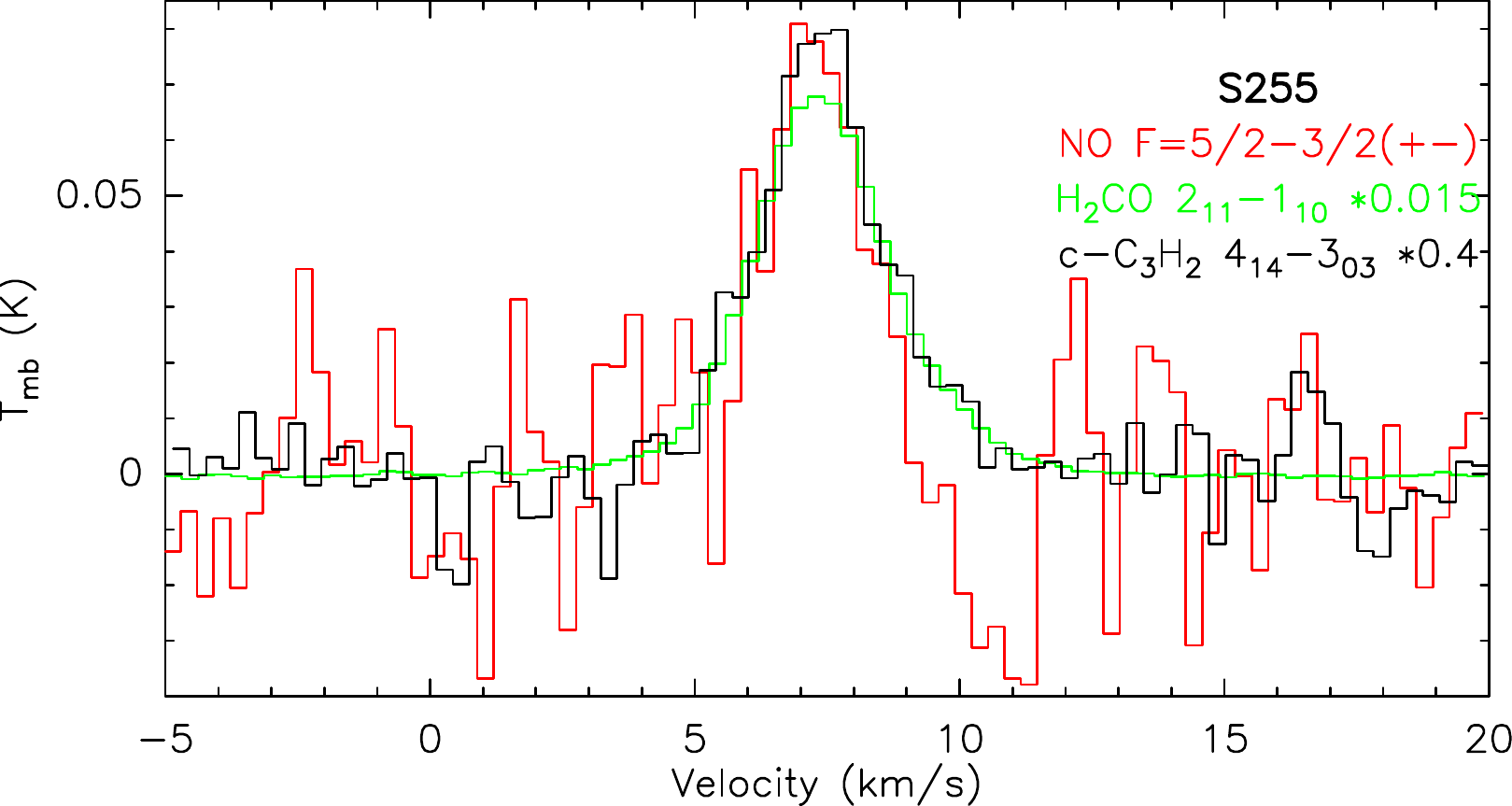}
\includegraphics[width=0.33\textwidth]{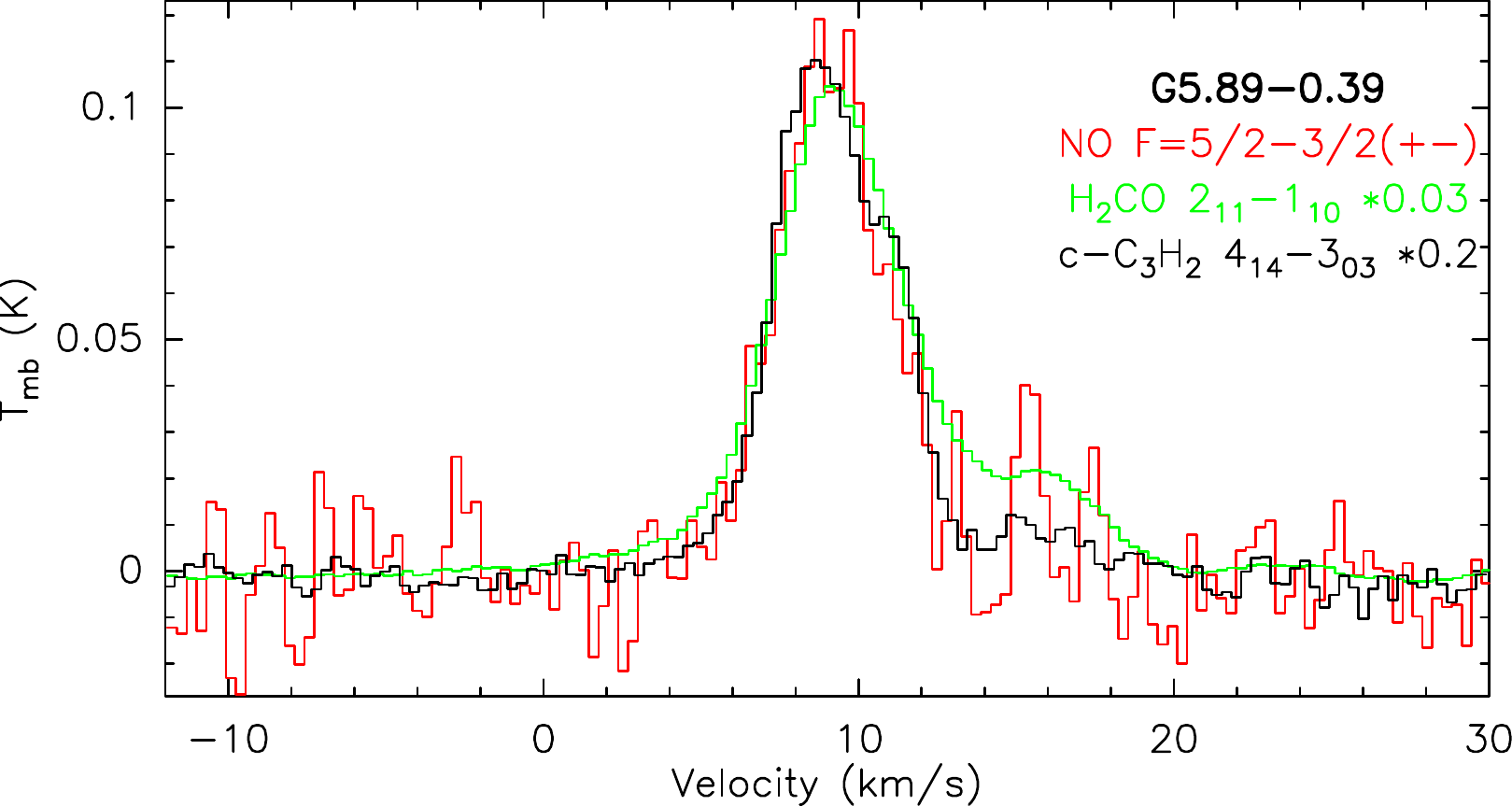}
\includegraphics[width=0.33\textwidth]{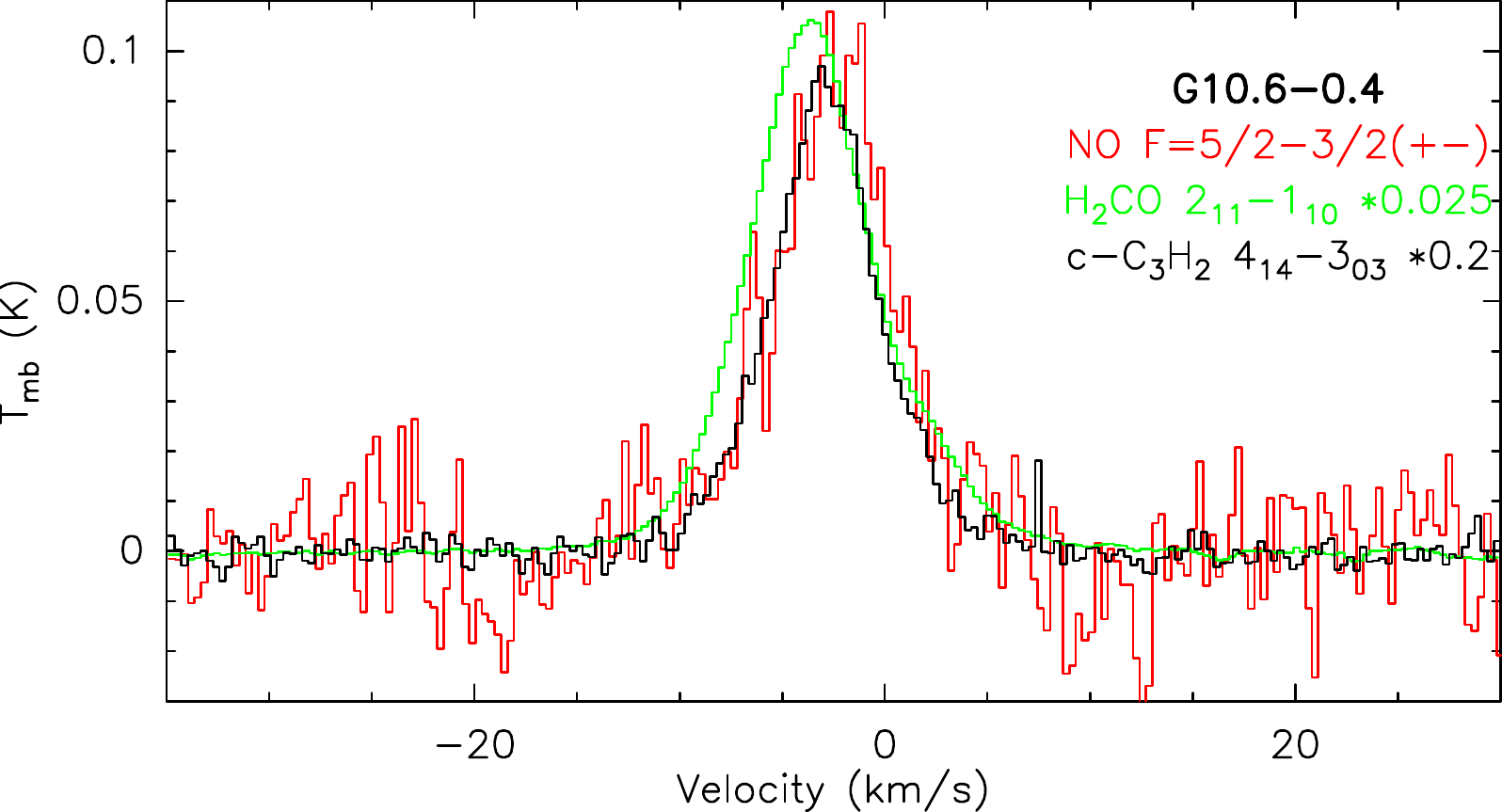}
\includegraphics[width=0.33\textwidth]{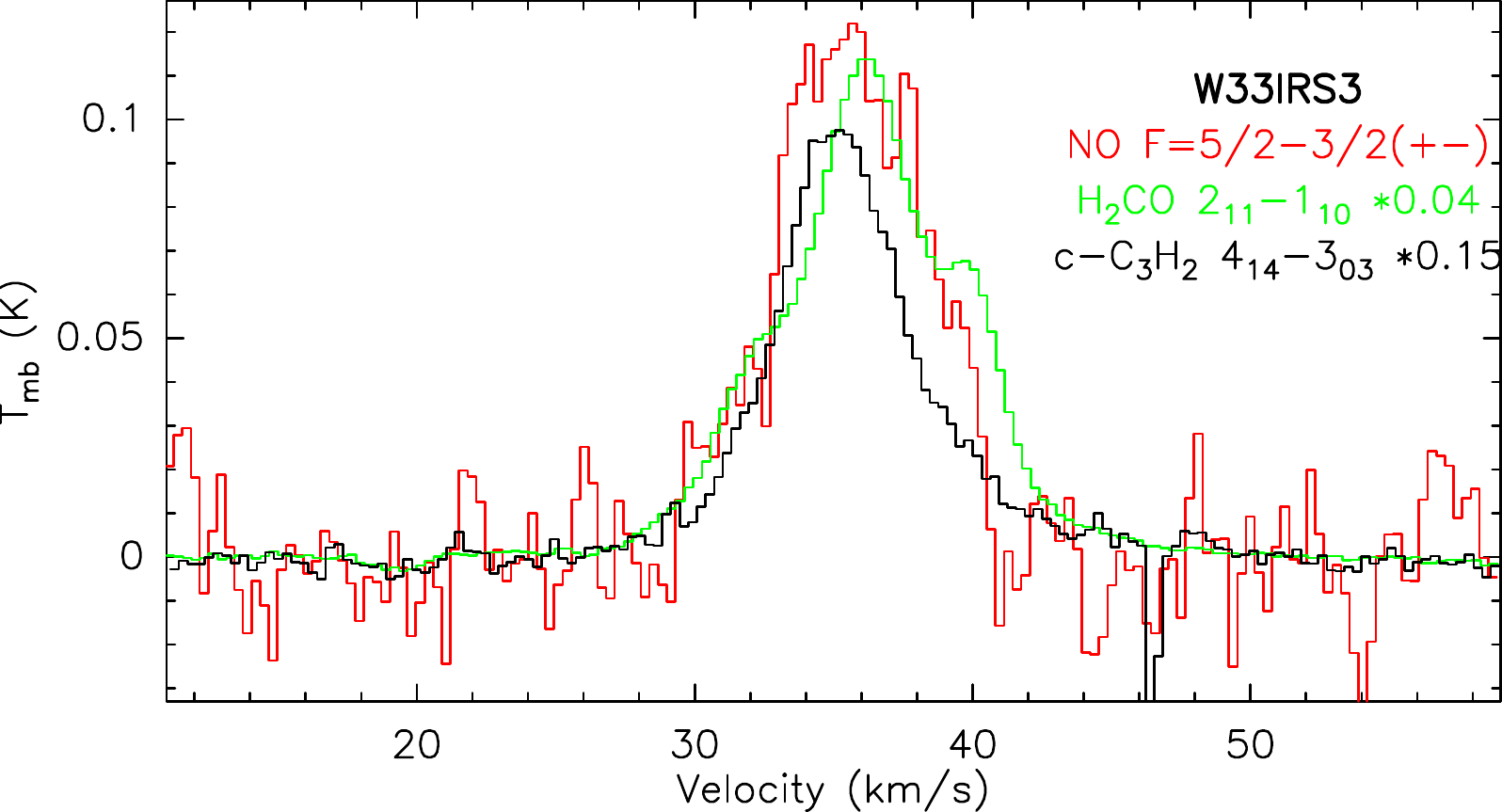}
\includegraphics[width=0.33\textwidth]{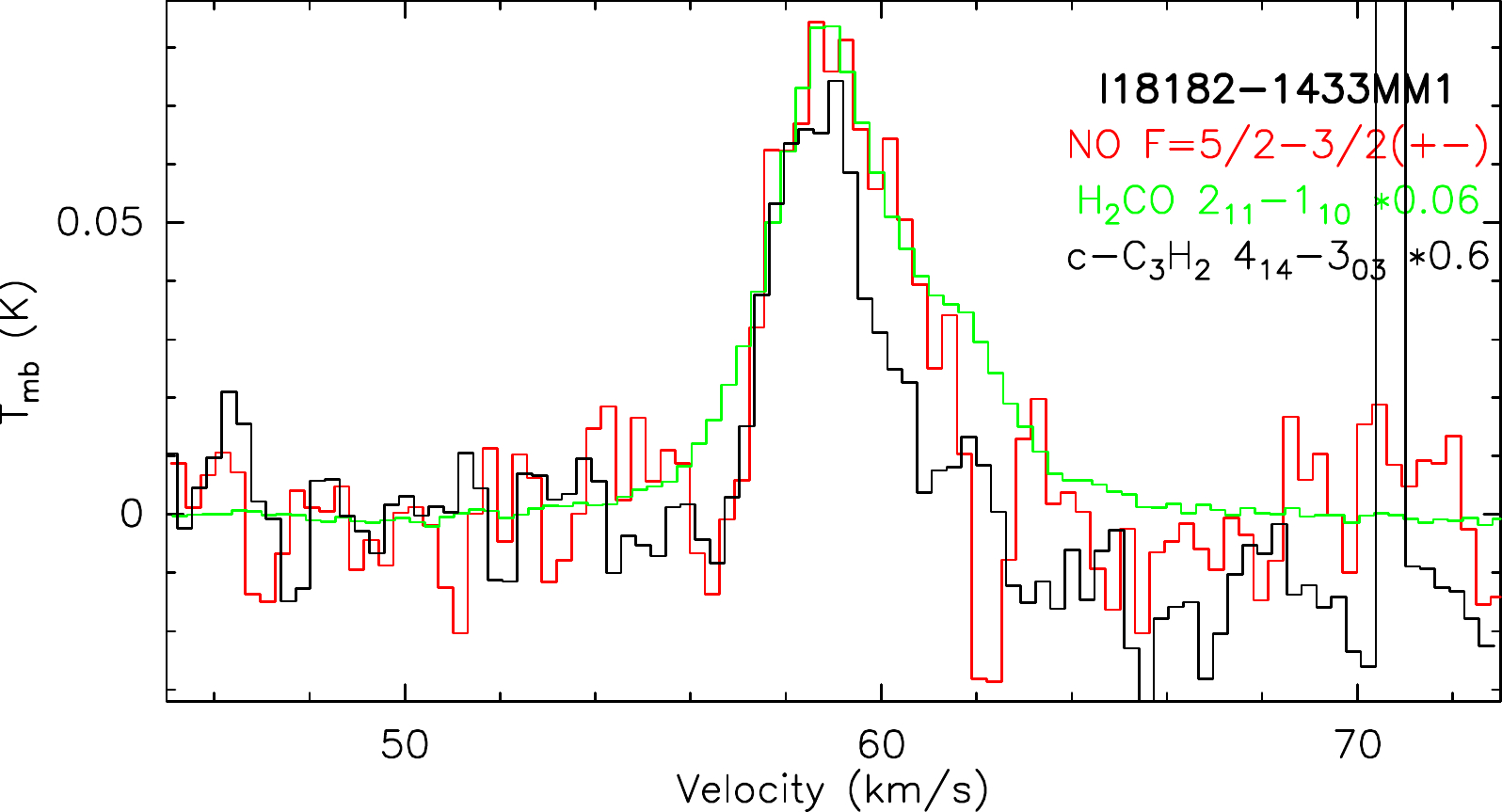}
\includegraphics[width=0.33\textwidth]{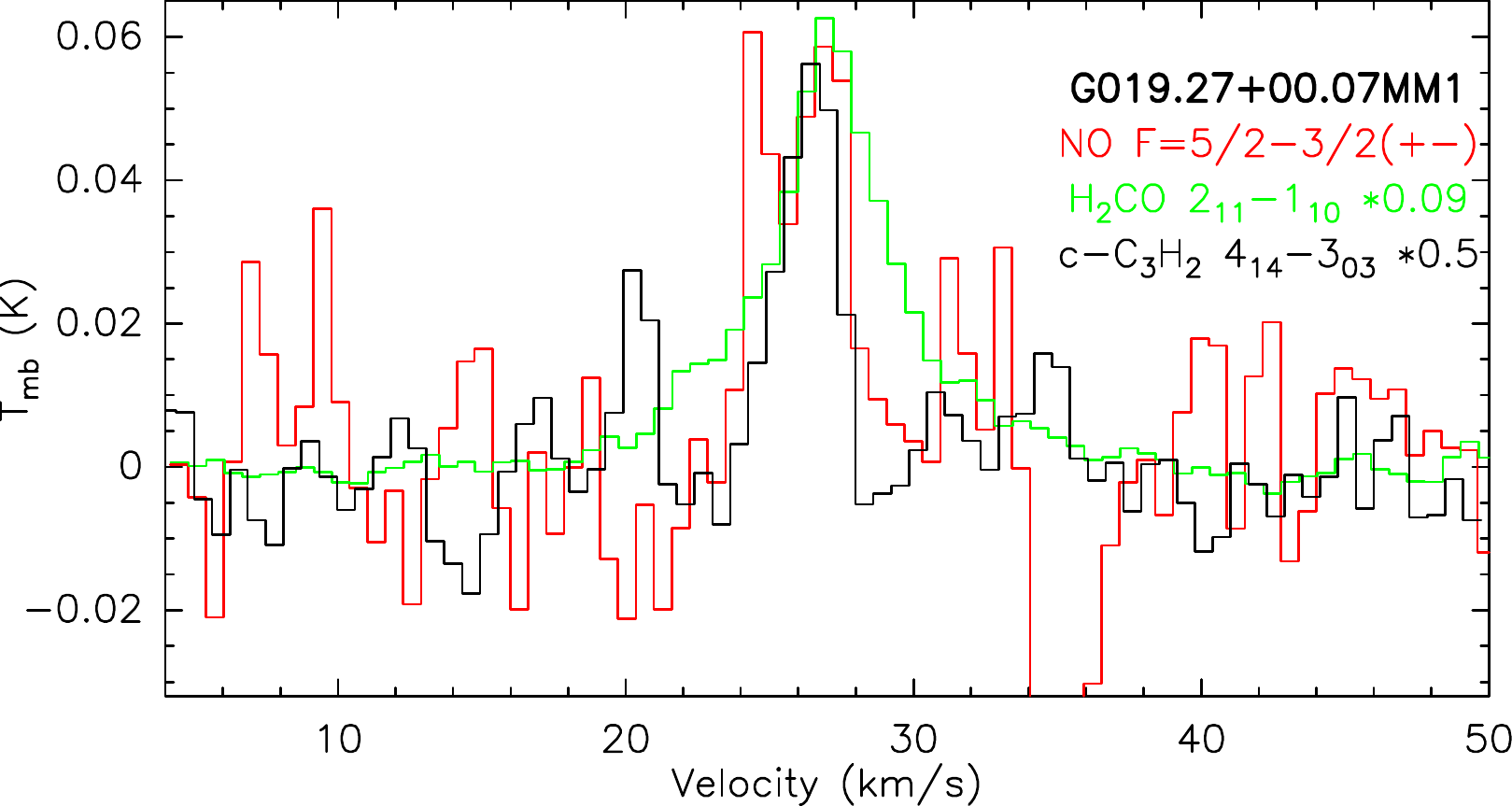}
\includegraphics[width=0.33\textwidth]{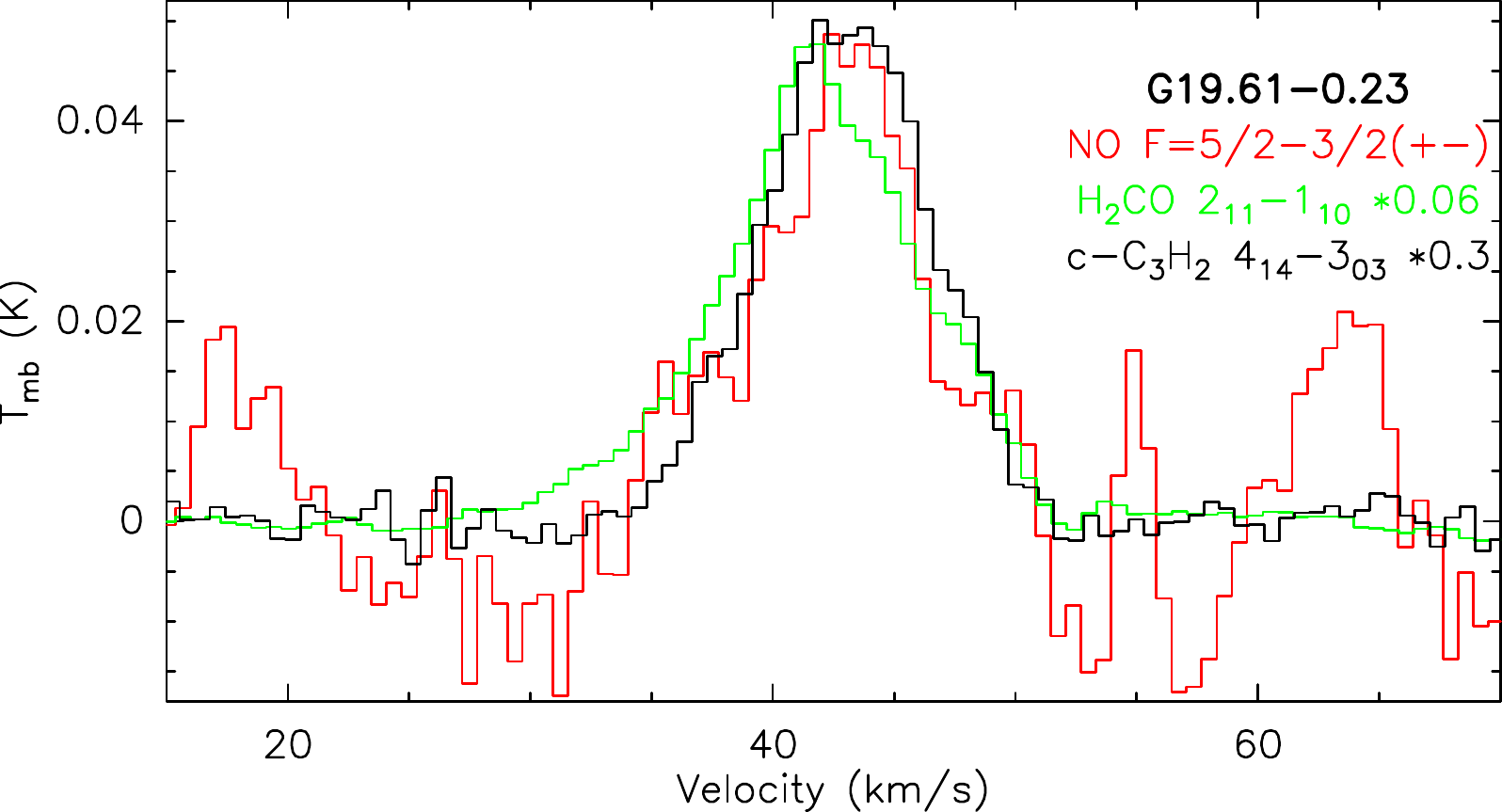}
\includegraphics[width=0.33\textwidth]{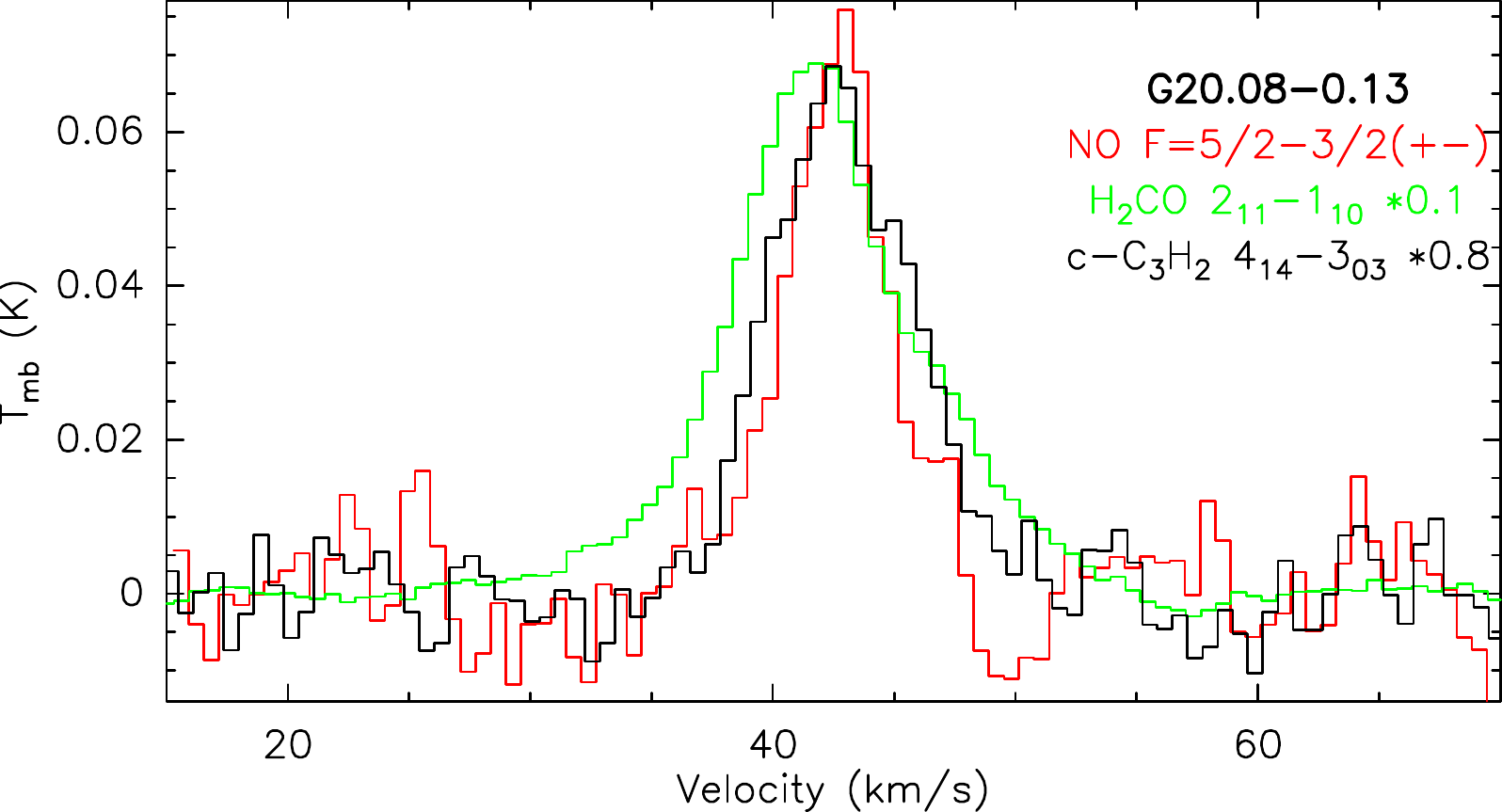}
\includegraphics[width=0.33\textwidth]{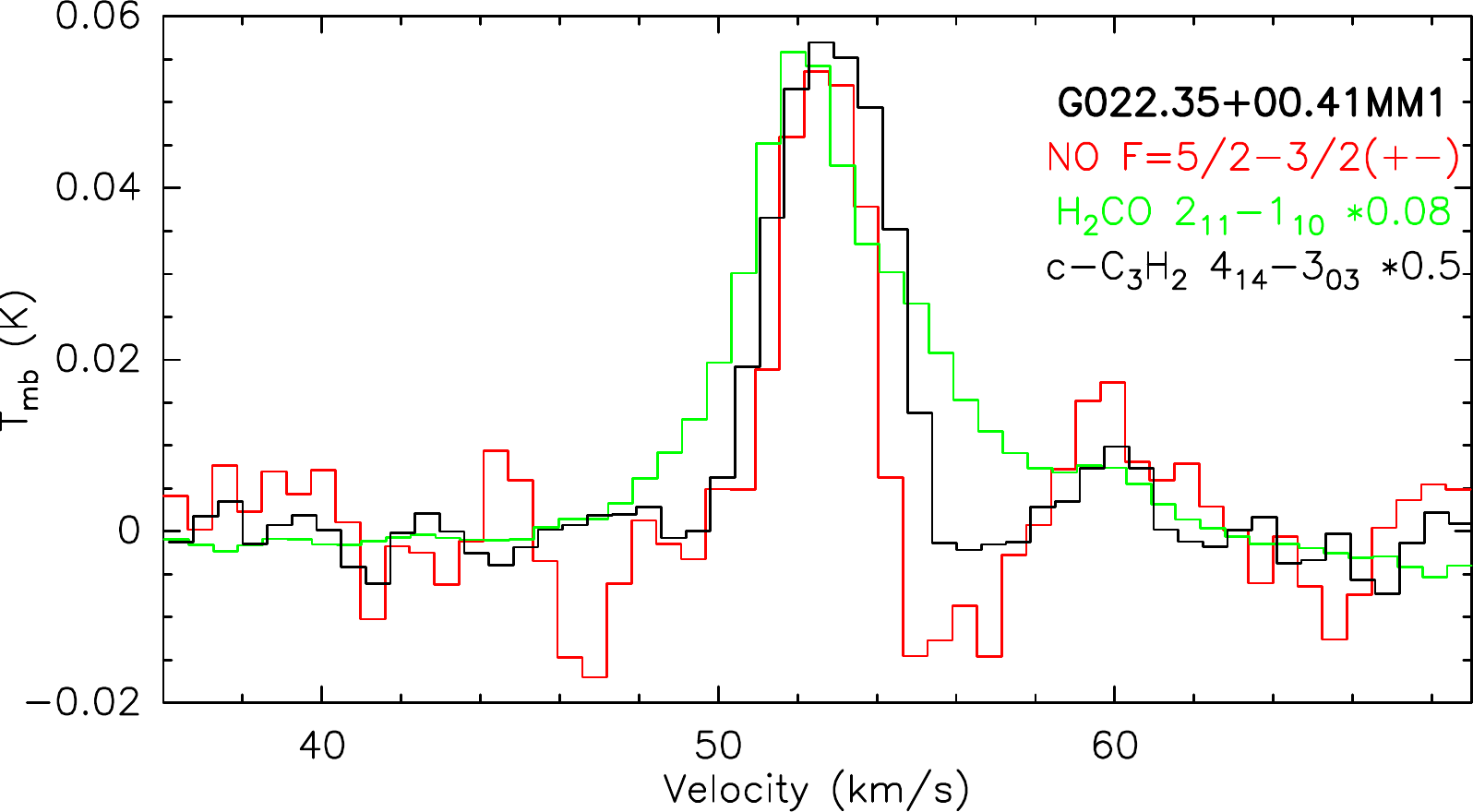}
\includegraphics[width=0.33\textwidth]{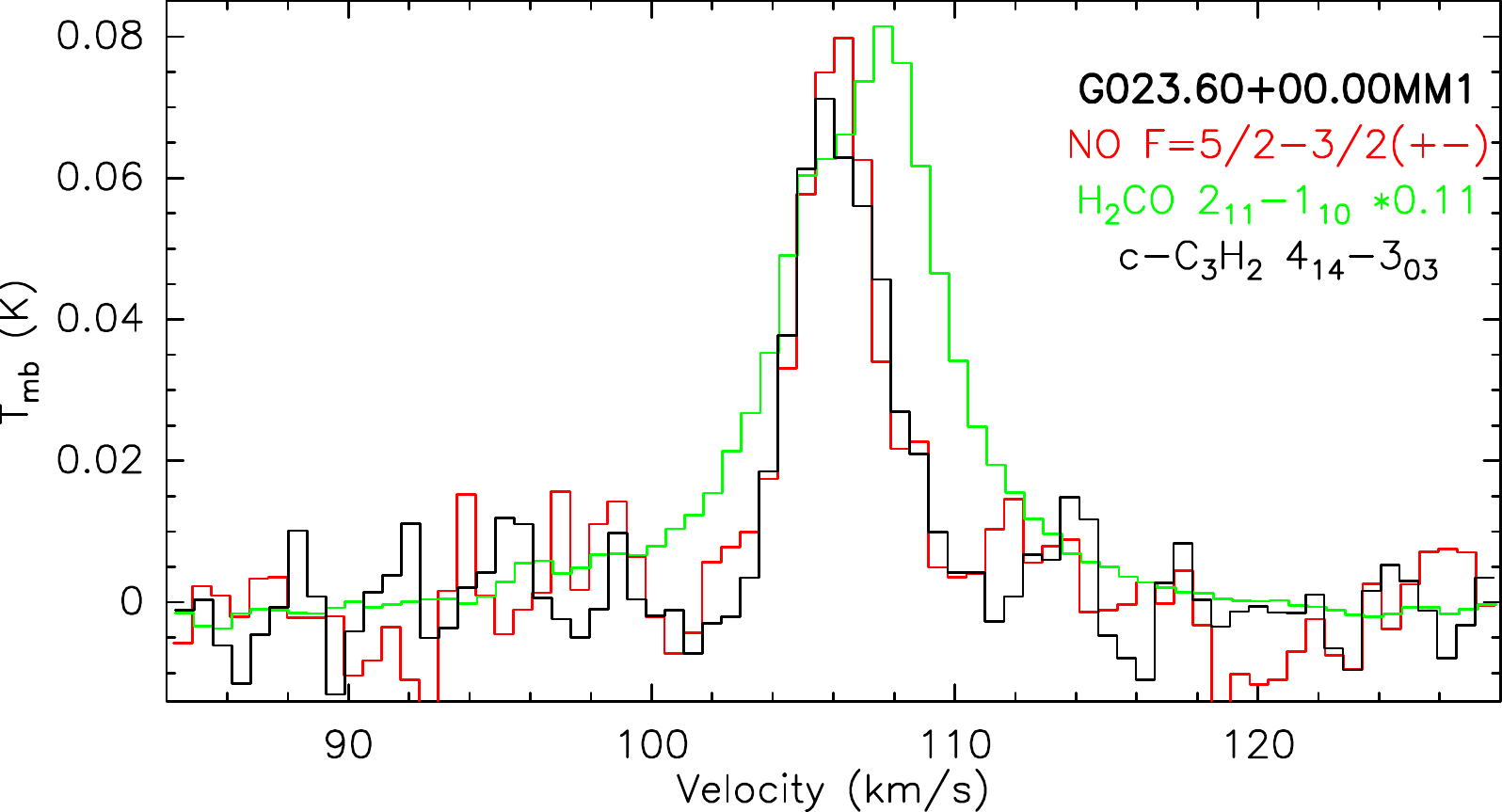}
\includegraphics[width=0.33\textwidth]{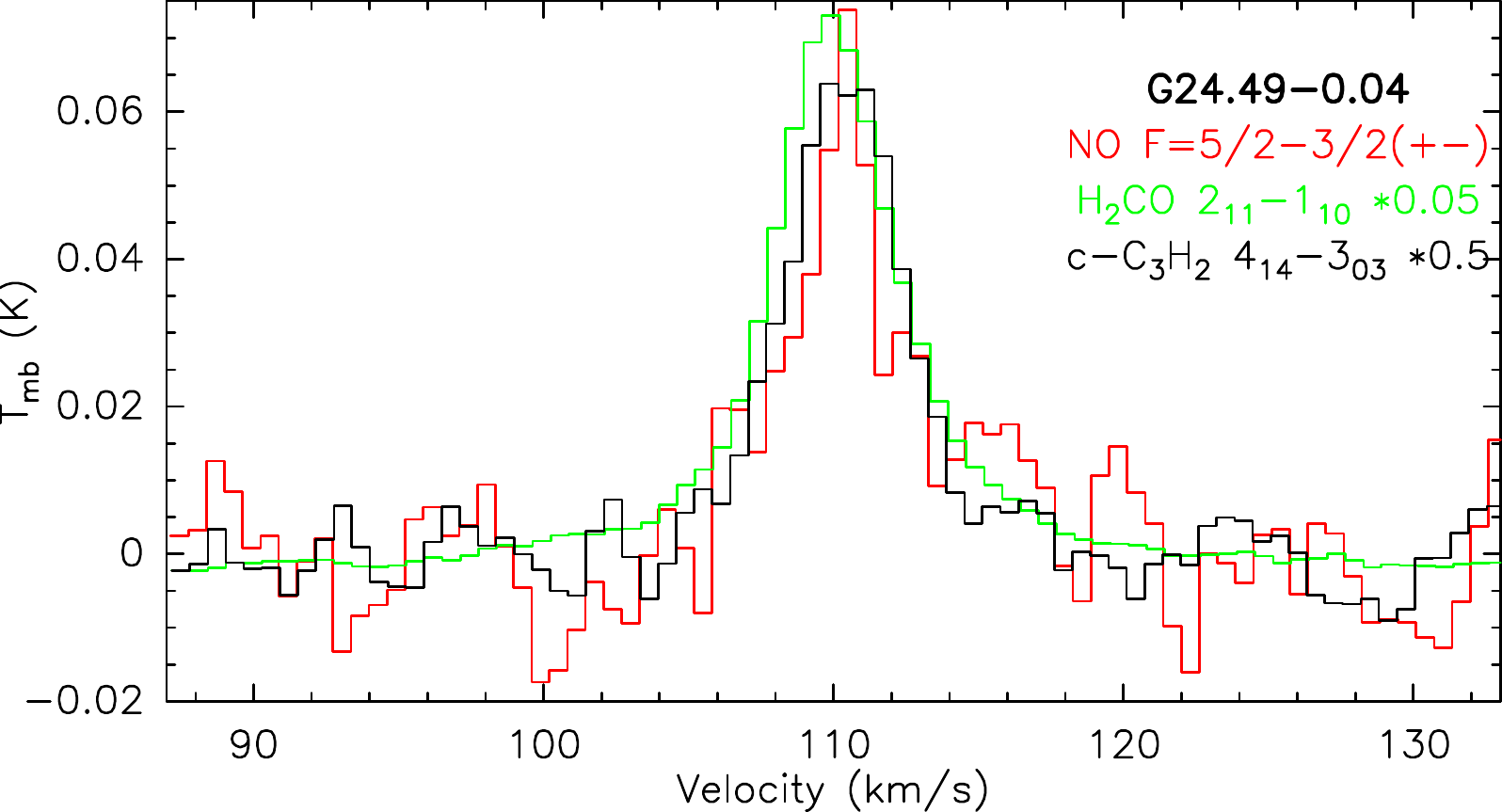}
\includegraphics[width=0.33\textwidth]{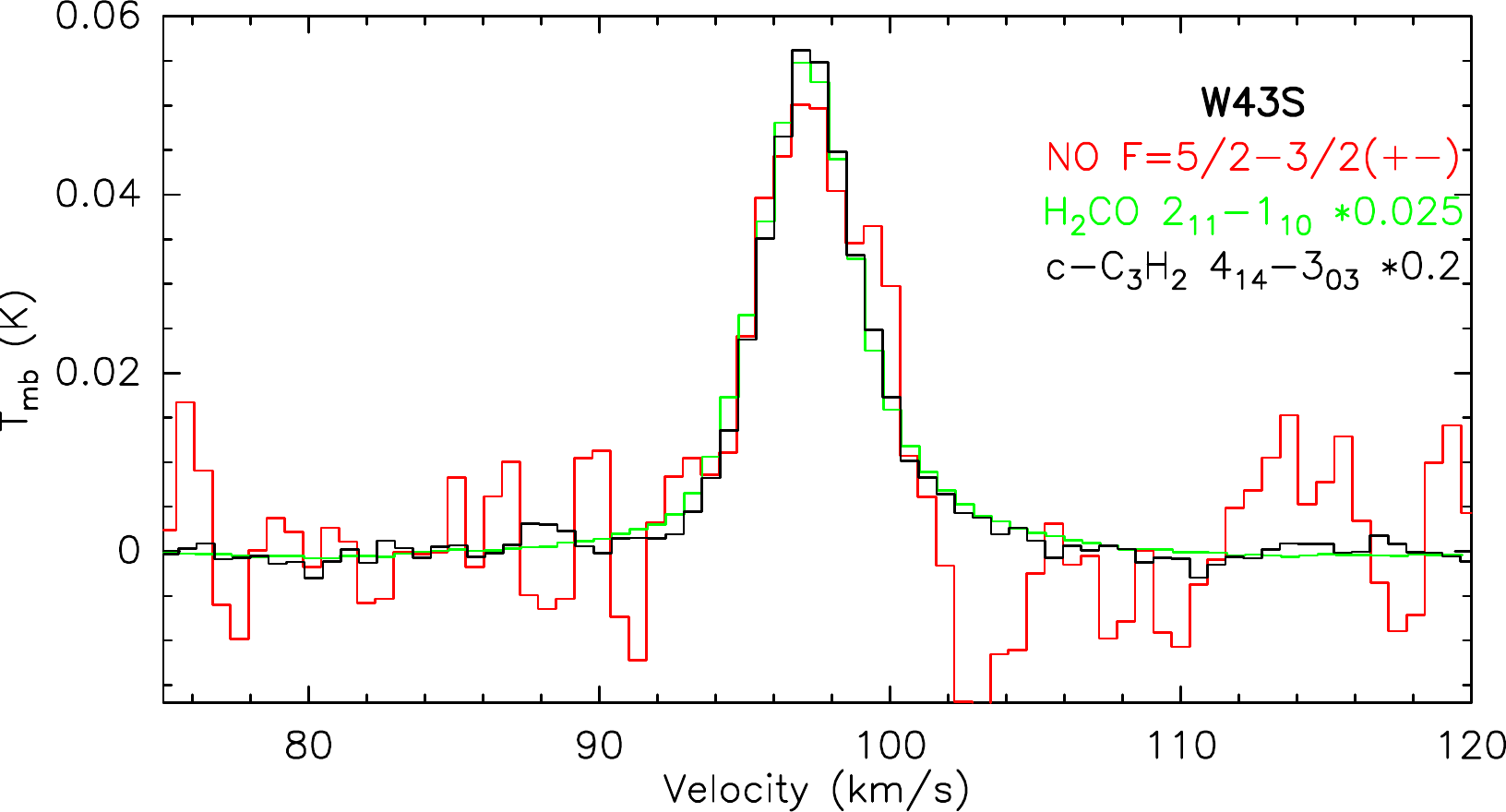}
\includegraphics[width=0.33\textwidth]{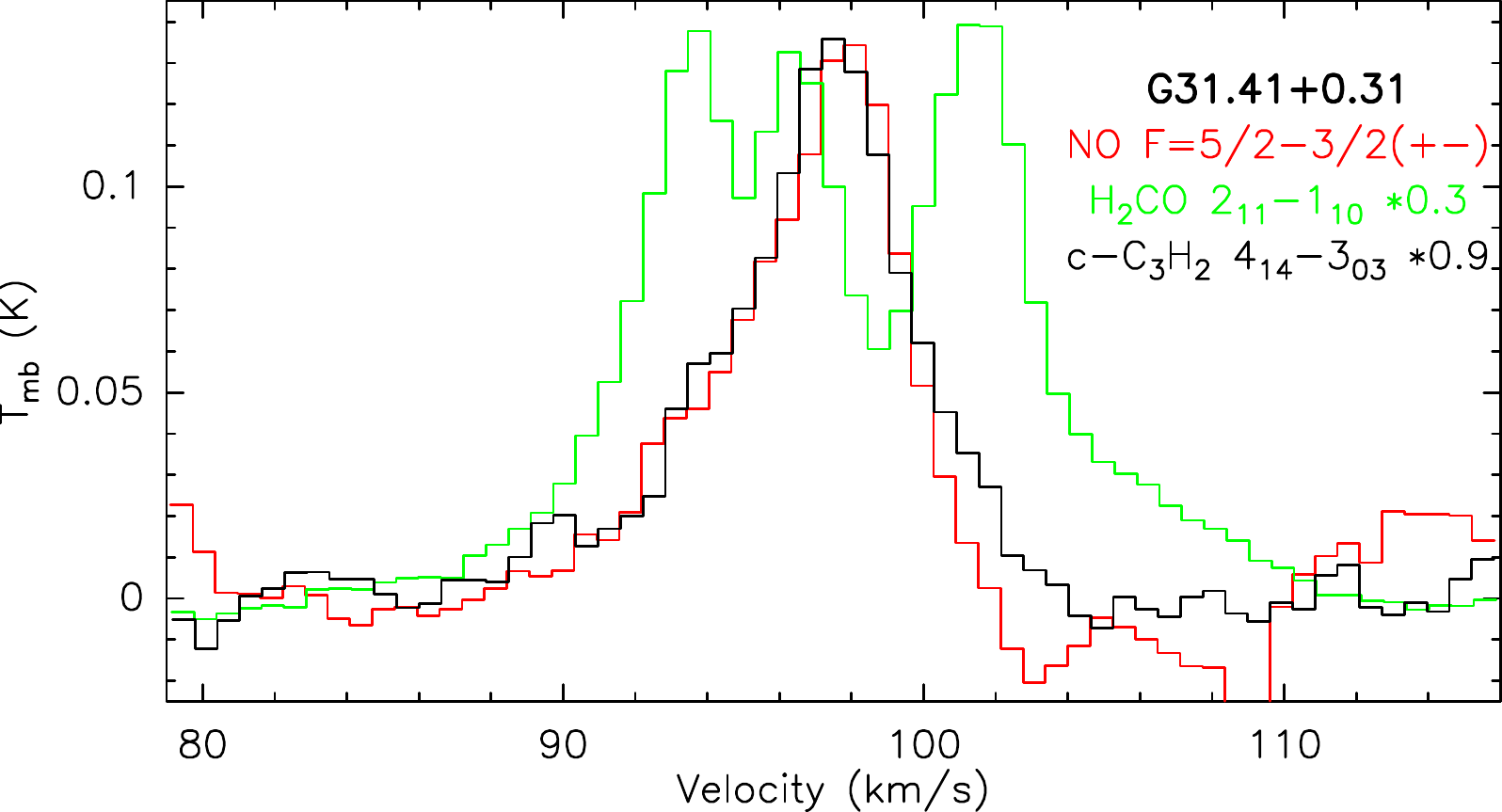}
\includegraphics[width=0.33\textwidth]{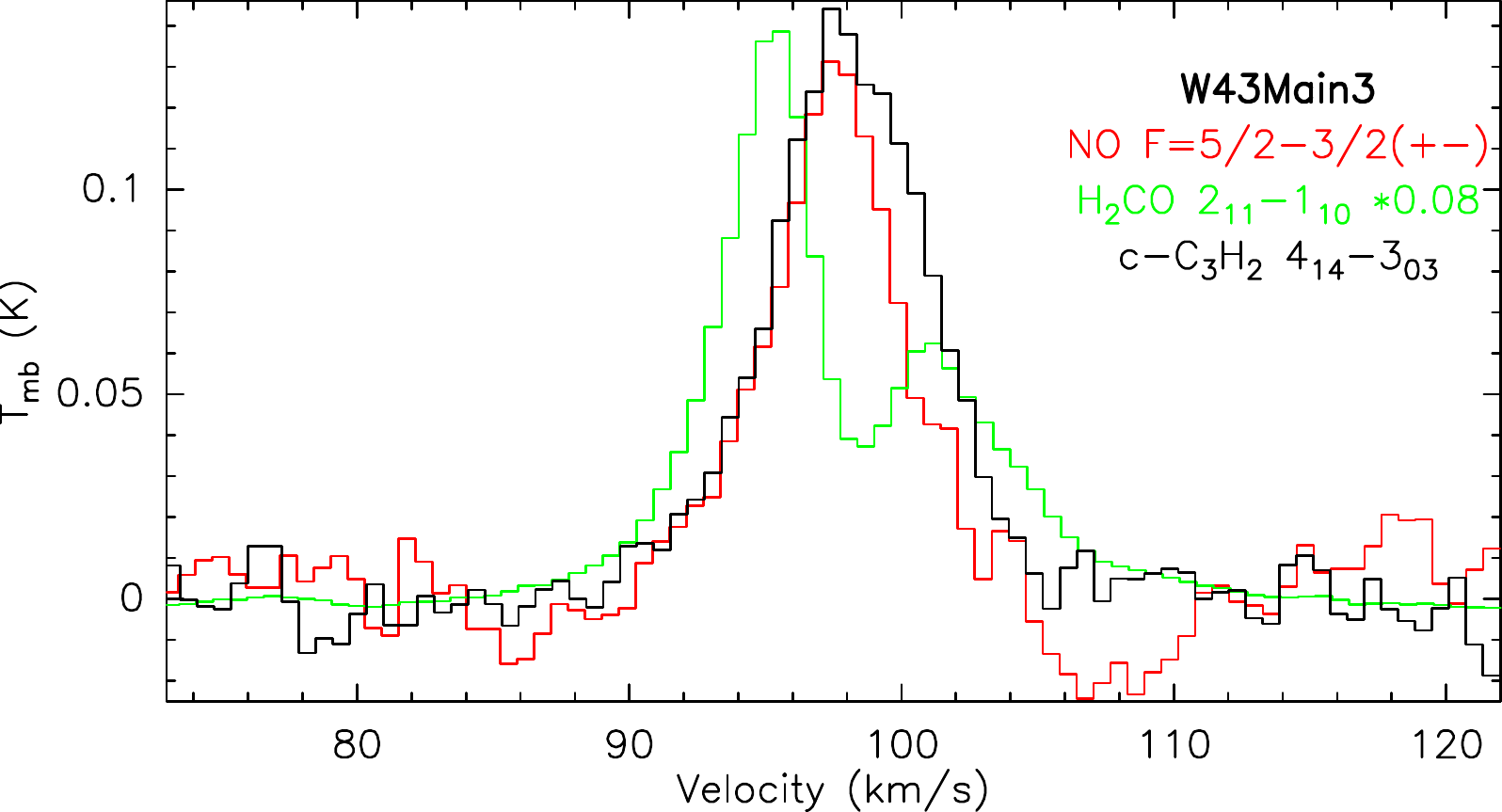}
\includegraphics[width=0.33\textwidth]{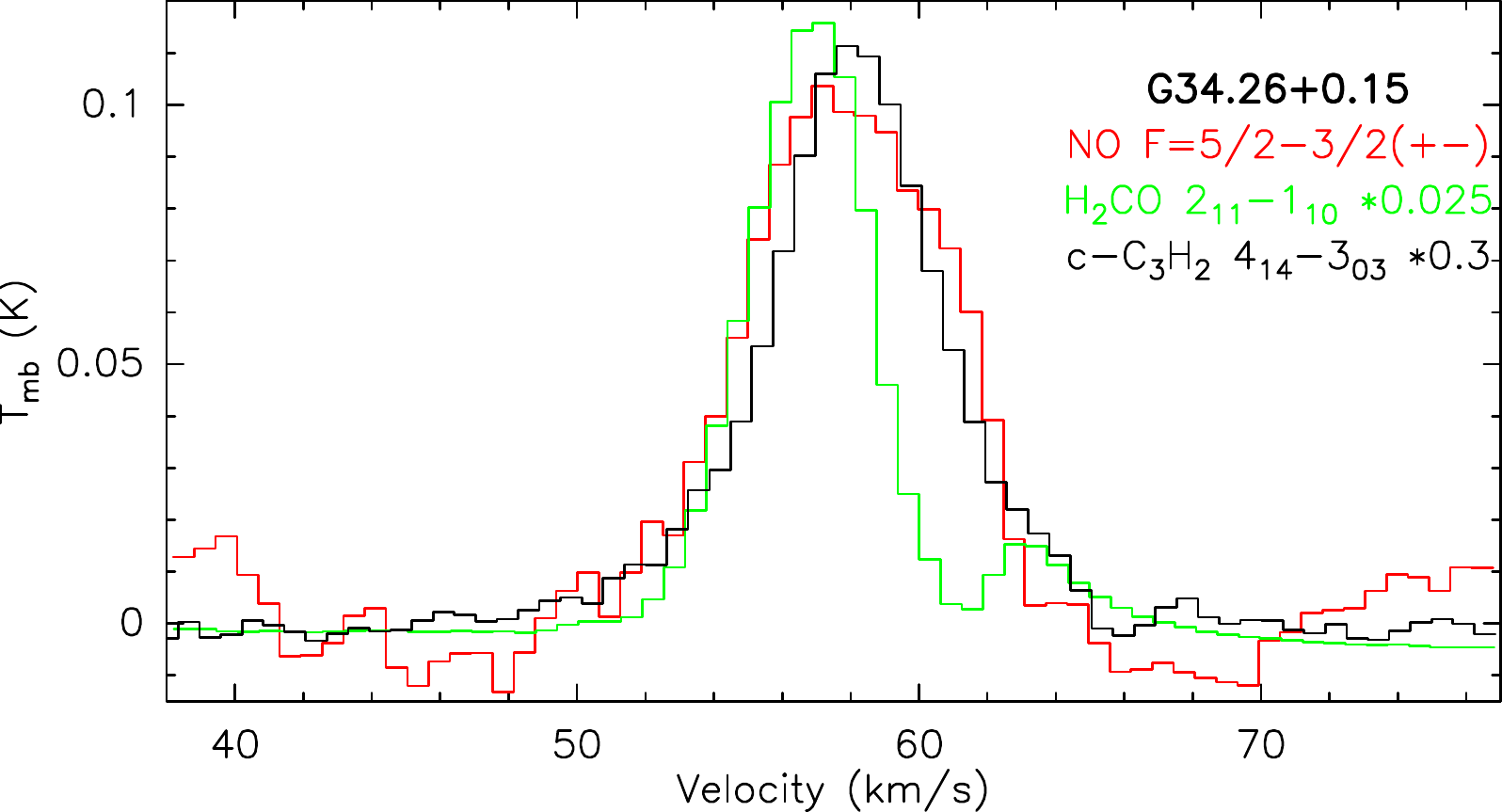}
\includegraphics[width=0.33\textwidth]{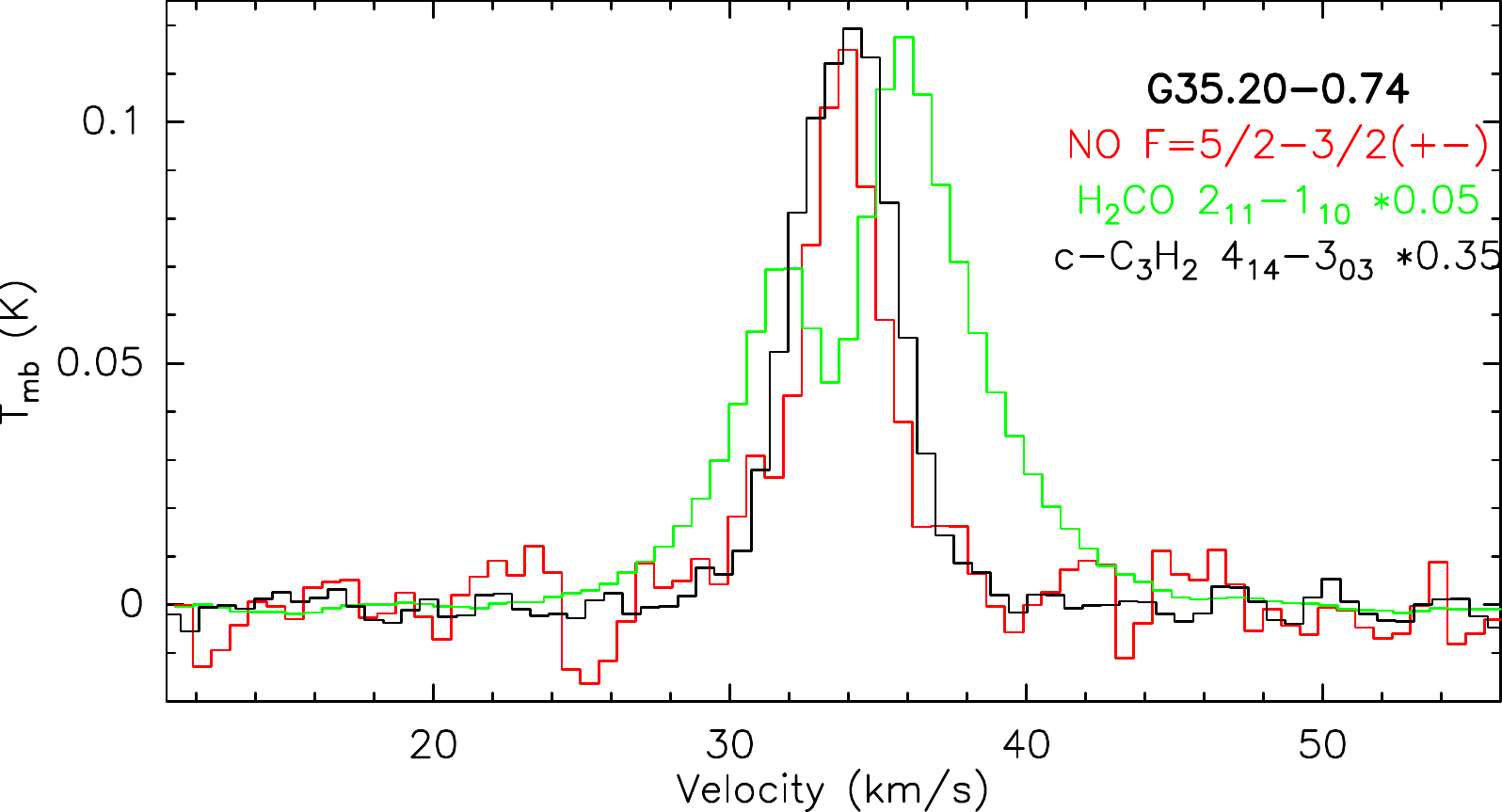}
\end{figure}

\onecolumn
\begin{figure}[!h]
\centering
\includegraphics[width=0.33\textwidth]{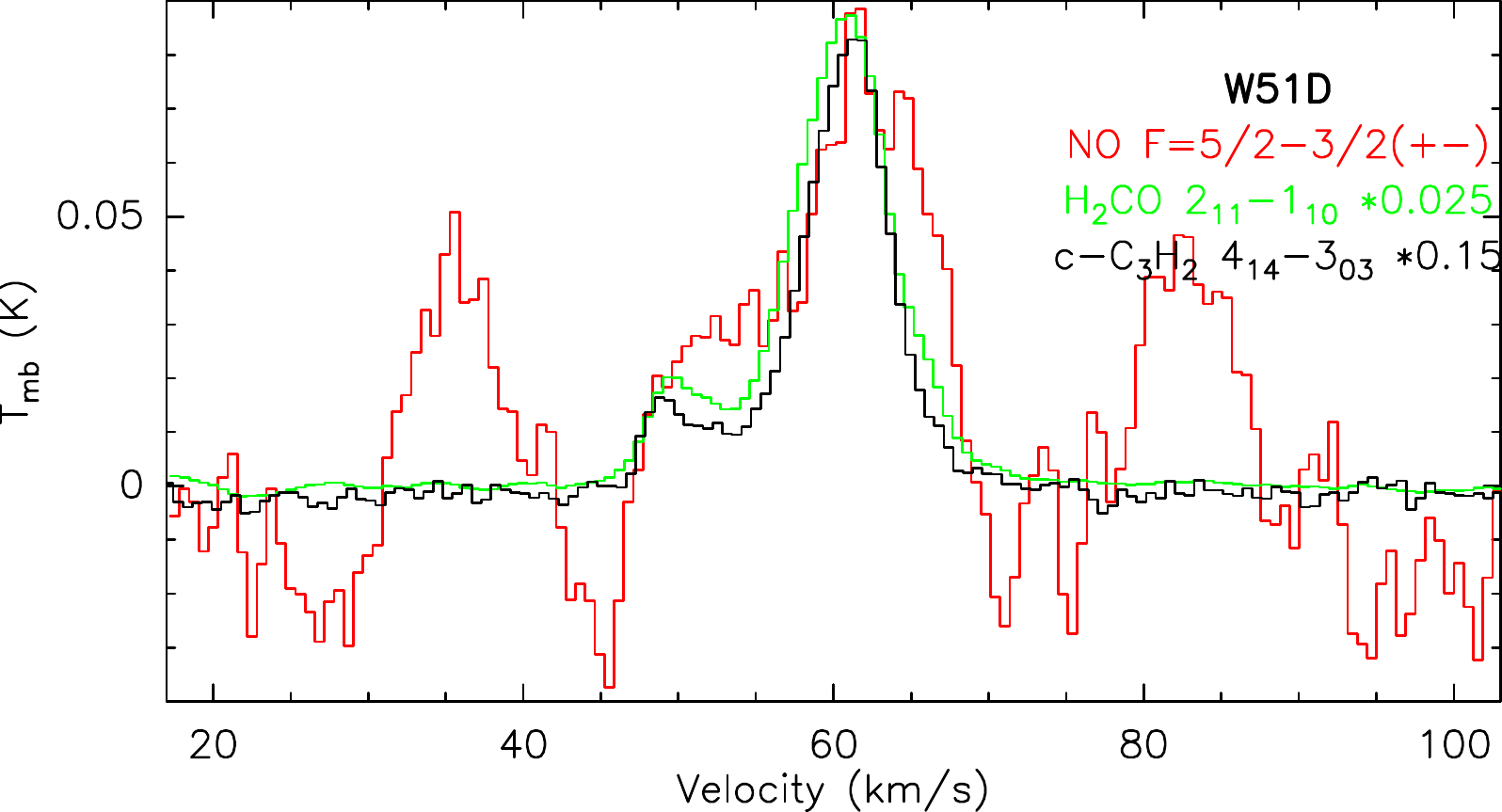}
\includegraphics[width=0.33\textwidth]{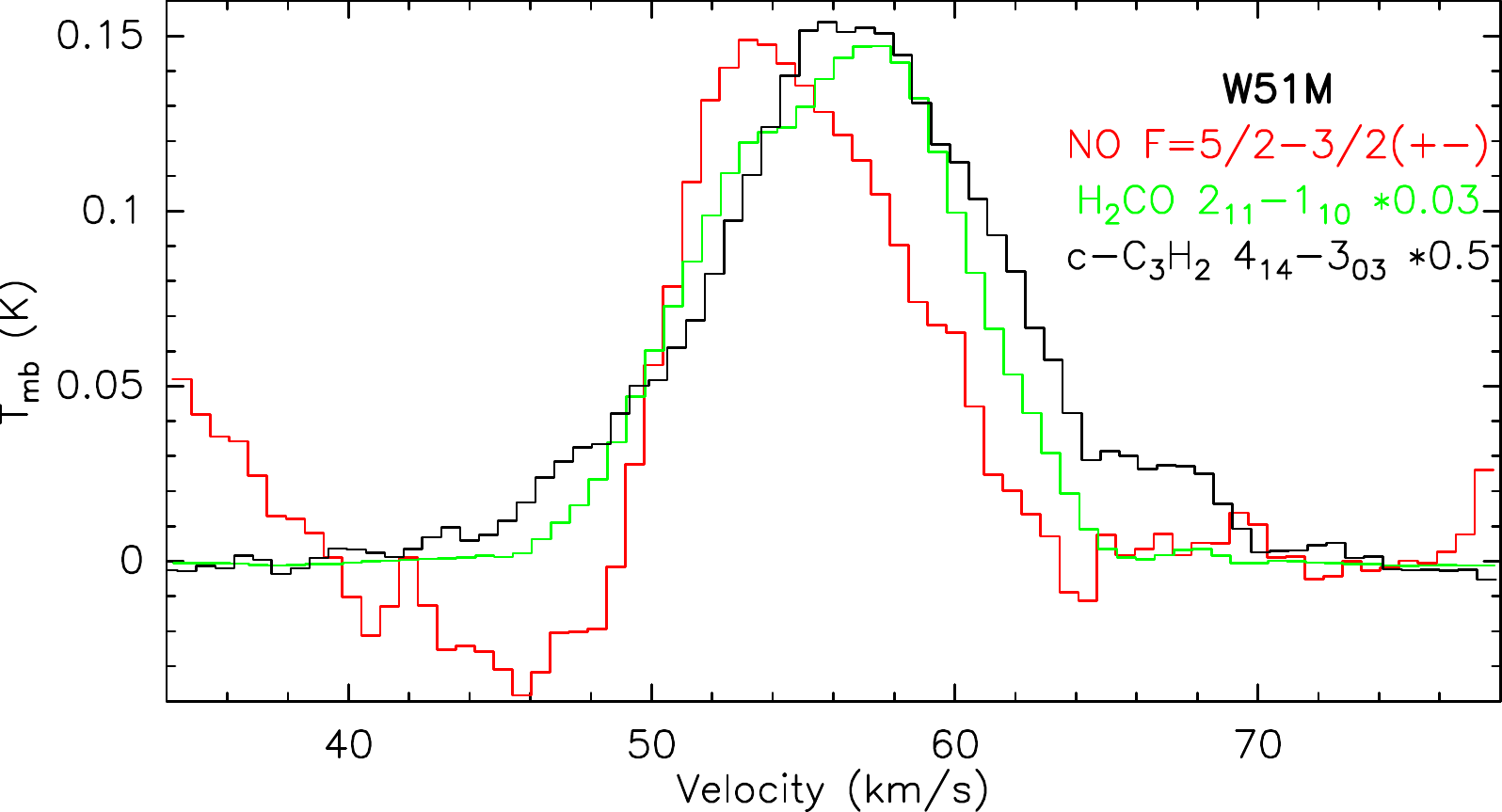}
\includegraphics[width=0.33\textwidth]{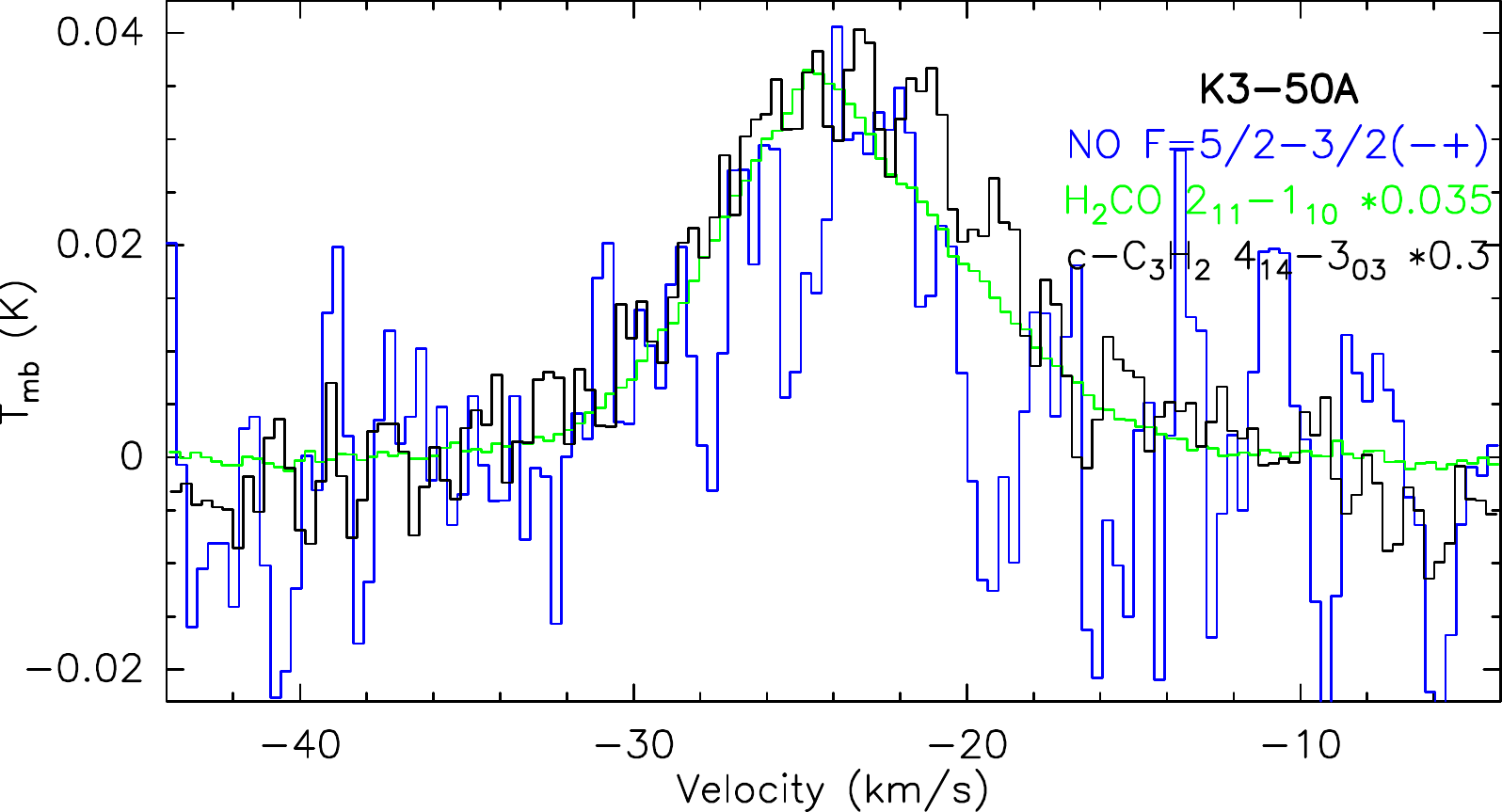}
\includegraphics[width=0.33\textwidth]{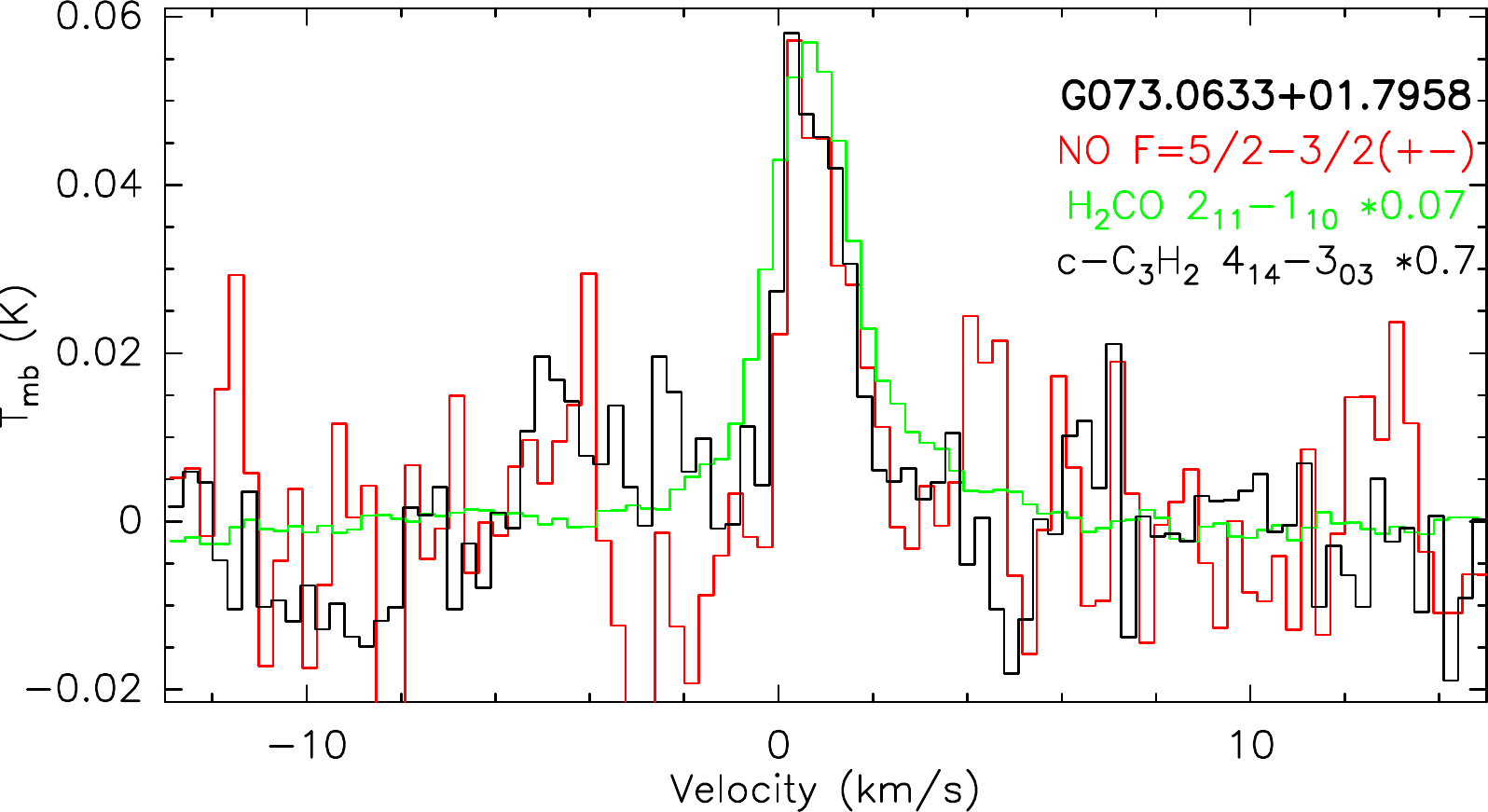}
\includegraphics[width=0.33\textwidth]{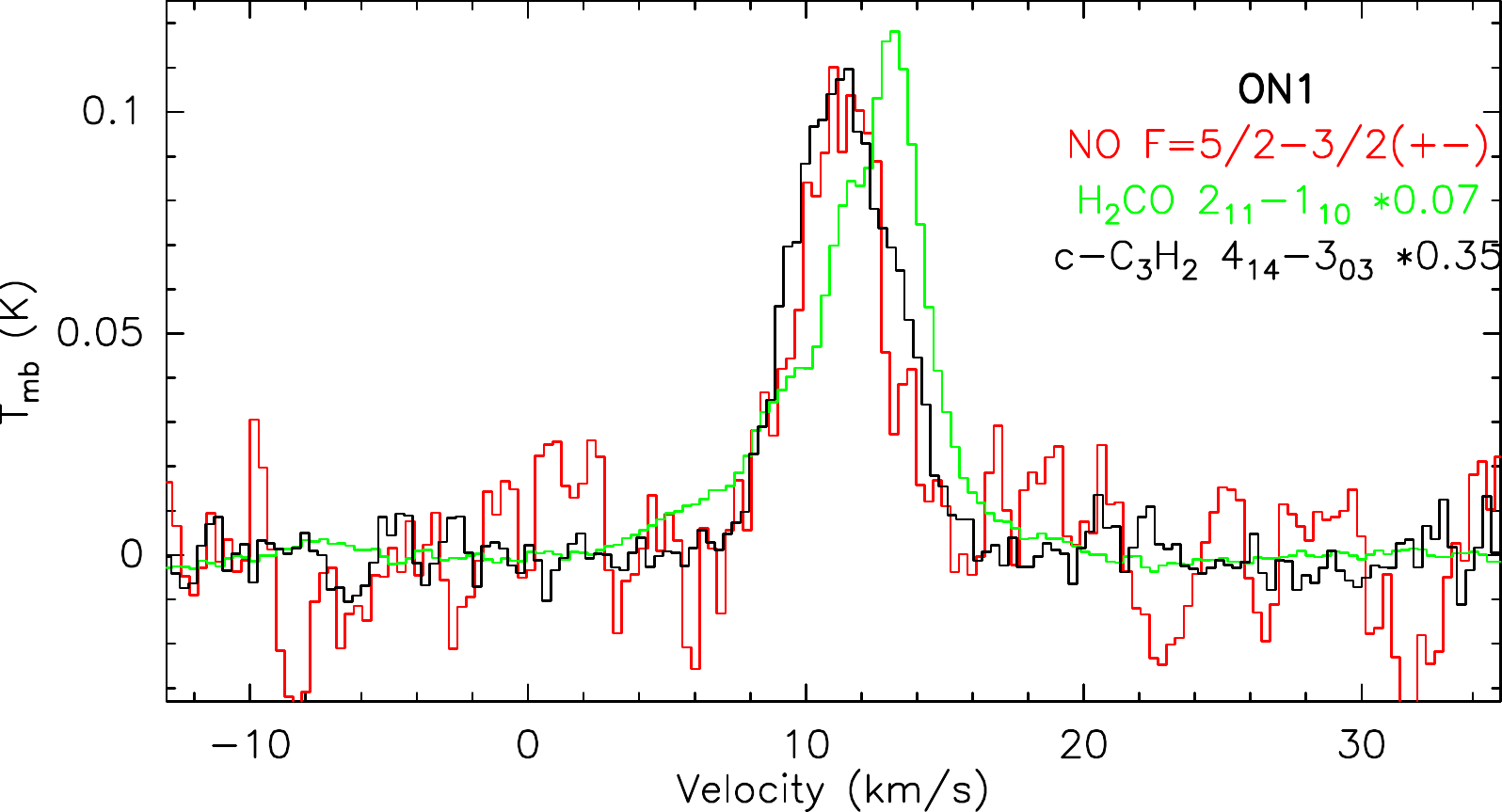}
\includegraphics[width=0.33\textwidth]{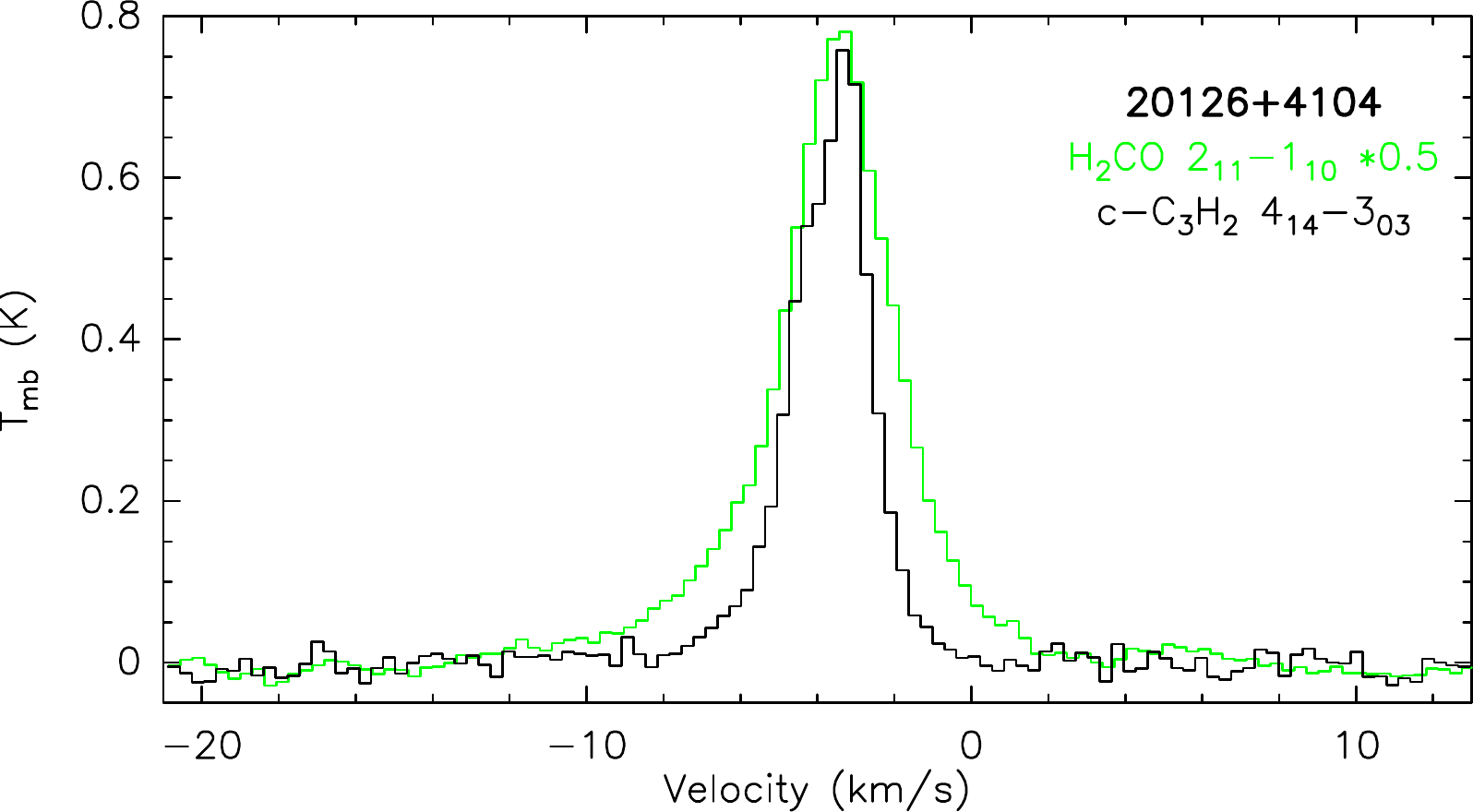}
\includegraphics[width=0.33\textwidth]{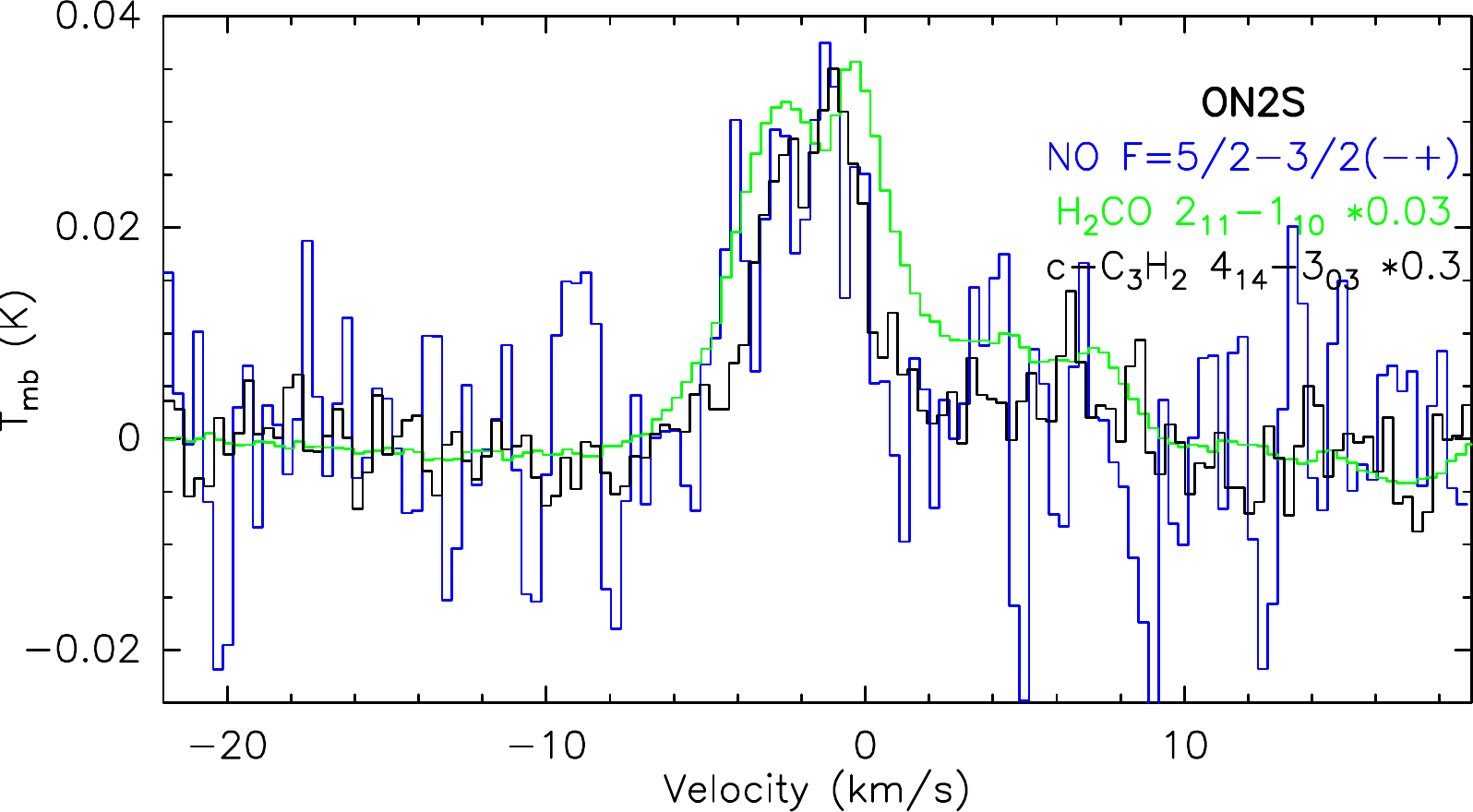}
\includegraphics[width=0.33\textwidth]{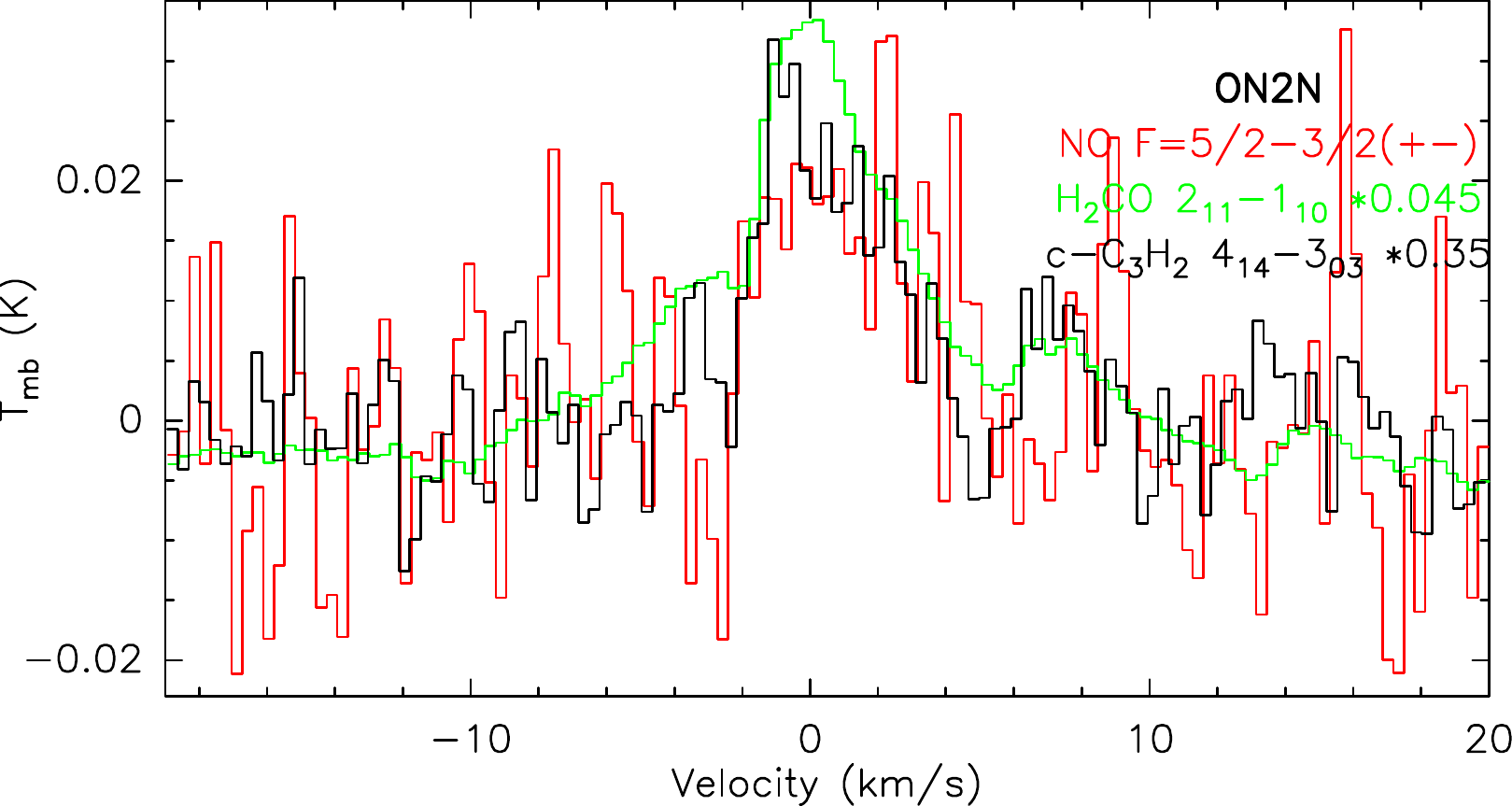}
\includegraphics[width=0.33\textwidth]{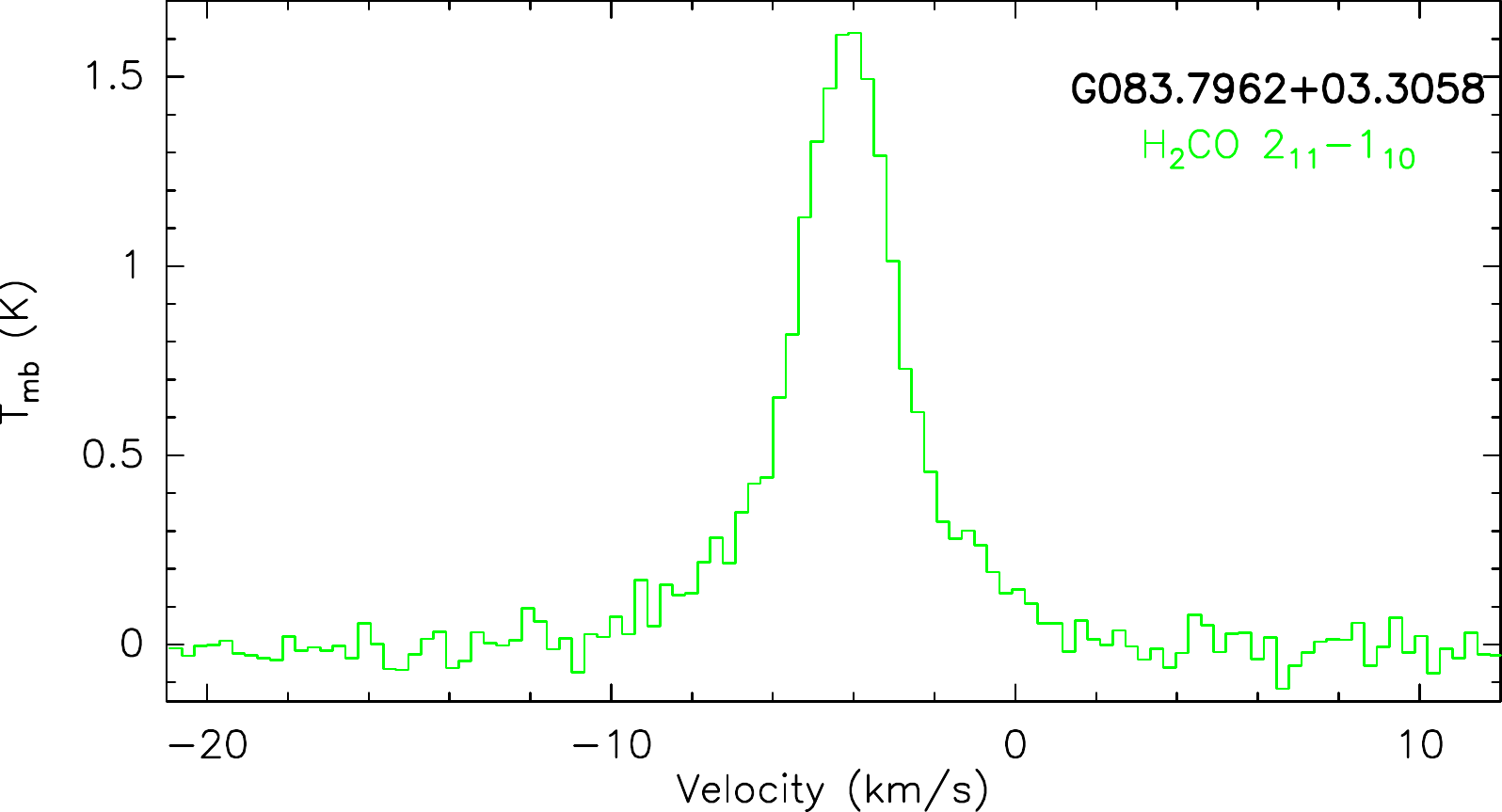}
\includegraphics[width=0.33\textwidth]{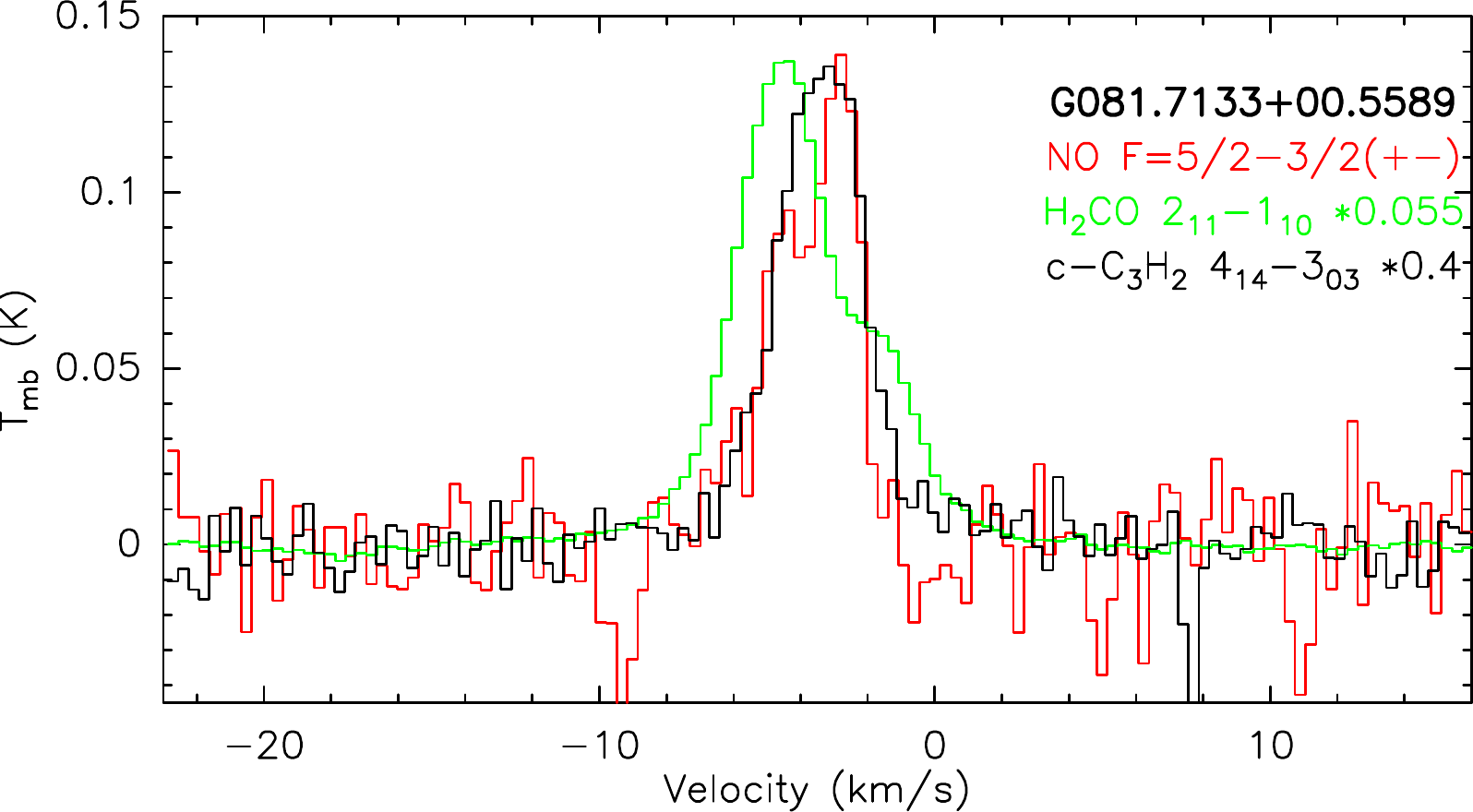}
\includegraphics[width=0.33\textwidth]{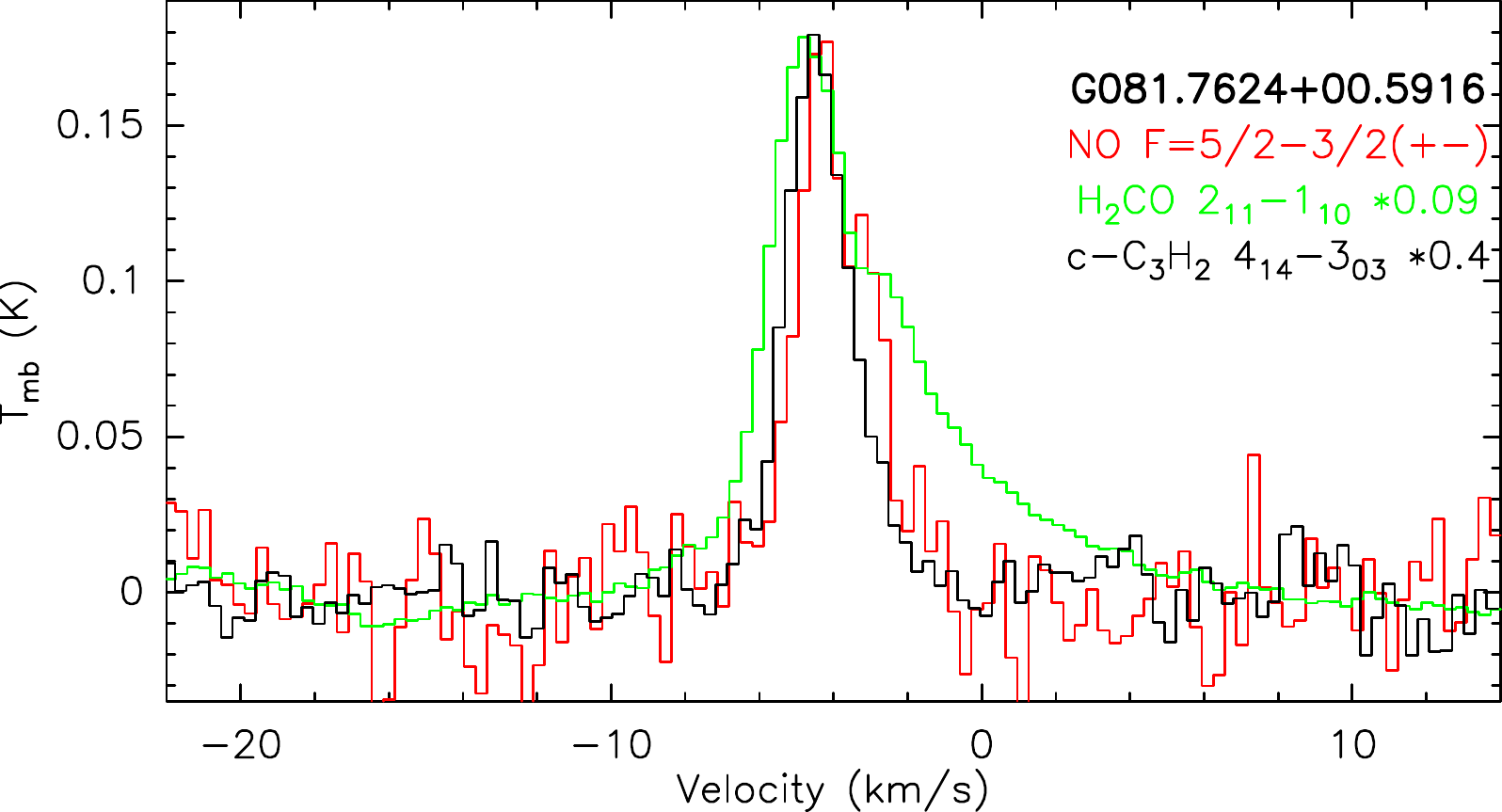}
\includegraphics[width=0.33\textwidth]{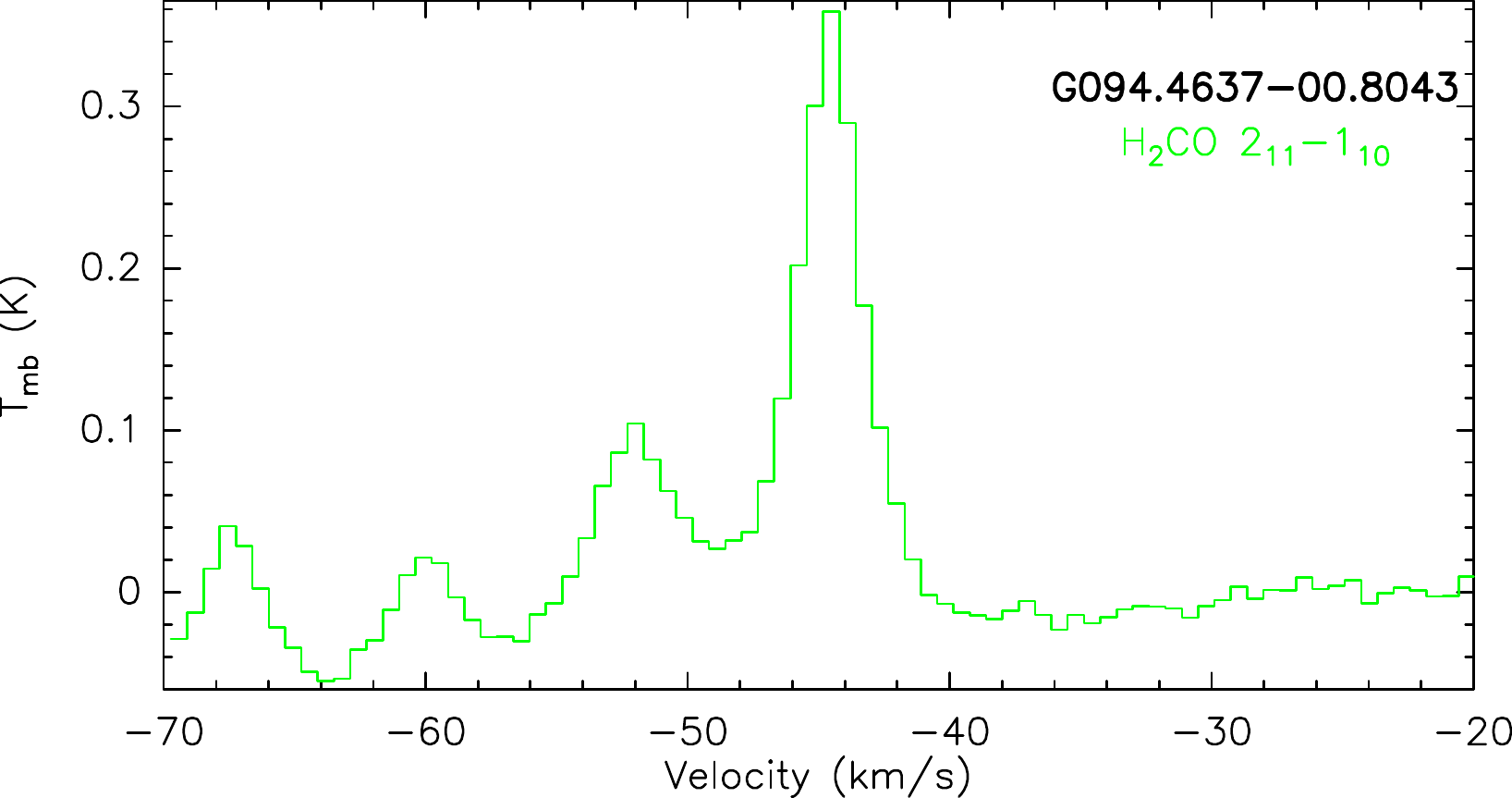}
\includegraphics[width=0.33\textwidth]{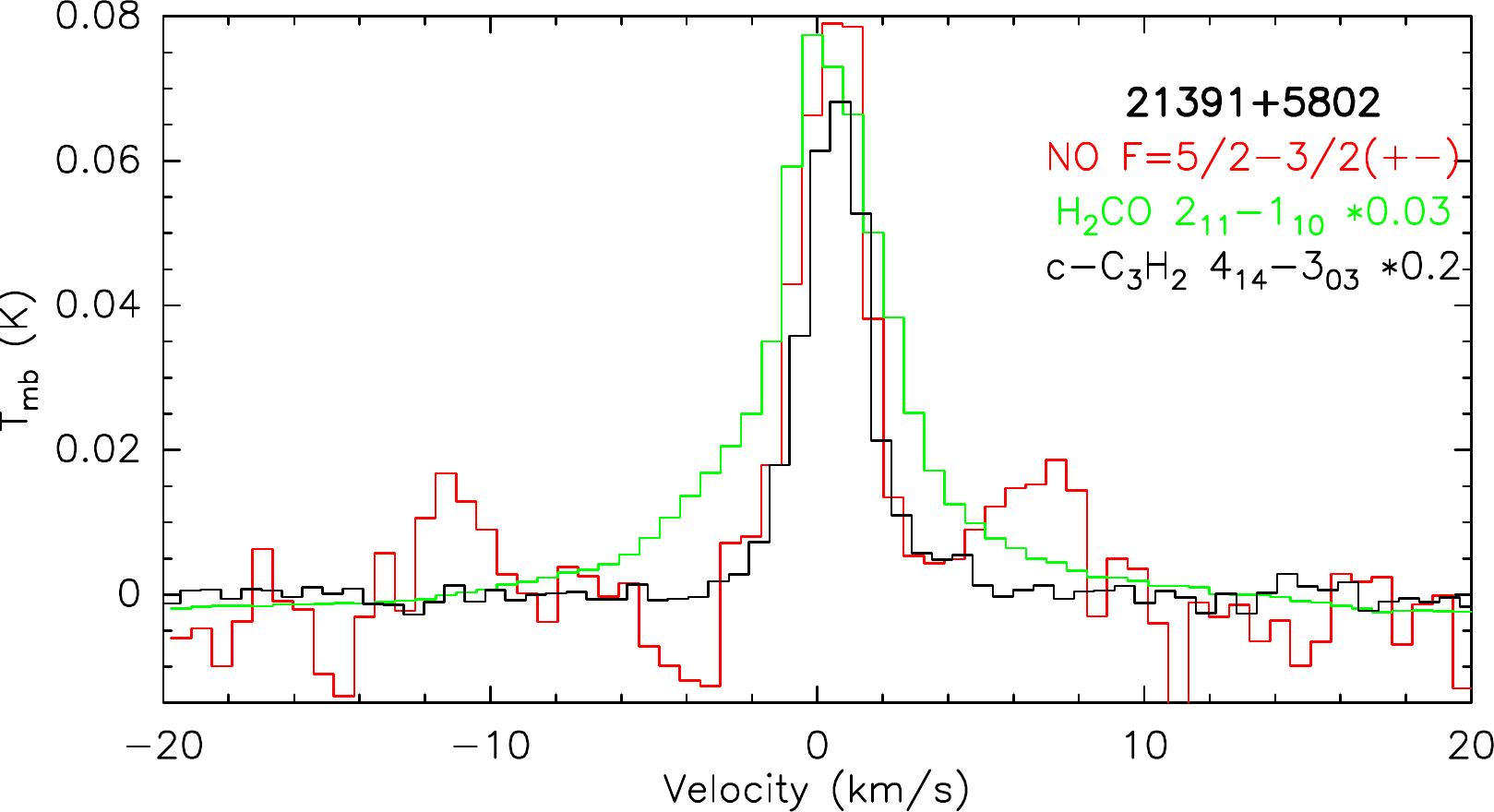}
\includegraphics[width=0.33\textwidth]{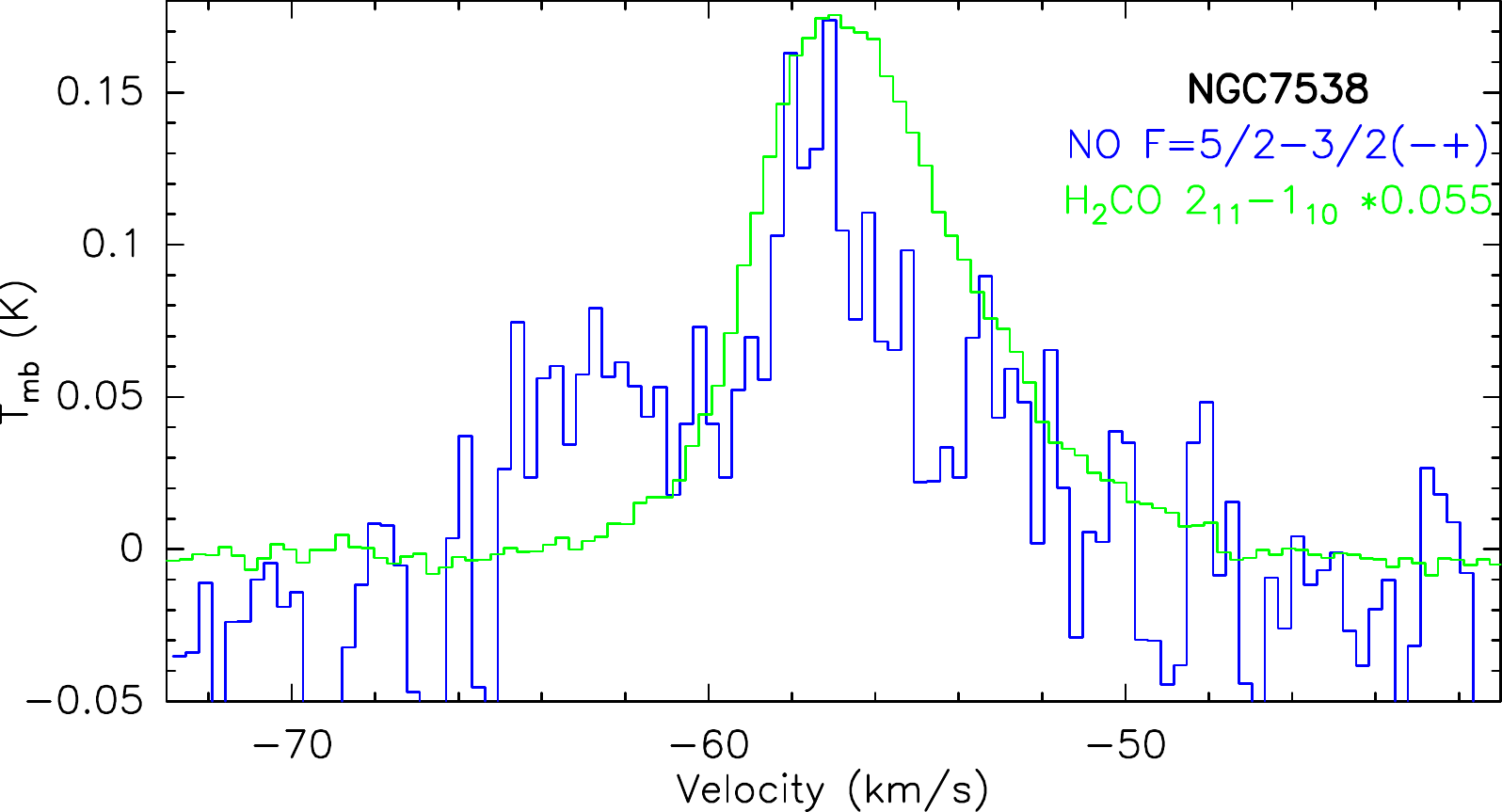}
\includegraphics[width=0.33\textwidth]{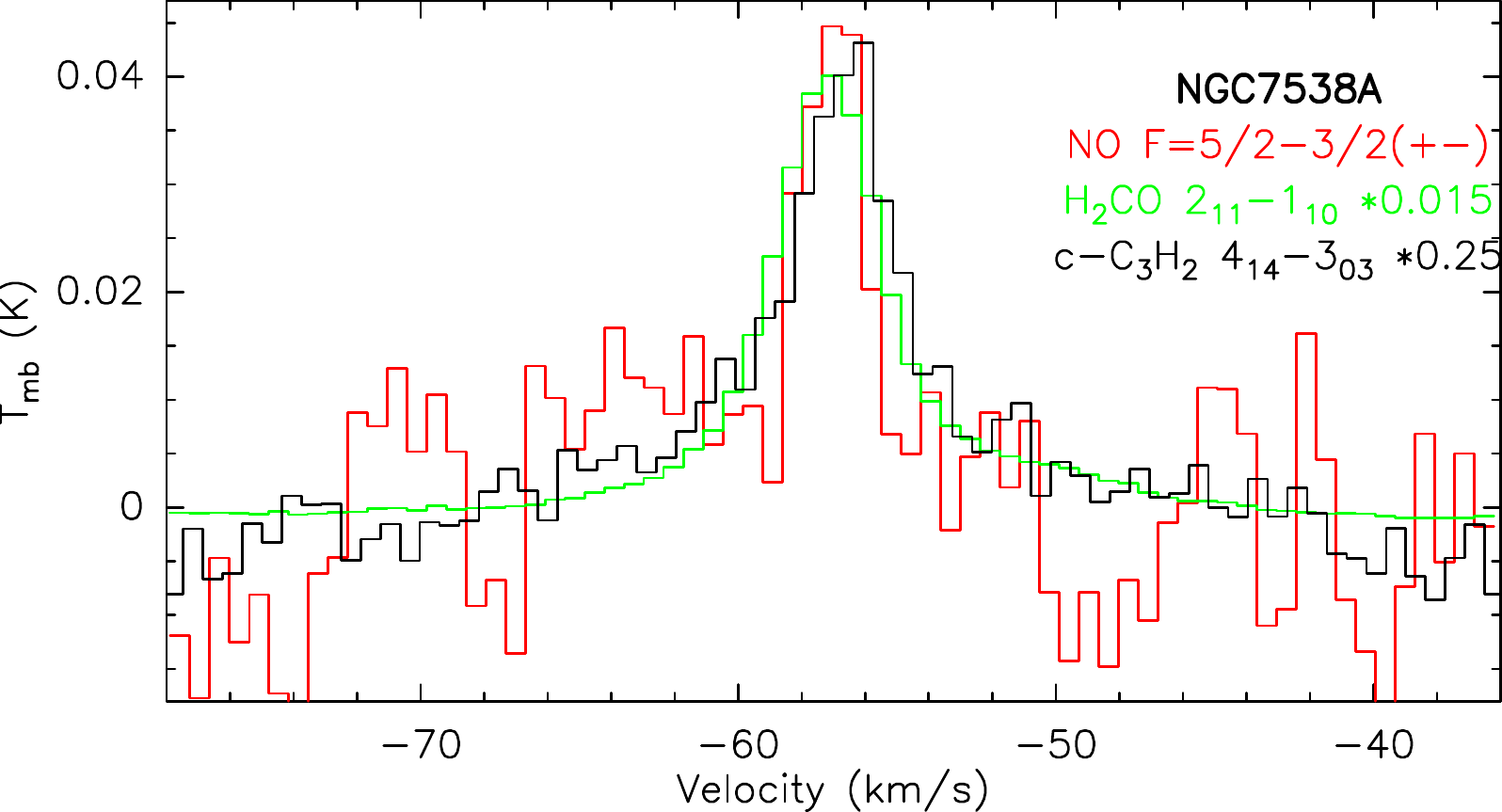}
\caption{Detected NO, $c$-C$_3$H$_2$ and H$_2$CO lines.}
\label{spectra of 3 molecules}
\end{figure}


\end{document}